\documentstyle[epsfig,12pt]{article}
\newcommand\beq{\begin{equation}}
\newcommand\eeq{\end{equation}}
\newcommand\bea{\begin{eqnarray}}
\newcommand\eea{\end{eqnarray}} 
\newcommand\jp{J^\prime }

\begin{document}

\begin{center}
{\Large Exact and Approximate Theoretical Techniques for Quantum Magnetism 
in Low Dimensions}
\end{center}

\vskip 1.0 true cm
\centerline{\bf Swapan K. Pati$^1$, S. Ramasesha$^2$, Diptiman Sen$^3$}
\vskip .5 true cm

\centerline{\it $^1$ Jawaharlal Nehru Centre for Advanced Scientific 
Research,}
\centerline{\it Jakkur, Bangalore 560064, India} 

\centerline{\it $^2$ Solid State and Structural Chemistry Unit,} 
\centerline{\it Indian Institute of Science, Bangalore 560012, India} 

\centerline{\it $^3$ Centre for Theoretical Studies,}
\centerline{\it Indian Institute of Science, Bangalore 560012, India} 
\vskip 1.5 true cm

\begin{abstract}

Quantum magnetism in low dimensions has been one of the central areas of
theoretical research for many decades now. One of the key reasons for
the long standing interest in this field has been the existence of
simplified models, which serve as paradigms for understanding the role
of strong interactions in many-electron systems. Although simple, 
these models quite often can not be solved exactly. In this review, we
discuss a variety of analytical and numerical methods, which treat
the system in a systematic and controlled manner. The central method
employed in all the studies is the density matrix renormalization group
(DMRG) method. This is supported by small scale numerical exact calculations,
and by analytical methods using field theoretic techniques. We have
considered a number of magnetic systems, from magnetic clusters to extended 
lattices, and have found some novel quantum ground states and low-energy 
elementary excitations. In some cases, we have also employed the 
finite-temperature DMRG method to accurately compute the low-temperature 
thermodynamic properties such as specific heat and magnetic susceptibility. 

\end{abstract}

\newpage
\clearpage

\section{\bf Introduction}

The nonrelativistic Schr\"odinger equation of a system of electrons
is spin independent. It therefore appears at first glance
that the solutions of the Schr\"odinger equation should also be
spin independent. However, the indistinguishability of the electrons
forces the total wave function which is a product of the spin wave function
and the spatial wave function to be antisymmetric. This in turn implies
that for two electrons, a spatially symmetric wave function should be 
associated with an antisymmetric spin wave function and {\it vice versa}.
The different charge distribution in the spatially symmetric and 
antisymmetric wave functions leads to different coulomb repulsions by virtue
of which the spin states which are symmetric and antisymmetric have 
different energies \cite{landau}. Dirac represented the splitting between 
the energies
of the two spin states by the spin operator $-2J {\hat {\bf S}}_i \cdot 
{\hat {\bf S}}_j$, where $J$ is the exchange integral involving the two 
spatial orbitals in which the two electrons are singly occupied. In most open
shell atomic systems the exchange integral $J$ is large enough to force
the total spin of the ground state configuration to be the largest
permissible value. This in essence is the Hund's rule of maximum 
multiplicity and is also the reason why we find transition 
and rare earth metal ions in the high spin states in nature.

In solids containing transition metal or rare earth ions surrounded by
ligands, the relative alignment of the unpaired spins at the metal site
is not at all obvious. In order to understand this, we should examine
the possible pathways for the delocalization of the valence electrons
in the system. If the favorable delocalization pathways involve 
antiparallel alignment of the metal ion spins then the nature of the
exchange interaction between the metal ion spins is antiferromagnetic
and ferromagnetic otherwise. This follows from the fact that 
delocalization of electrons lowers kinetic energy of the electrons and
therefore the ground state corresponds to an alignment of spins that
allows for maximum delocalization. This is indeed the reason why the
ground state of a hydrogen molecule is a spin singlet. In a system with
degenerate partially occupied orbitals, the Hund's coupling favors
high-spin alignment of the electrons on an ion. If delocalization 
pathways exist that allow for these high-spin states in the process of
delocalization, then the alignment of the spins on two neighboring
centers will be ferromagnetic. If delocalization pathways exist
only when these high-spin states are aligned antiparallel, one would have 
an antiferromagnetic alignment of the spin. Thus, the overall nature of 
the spin alignment is governed by a competition between the Hund's coupling
and electron delocalization \cite{mattis,sinha}.

Given a collection of spins, the exchange Hamiltonian for the system is
written as
\beq
H~=~\sum_{ij} J_{ij} {\hat {\bf S}}_i \cdot {\hat {\bf S}}_j ~,
\eeq
where $J_{ij}$ is the effective exchange integral for the interaction
between the spins at sites 'i' and 'j'. Since the spins in the cluster
arise from unpaired electrons of a transition metal ion in a crystal field,
it is natural to expect that the spin-orbit as well as spin-spin interactions
of the electrons in the ion could alter the nature of the total spin on the 
ion by giving the net spin a preferred direction of orientation. Such a 
situation can be easily handled by treating each $J_{ij}$ as a vector
and generalizing the exchange Hamiltonian as
\beq
H~=~\sum_{ij} ~( ~J_{ij}^x {\hat S_i^x}{\hat S_j^x}~+~
J_{ij}^y {\hat S_i^y}{\hat S_j^y}~+~J_{ij}^z {\hat S_i^z}{\hat S_j^z} ~) ~.
\eeq
Such a model is often referred to as the XYZ spin model \cite{mattis}. 
Two extreme cases
are often studied, (i) the spin is assumed to have no projection on the X-Y 
plane, in which case the resulting model is the Ising model and corresponds
to scalar spins, and (ii) the spin is assumed to have no projection on the 
z-axis, in which case we have an XY model or a planar spin model. The Ising
model is a discrete classical model as it consists of no noncommuting 
operators in its Hamiltonian while the XY model could be classical or quantum 
mechanical. Usually, while dealing with large site spin systems, it is not 
uncommon to assume that the spins are classical, in the spirit of Bohr's 
correspondence principle.

In the crystalline state, the spins in the solid would be arranged on a 
lattice. If the exchange interaction is predominant between spins 
along a single crystalline direction, the model could be treated as a one
dimensional array of spins. There are many examples of solids in which
this holds \cite{bray}. 
Likewise, it is also possible that the interactions amongst
spins is large along two crystallographic directions and weak along a third
direction and this would result in two-dimensional spin system \cite{rado}. 

In this review article, we will mainly concern ourselves with the study of 
isotropic spin clusters and one-dimensional spin systems, sometimes in the
presence of an external magnetic field. We will be mostly interested in 
properties of the ground state and the low-lying excitations, since these are
the states which govern the low-temperature properties of systems such
as the specific heat and magnetic susceptibility. The 
symmetries of a system often enable us to characterize the energy eigenstates 
in terms of quantum numbers such as the total spin $S_{tot}$, the component 
of the total spin along some particular direction, say, $S_{tot,z}$, spin 
parity (which is a symmetry for states with $S_{tot,z} = 0$), the wave 
number $k$ for a translation invariant system, and possibly other spatial
symmetries depending on the structure. We will discuss below how the use of 
symmetries can help to lessen the numerical effort required to study the 
low-energy states.

In the next two sections we introduce some exact numerical
methods, and describe the application of these methods to magnetic
clusters. In section 4, we discuss two analytical methods which 
use field theoretic approximations. In section 5, we describe an innovative 
way of solving the spin Hamiltonians, which goes beyond the conventional 
techniques and is based on the density matrix renormalization group (DMRG) 
theory. Various applications of DMRG to the properties of 
low-dimensional extended chains are described in sections 6 to 8.

\section{\bf Exact Calculations}

The properties of a spin Hamiltonian can be computed from the eigenstates 
of the Hamiltonian which are in turn obtained by setting up the Hamiltonian
matrix in a suitable basis and diagonalizing it thereafter. While the
procedure itself is quite straightforward, the space spanned by the 
Hamiltonian rapidly increases with the number of the spins in the system.
The Fock space dimensionality of a system of $n$ spins with spin $s_i$ is
given by
\beq
D_{F}~=~ \prod_{i=1}^n ~(2s_i+1) ~.
\eeq
The Hamiltonian matrix is block-diagonal in structure with each block
corresponding to specified values of the quantities conserved by the 
Hamiltonian. Thus, for an isotropic spin system, the $z$-component of the
total spin, $M_S$, and the total spin $S$ are conserved. Restricting the
Fock space to specified values of $M_S$ and $S$ gives Hilbert spaces
whose dimensionalities are smaller than the Fock space dimensionality. 

While constructing spin basis functions which are eigenstates of the total
${\hat S}^z$ operator is quite simple, construction of spin adapted functions
(SAF, eigenstates of the total ${\hat {\bf S}}^2$ operator) is not direct. 
Perhaps the simplest and chemically appealing way of constructing SAF is by 
the valence bond (VB) method which uses the Rumer-Pauling rules. This 
method is best illustrated by applying it
to a system of $2n$ spins, each possessing a spin
of half, in the total spin $S=0$ sector. A total spin singlet can be formed by
choosing pairs of sites and spin coupling each of them to obtain a singlet. 
The product of these singlet pairs will be a spin eigenstate of the operator 
${\hat {\bf S}}^2_{total}$. This is illustrated in Fig. 1. 
However, there are more ways to spin couple in pairs than the number of 
linearly independent singlet states, {\it e.g.}, the state $|3>$ in Fig. 
1 can be expressed as a linear combination of the states $|1>$ and
$|2>$. The overcompleteness can be avoided by resorting to the Rumer-Pauling 
rules. To implement this rule, we arrange the $2n$ spins at the vertices of a 
regular $2n$-gon. We draw lines between pairs of sites that are singlet 
paired. According to the Rumer-Pauling rule, the subset of these, encompassing
all diagrams (to be called 'legal' diagrams) with no crossing lines, forms a 
complete and linearly independent set of states \cite{klein}. 

The Rumer-Pauling rules can be easily extended to construct complete and 
linearly independent basis sets in higher spin Hilbert spaces involving 
spin-$1/2$ objects. This is done most easily with the help of phantom sites.
If we wish to construct VB diagrams for total spin $S$ subspace involving
$n$ spin-$1/2$ objects, then we introduce $2S$ additional sites to be called 
phantom sites. Besides imposing the Rumer-Pauling rules on the diagrams 
with $n+2S$ sites, we impose the additional constraint that there should
be no singlet lines amongst the $2S$ phantom sites. In Fig. 2, we show
a few examples of VB diagrams with higher total spin. 

It is also quite simple to extend the VB rules to spin clusters made up of 
different site spins \cite{raghu}. 
If the spin at a site is $s_i$, then we replace
this site by a set of $2s_i$ sites, each with spin-$1/2$. We then proceed
with constructing the VB basis, as though the system is made up entirely
of spin-$1/2$ objects, with one difference, namely, we impose the additional
constraint that there should be no singlet lines within the subset of
$2s_i$ sites which replace the spin $s_i$ at site $i$. The VB diagrams
with total spin $S \ge {{1} \over {2}}$ are constructed as before with
the help of phantom sites. An example of a legal VB diagram involving
higher site spins is shown in Fig. 2.

Generating and storing the VB diagrams on a computer is also quite simple.
We associate one bit with every site. The state of the bit is '1', if in
the VB diagram a line begins at the site and the state is '0' if a
line ends at the site. Thus, we can associate an integer of $n$ bits
with every VB diagram involving $n$ spin-$1/2$ objects. This association
is unique if we decipher the bit pattern of the integer corresponding
to the diagram from inside-out, much like expanding an algebraic expression
with multiple parentheses. In Fig. 2, we have also shown the bit
pattern and the associated integer for each VB diagram. The VB diagrams
are generated on a computer by checking the bit pattern in all $n$-bit 
integers to see if they satisfy the criterion for representing the desired
VB diagram. This also allows us to generate the VB diagrams as an ordered 
sequence of the integers that represent them, a fact that helps in rapid
generation of the Hamiltonian matrix.

The Hamiltonian matrix in the VB basis can be easily constructed by knowing
the action of the operator ${\hat {\bf S}}_i \cdot {\hat {\bf S}}_j$ for 
spin-$1/2$ particles (i) on a singlet line joining sites $i~ {\rm and} ~j$ 
and (ii) on the pair of sites $i~ {\rm and} ~j$ singlet paired to two 
different sites $i^{\prime} ~{\rm and} ~j^{\prime}$ (Fig. 3). In the
case of Hilbert spaces with 
nonzero total $S$, the Hamiltonian involves spin exchange between the real 
sites only. However, these exchange operators could lead to VB diagrams
in which the phantom sites are interconnected. In this event, simply
neglecting these resultant states is sufficient to ensure that we are
dealing exactly with the spin $S$ Hilbert space. The Hamiltonian for spin
clusters with arbitrary spins can be treated as consisting of operators with 
only spin-$1/2$ objects. This is done by replacing the spin exchange operator 
between sites $i$ and $j$, $~{\hat {\bf S}}_i \cdot {\hat {\bf S}}_j$ by the
operator $(\Sigma_{k=1}^{2s_i} {\hat \tau}_k) \cdot (\Sigma_{l=1}^{2s_j} 
{\hat \tau}_l)$, where the operators ${\hat \tau}_k$ and ${\hat \tau}_l$ 
are the usual spin-$1/2$ operators.

The matrix representing the Hamiltonian in the VB basis is in general 
nonsymmetric since the VB basis is nonorthogonal. However, the matrix itself
is sparse. There exist efficient numerical algorithms \cite{rettrup}
for obtaining the low-lying eigenstates of sparse nonsymmetric matrices, and 
it is possible to solve a nonsymmetric matrix eigenvalue problem for a 
million by million matrix with about a 100 million nonzero matrix elements 
on a powerful PC based workstation.

While VB theory guarantees spin purity in the computed eigenstates of the
spin conserving Hamiltonians, it has several drawbacks. Computation of
quantities such as spin-spin correlation functions and spin densities 
are not easy because a site-spin operator operating on a VB diagram spoils
the spin purity of the diagram. This could be overcome by converting the 
eigenstate of the Hamiltonian in the VB basis to the constant $M_S$ basis. 
Another difficulty with the VB procedure is exploiting the spatial symmetries 
of the problem. Operation by a spatial symmetry operator on a legal VB
diagram could lead to illegal VB diagrams. Disentangling these illegal 
diagrams into legal diagrams can be computationally prohibitive \cite{srzgs}.

In more general spin problems, it is often advantageous to use the constant
$M_S$ basis and exploit all the spatial symmetries. Partial spin symmetry
adaptation in these cases is also possible by using the spin parity operator.
The effect of the spin parity operator on a basis state is to flip all the
spins in the state. In the $M_S=0$ sector, it is possible to factor the
Hilbert space into odd and even parity Hilbert spaces. The odd (even) parity 
Hilbert space is spanned by basis vectors with odd (even) total spin. This 
also has the effect of reducing the dimensionality of the Hilbert space 
besides providing partial spin symmetry adaptation. It is rather simple to 
set up the Hamiltonian matrix in the symmetry adapted basis. The Hamiltonian 
matrix is symmetric and usually very sparse. The lowest few eigenstates can be
easily computed by employing the Davidson algorithm. Given these eigenstates, 
the computation of properties can proceed by converting an eigenstate in the 
symmetrized basis into that in the unsymmetrized basis. The orthogonality of 
the basis states as well as the simple rules involved in obtaining the 
resultant when a basis state is operated upon by any type of spin operators 
in any combination affords easy computation of a variety of properties of a
magnetic system.

The exact diagonalization techniques discussed above are in general applicable
to systems whose Hilbert space dimensionality is about 10 million. 
The major problem with exact diagonalization methods is the exponential
increase in dimensionality of the Hilbert space with the increase in the
system size. Thus, the study of larger systems becomes not only CPU
intensive but also memory intensive as the number of nonzero elements of
the matrix increases rapidly with system size. With increasing power of the
computers, slightly larger problems have been solved every few years. To
illustrate this trend, we consider the case of the spin-$1$ Heisenberg chain.
In 1973, ten years before the Haldane conjecture, De Neef \cite{deneef} used 
the exact diagonalization procedure to solve a $8$-site spin-$1/2$ chain. In 
1977, Blote \cite{blote} diagonalized the Hamiltonian of a chain of 10 sites. 
In 1982, Botet and Jullien \cite{botet} increased this to 12 sites. In 1984,
Parkinson and Bonner \cite{bonner} solved the $14$-site spin-$1$ problem. In
the same year, Moreo \cite{moreo} solved the $16$-site spin-$1$ chain. In
1990, Takahashi \cite{taka1} pushed this up to $18$ sites. And in 1994,
Golinelli {\it et al.} \cite{goli} have solved for the low-lying
states of a $22$-site spin-$1$ chain. The growth in chain length of the
longest spin-$1$ chain solved is almost linear with time, roughly increasing
by $2$ sites in every three years. Just to remind ourselves, the Fock space
dimensionality in this case increases as $3^N$ with chain length N. The size 
of the matrix also increases similarly and the CPU and storage scales 
quadratically with the size of the matrix, if we are targeting only a few 
eigenstates. For this reason, for systems which span much larger spaces, 
the focus has shifted to approximate techniques.

\section{Applications to Spin Clusters}

In recent years, some of the magnetic clusters that have been studied 
extensively are the Mn$_{12}$ \cite{chud}, Fe$_8$ \cite{wern} and 
V$_{15}$ \cite{chio} clusters. These clusters show many interesting 
phenomena such as quantum resonant tunneling and quantum interference
\cite{qtm}. Basic to a proper understanding of these phenomena is 
a knowledge of the low-energy excitation spectrum in these systems. The 
methods discussed under exact diagonalization schemes allow us to calculate 
the low-energy excitation spectrum, given a set of exchange constants. 
However, the exchange constants themselves are not known with any certainty. 
Therefore, it is all the more important to be able to carry out exact 
diagonalization studies of low-lying states to infer the possible sign and 
magnitude of the exchange constants \cite{raghu}.

For the Mn$_{12}$ cluster, we show the geometry and the exchange parameters 
in Fig. 4. The crystal structure suggests that the exchange 
constant $J_1$ is 
largest and antiferromagnetic in nature \cite{mn12ac}. Based on magnetic 
measurements, it has been suggested that $J_1$ has a magnitude of 215K. The 
magnitude and sign of the other exchange constants are based on comparisons 
with manganese systems in smaller clusters. It has been suggested that the
exchange constant $J_2$ and $J_3$ are antiferromagnetic and have a magnitude
of about 85K. However, for the exchange constant $J_4$, there is no concrete
estimate, either of the sign or of the magnitude. In an earlier study, the
Mn$^{III}$ - Mn$^{IV}$ pair with the strongest antiferromagnetic exchange
constant was replaced by a composite spin-$1/2$ object \cite{gat1} and the
exchange Hamiltonian of the cluster solved for three different sets of
parameters. It was found that the ordering of the energy levels were very
sensitive to the relative strengths of the exchange constants. In these
studies, $J_4$ was set to zero and the low-lying excited states were computed.
Besides, only states with spin $S$ up to ten could be obtained because of the
replacement of the higher spin ion pairs by the composite spin-$1/2$ objects.

The technique described earlier, however, allows an exact computation of 
the low-lying states of Mn$_{12}$. The results of the exact calculations
are presented in Table 1. We note that none of the three sets of parameters 
studied using an effective Hamiltonian, gives the correct ground and excited 
states, when an exact calculation is performed. It appears that setting the 
exchange constant $J_4$ to zero, cannot yield an $S=10$ ground state (Table 1, 
cases A, B and C). When $J_3$ is equal to or slightly larger than $J_2$ (cases 
A and B, Table 1), we find a singlet ground state, unlike the result of the 
effective Hamiltonian in which the ground state has $S=8$ and $S=0$ 
respectively. The ground state has spin $S=6$, when $J_3$ is slightly smaller 
than $J_2$ (case C, Table 1). In all these case, the first few low-lying 
states are found to lie within 20K of the ground state.

When we use the parameters suggested by Chudnovsky \cite{chud} (case D,
Table 1), we obtain an $S=10$ ground state separated from an $S=9$ first
excited state by 223K. This is followed by another $S=9$ excited state at
421K. Only when the exchange constant $J_4$ is sufficiently strongly
ferromagnetic (case E, Table 1), do we find an $S=10$ ground state with an
$S=9$ excited state separated from it by a gap of 35K, which is close to the
experimental value \cite{egap}. The second higher excited state has $S=8$, and
is separated from the ground state by 62K.

In Fig. 5, we show the spin density \cite{mn12spnden} for the 
Mn$_{12}$ cluster in the ground state for the $S=10,M_S=10$ state. While the 
manganese ions connected by the strong antiferromagnetic exchange show 
opposite spin densities, it is worth noting that the total spin density on 
these two ions is $0.69$, well away from a value of $0.5$ expected if these 
ions were indeed to form a spin-$1/2$ object.

The Fe$_8$ cluster is shown in Fig. 6. Each of the Fe ions has a spin
of $2$ and the ground state of the system has a total spin $S=10$, with $S=9$
excited state separated from it by about 20K. All the exchange interactions
in this system are expected to be antiferromagnetic. While the structure of 
the complex dictates that the exchange interaction $J_2$ along the back of the
butterfly should be small in comparison with the interaction
$J_1$ across the wing \cite{fe8butterfly}, in earlier studies it was reported
that such a choice of interaction parameters would not provide a $S=10$ ground
state \cite{delfs}.

Results from the exact calculations of the eigenstates of the Fe$_8$ cluster
using three sets of parameters is shown in Table 2. In two of these cases, 
$J_2$ is very much smaller than $J_1$. We find that in all these cases, 
the ground state has a spin $S=10$ and the lowest excited state has spin, 
$S=9$. One of the main difference we find amongst the three sets of parameters 
is in the energy gap to the lowest excited state (Table 2). For the set of 
parameters used in the earlier study, this gap is the lowest at 3.4K. For 
the parameter sets 1 and 3 \cite{fe8jval}, this gap is respectively 13.1K 
and 39.6K. While in cases 1 and 2, the second excited state is an $S=8$ 
state, in case 3, this state also has spin 9.

The spin densities in all the three cases for the ground state are shown in 
Fig. 7. The spin densities in all cases are positive at the corners. 
In cases 1 and 2, the spin density on the Fe ion on the backbone is positive 
and negative on the remaining two Fe sites \cite{fe8spnden}. However, in 
case 3, the negative and positive spin density sites for Fe ions in the middle 
of the edges is interchanged. This is perhaps due to the fact that in cases 1 
and 2, the exchange constant $J_3$ is less than $J_4$, while in case 3, this 
is reversed. Thus, a spin density measurement can provide relative strengths 
of these two exchange constants. In all the three case, the difference between 
the spin densities in the ground and excited states is that the decrease in 
the spin density in the excited state is mainly confined to the corner Fe 
sites.

\section{\bf Field Theoretic Studies of Spin Chains}

One-dimensional and quasi-one-dimensional quantum spin systems have been
studied extensively in recent years for several reasons. Many such systems
have been realized experimentally, and a variety of theoretical techniques, 
both analytical and numerical, are available to study the relevant models. 
Due to large quantum fluctuations in low dimensions, such systems often 
have unusual properties such as a gap between the ground state and the 
excited states. The most famous example of this is the Haldane gap which was
predicted theoretically in integer spin Heisenberg antiferromagnetic chains 
\cite{hald1}, and then observed experimentally in a spin-$1$ system $\rm 
Ni(C_2 H_8 N_2)_2 NO_2 (Cl O_4)$ \cite{buye}. Other examples include the
spin ladder systems in which a small number of one-dimensional spin-$1/2$
chains interact amongst each other \cite{dago}. It has been observed that if 
the number of chains is even, {\it i.e.}, if each rung of the ladder (which is
the unit cell for the system) contains an even number of spin-$1/2$ sites, 
then the system effectively behaves like an integer spin chain with a gap in 
the low-energy spectrum. Some two-chain ladders which show a gap are $\rm 
(VO)_2 P_2 O_7$ \cite{eccl}, $\rm Sr Cu_2 O_3$ \cite{azum} and $\rm 
Cu_2 (C_5 H_{12} N_2)_2 Cl_4$ \cite{chab1}. Conversely, a three-chain
ladder which effectively behaves like a half-odd-integer spin chain and does 
{\it not} exhibit a gap is $\rm Sr_2 Cu_3 O_5$ \cite{azum}. A related 
observation is that the quasi-one-dimensional system $\rm CuGeO_3$ 
spontaneously dimerizes below a spin-Peierls transition temperature 
\cite{hase}; then the unit cell contains two spin-$1/2$ sites and the 
system is gapped.

The results for gaps quoted above are all in the absence of an external 
magnetic field. The situation becomes more interesting in the presence of a 
magnetic field \cite{chab2}. Then it is possible for an integer spin chain to 
be gapless and a half-odd-integer spin chain to show a gap above the ground 
state for appropriate values of the field [37-45].
This has been 
demonstrated in several models using a variety of methods such as exact 
diagonalization of small systems and bosonization \cite{schu,affl1}. In 
particular, it has been shown that the magnetization of the system can 
exhibit plateaus at certain nonzero values for some finite ranges of the 
magnetic field. Further, for a Hamiltonian which is invariant under 
translation by one unit cell, the value of the magnetization per unit cell is 
quantized to be a rational number at each plateau \cite{oshi}. In section 8, 
we will study the magnetization plateau which can occur in a three-chain 
ladder.

In the next two subsections, we will discuss some field theoretic methods
which can be used for studying spin chains and ladders. These methods rely on 
the idea that the low-energy and long-wavelength modes of a system ({\it i.e.},
wavelengths much longer than the lattice spacing $a$ if the system is defined 
on a lattice at the microscopic level) can often be described by a continuum 
field theory. 

\subsection{\bf Nonlinear $\sigma$-model}

The nonlinear $\sigma$-model (NLSM) analysis of antiferromagnetic spin chains 
with the inclusion of $J_2$ (next-nearest neighbor coupling) and $\delta$ 
(dimerization) proceeds as follows \cite{rao2}. The Hamiltonian for the 
frustrated and dimerized spin chain can be written as
\beq
{\hat H} ~=~ \sum_i ~[ ~1 ~-~ (-1)^i ~\delta ~] ~ {\hat {\bf S}}_i \cdot 
{\hat {\bf S}}_{i+1} ~ + ~J_2 ~\sum_i ~{\hat {\bf S}}_i \cdot {\hat {\bf 
S}}_{i+2} ~. 
\label{ham1}
\eeq
The interactions are schematically shown in Fig. 8. The region of interest is
defined by $J_2 ~\ge 0$ and $0 \le \delta \le 1$. We first do a classical 
analysis in the $S \rightarrow \infty$ limit to find the ground state 
configuration of the spins. Let us make the general {\it ansatz} that the 
ground state is a coplanar configuration of the spins with the energy per 
spin being equal to
\beq
e_0 ~=~ S^2 ~\left[ {J_1 \over 2} ~ (1 +\delta) \cos \theta_1 ~+~
{J_1 \over 2} ~ (1-\delta ) \cos \theta_2 ~+~ J_2 \cos (\theta_1
+ \theta_2) \right]~, 
\eeq
where $\theta_1$ is the angle between the spins ${\bf S}_{2i}$
and ${\bf S}_{2i+1}$ and $\theta_2$ is the angle between the
spins ${\bf S}_{2i}$ and ${\bf S}_{2i-1}$. Minimization of the classical 
energy with respect to the $\theta_i ~$ yields the following three phases.

\noindent (i) Neel phase: This phase has $\theta_1 = \theta_2 = \pi$; 
hence all the spins point along the same line and they go as $\cdots
\uparrow \downarrow \uparrow \downarrow \cdots$ along the chain. This 
phase is stable for $1-\delta^2 > 4J_2/J_1$. 

\noindent (ii) Spiral phase: Here, the angles $\theta_1$ and $\theta_2$ are 
given by
\bea
\cos \theta_1 &=& - {1\over{1+\delta}}~ {\left[~ {{1 - \delta^2}
\over{4 J_2/J_1}} ~+~ {\delta \over {1 + \delta^2}}~ {4 J_2
\over J_1} ~ \right]} \nonumber \\ 
{\rm and} \quad \cos \theta_2 &=& - {1\over{1 -\delta}}~ {\left[~ {{1 - 
\delta^2} \over {4 J_2/J_1}} ~-~ {\delta \over {1 - \delta^2}}~ {4 J_2 \over 
J_1} ~\right]},
\eea
where $\pi/2 < \theta_1 <\pi$ and $0<\theta_2 < \theta_1$. Thus the spins
lie on a plane. This phase
is stable for $1-\delta^2 < 4 J_2/J_1 < (1-\delta^2) /\delta$.

\noindent (iii) Colinear phase: This phase (which needs both dimerization and 
frustration) is defined to have $\theta_1 = \pi$ and $\theta_2 = 0$; hence all
the spins again point along the same line and they go as $\cdots \uparrow 
\uparrow \downarrow \downarrow \cdots$ along the chain. It is stable for 
$(1-\delta^2) / \delta < 4 J_2/J_1$. 

\noindent These phases along with their boundaries are depicted in Fig. 
9. Thus even in the classical limit $S \rightarrow \infty$, the
system has a rich ground state `phase diagram' \cite{fn}.

We can now go to the next order in $1/S$, and study the spin wave spectrum 
about the ground state in each of the phases. The main results are as follows.
In the Neel phase, we find two zero modes, {\it i.e.}, modes for 
which the energy 
$\omega_k$ vanishes linearly at certain values of the momentum $k$, with the 
slope $d\omega_k /dk$ at those points (the velocity) being the same for the 
two modes. In the spiral phase, we have three zero modes, two with the 
same velocity describing out-of-plane fluctuations, and one with a higher 
velocity describing in-plane fluctuations. In the colinear phase, we get two 
zero modes with equal velocities just as in the Neel phase. The three phases 
also differ in the behavior of the spin-spin correlation function $S(q) = 
\sum_n \langle {\hat {\bf S}}_0 \cdot {\hat {\bf S}}_n \rangle \exp (-iqn)$ 
in the classical limit. $S(q)$ is peaked at $q = (\theta_1 + \theta_2)/2$, 
{\it i.e.}, at $q=\pi$ in the Neel phase, at $\pi/2 < q < \pi$ in the spiral
phase and at $q=\pi/2$ in the colinear phase. 

To study the interactions between the spin waves, it is convenient to derive
a semiclassical NLSM field theory which can describe the low-energy and 
long-wavelength excitations. The field theory in the Neel phase is given by 
a $O(3)$ NLSM with a topological term \cite{hald1,affl1}. The field variable 
is a unit vector $\vec \phi$ with the Lagrangian density
\beq
{\cal L} ~=~ {1 \over {2 c g^2}} ~{\dot {\vec \phi}}^2 ~-~ {c \over {2 
g^2}} ~{\vec \phi}^{\prime 2} ~+~ {\theta \over {4 \pi}} ~{\vec \phi} 
\cdot {\vec \phi}^{\prime} \times {\dot {\vec \phi}} ~,
\label{lag1}
\eeq
where $c = 2 S (1 - 4J_2 - \delta^2 ~)^{1/2} ~$ is the spin wave 
velocity, $g^2 = 2 /[S (1 - 4 J_2 - \delta^2 ~)^{1/2} ]~$ is the coupling 
constant (which describes the strength of the interactions between the spin
waves), and $\theta = 2 \pi S (1 - \delta)$ is the coefficient of the
topological term (the integral of this term is an integer which defines the
winding number of a field configuration ${\vec \phi} (x,t)$). Note that 
$\theta$ is independent of $J_2 ~$ in the NLSM. (Time and space derivatives 
are denoted by a dot and a prime respectively). For $(\theta ,{\rm mod} 2\pi )
= \pi$ and $g^2 ~$ less than a critical value, it is known that the
system is gapless \cite{affl1,affl2}. For any other value of $\theta$, the 
system is gapped. For $J_2 = \delta =0$, one therefore expects that integer 
spin chains should have a gap while half-odd-integer spin chains should 
be gapless. This is known to be true even for small values of $S$ like 
$1/2$ (analytically) and $1$ (numerically) although the field theory is 
only derived for large $S$. In the presence of dimerization, one expects 
a gapless system at certain special values of $\delta$. For $S=1$, the 
special value is predicted to be $\delta_c =0.5$. We see that the {\it 
existence} of a gapless point is correctly predicted by the NLSM. However, 
as we will see later, according to reliable numerical results from DMRG, 
$\delta_c$ is $0.25$ for $J_2 =0$ 
\cite{kato} and it decreases with $J_2 ~$ as shown in Fig. 10. 
These deviations from field theory are probably due to higher order 
corrections in $1/S$ which have not been studied analytically so far.
 
In the spiral phase, it is necessary to use a different NLSM which is known 
for $\delta =0$ \cite{rao1,alle}. The field variable is now an $SO(3)$ matrix 
${\bf R}$. The Lagrangian density is
\beq
{\cal L} ~=~ {1 \over {2 c g^2}} ~{\rm Tr} ~\Bigl( ~{\dot 
{\bf R}^T} {\dot {\bf R}} ~{\bf P}_0 ~\Bigr) ~-~ 
{c \over {2 g^2}} ~{\rm Tr} ~\Bigl( ~
{\bf R}^{\prime T} {\bf R}^{\prime} ~{\bf P}_1 ~\Bigr) ,
\eeq
where $c = S (1 + y) {\sqrt {1 - y^2}} / y$, $g^2 = 2 {\sqrt {(1 + y) /(1 - 
y)}} /S$ with $1/y = 4 J_2 ~$, and ${\bf P}_0 ~$ and ${\bf P}_1 ~$ are 
diagonal matrices with diagonal elements $(1,1,2 y (1 - y) / (2 y^2 - 2 y + 
1))$ and $(1,1,0)$ respectively. Note that there is no topological term.
Hence there is no apparent difference between integer and half-odd-integer 
spin chains in the spiral phase. A one-loop renormalization group \cite{rao1} 
and large $N$ analysis \cite{alle} indicate that the system should have a gap 
for all values of $J_2 ~$ and $S$, and that there is no reason for a 
particularly small gap at any special value of $J_2 ~$. 

Finally, in the colinear phase, the NLSM is known for $\delta =1$, {\it 
i.e.}, for 
the spin ladder. The Lagrangian is the same as in (\ref{lag1}), with $c= 4S 
{\sqrt {J_2 (J_2 + 1)}}$, $g^2 = {\sqrt {1 + 1/J_2}}/S$ and $\theta =0$. 
There is {\it no} topological term for any value of $S$, and the model is 
therefore gapped.

The field theories for general $\delta$ in both the spiral and 
colinear phases are still not known. Although the results are 
qualitatively expected to be similar to the $\delta=0$ case in the spiral
phase and the $\delta=1$ case in the colinear phase, quantitative features
such as the dependence of the gap on the coupling strengths will require the
explicit form of the field theory.

\subsection{Bosonization}

Another field theoretic method for studying spin systems in one dimension is 
the technique of bosonization \cite{schu,affl1,gogo,shan1,vond}. This 
technique consists of mapping bosonic operators into fermionic ones, and then
using whichever set of operators is easier to compute with. 
For instance, consider a model with a single species of fermion with a linear 
dispersion relation $E (k) = \pm v k$, where the $\pm$ denotes the right-
and left-moving fermions respectively (with the corresponding 
fields being denoted by ${\hat \psi}_R$ and ${\hat \psi}_L$), and $v$ denotes 
the velocity. Similarly, consider a model with a single species of boson 
with the dispersion relation $E (k) = v |k|$; the right- and left-moving 
fields are denoted by ${\hat \phi}_R$ and ${\hat \phi}_L$ respectively. Then 
it can be shown that these operators are related to each other as
\bea
{\hat \psi}_R ~& \sim &~ \frac{1}{\sqrt {2\pi \alpha}} ~e^{-i 2 {\sqrt \pi} 
{\hat \phi}_R} ~, \nonumber \\
{\hat \psi}_L ~& \sim &~ \frac{1}{\sqrt {2\pi \alpha}} ~e^{i 2 {\sqrt \pi} 
{\hat \phi}_L} ~.
\eea
The length parameter $\alpha$ is a cut-off which is required to ensure that 
the contribution from high-momentum modes do not produce divergences when 
computing correlation functions. It is convenient to define the two 
bosonic fields
\bea
{\hat \phi} ~&=&~ {\hat \phi}_R ~+~ {\hat \phi}_L ~, \nonumber \\
{\hat \theta} ~&=&~ -~ {\hat \phi}_R ~+~ {\hat \phi}_L ~.
\eea
Then the fermionic density is given by
\beq
{\hat \rho} ~-~ \rho_0 ~=~ {\hat \psi}_R^\dagger {\hat \psi}_R ~+~ {\hat 
\psi}_L^\dagger {\hat \psi}_L ~=~ -~ \frac{1}{\sqrt \pi} ~\frac{\partial 
{\hat \phi}}{\partial x} ~,
\eeq
where $\rho_0$ is the background density; fluctuations around this density are
described by the quantum fields $\hat \psi$ or $\hat \phi$.

Although the dispersion relation is generally not linear for all the modes of
a given system, it often happens that the low-energy and long-wavelength 
modes can be studied using bosonization. For a 
fermionic system in one dimension, these modes are usually the ones lying 
close to the two Fermi points with momenta $\pm k_F$ respectively. One can 
define right- and left-moving fields ${\hat \psi}_R$ and ${\hat \psi}_L$ 
which vary slowly on the scale length $a$,
\beq
{\hat \psi} (x) ~=~ {\hat \psi}_R (x) ~e^{ik_F x} ~+~ {\hat \psi}_L (x) ~
e^{-ik_F x} ~.
\eeq
Quantities such as the density will generally contain terms which vary
slowly as well as terms varying rapidly on the scale of $a$,
\bea
{\hat \rho} - \rho_0 ~&=&~ {\hat \psi}^\dagger {\hat \psi} ~=~ {\hat 
\psi}_R^\dagger {\hat \psi}_R ~+~ {\hat \psi}_L^\dagger {\hat \psi}_L ~+~ 
e^{-i2k_F x} ~{\hat \psi}_R^\dagger {\hat \psi}_L ~+~ e^{i2k_F x} ~ {\hat 
\psi}_L^\dagger {\hat \psi}_R \nonumber \\
&=& ~- ~\frac{1}{\sqrt \pi} ~\frac{\partial {\hat \phi}}{\partial x} ~+~
\frac{1}{2\pi \alpha} ~{[} e^{i(2 {\sqrt \pi} {\hat \phi} - 2k_F x)} ~+~ 
e^{-i(2 {\sqrt \pi} {\hat \phi} - 2k_F x)} ~{]} .
\label{den}
\eea

One can compute various correlation functions in the bosonic language. 
Consider an operator of the form
\beq
{\hat O}_{m,n} ~=~ e^{i2 {\sqrt \pi} (m {\hat \phi} + n {\hat \theta})} ~.
\eeq
We find the following 
result for the two-point equal-time correlation function at spatial 
separations which are much larger than the microscopic lattice spacing $a$,
\beq
\langle 0 \vert ~T {\hat O}_{m,n} (x) {\hat O}_{m^\prime , n^\prime }^\dagger
(0) ~ \vert 0 \rangle ~\sim ~\delta_{mm^\prime } \delta_{nn^\prime } ~\Bigr( 
\frac{\alpha}{x} \Bigl)^{2(m^2 K + n^2 /K)} ~, 
\eeq
where $K$ denotes an interaction parameter which will be described below.
Note that the correlation function decays as a power law. In the language of 
the renormalization group, the scaling dimension of ${\hat O}_{m,n}$ is given 
by $m^2 K ~+~ n^2 /K$.

We can now study a quantum spin chain using bosonization. To be specific, 
let us consider a spin-$1/2$ chain described by the anisotropic Hamiltonian 
\beq
{\hat H} ~=~ \sum_{i=1}^N ~[~ \frac{J}{2} ~({\hat S}_i^+ {\hat S}_{i+1}^- 
+ {\hat S}_i^- {\hat S}_{i+1}^+) ~+~ J_z {\hat S}_i^z {\hat S}_{i+1}^z ~-~ 
h {\hat S}_i^z ~] ~,
\label{xxzh}
\eeq
where the interactions are only between nearest neighbor spins, and
$J > 0$. ${\hat S}_i^+= {\hat S}_i^x + i {\hat S}_i^y$ and ${\hat S}_i^- =
{\hat S}_i^x -i {\hat S}_i^y$ are the spin 
raising and lowering operators, and $h$ denotes a magnetic field. Note that 
the model has a $U(1)$ invariance, namely, rotations about the $S^z$ axis. 
When $J_z=J$ and $h=0$, the $U(1)$ invariance is enhanced to an $SU(2)$ 
invariance, because at this point the model can be written simply as ${\hat H}
=J \sum_i {\hat S}_i \cdot {\hat S}_{i+1}$. Although the 
model in (\ref{xxzh}) can be exactly solved using the Bethe {\it ansatz}, 
and one has the explicit result that the model is gapless for a certain
range of values of $J_z /J$ and $h/J$ (see Ref. \cite{cabr2}), it is not 
easy to compute explicit correlation functions in that approach. We therefore
use bosonization to study this model.

We first use the Jordan-Wigner transformation to map the spin model to a model 
of spinless fermions. We map an $\uparrow$ spin or a $\downarrow$ spin at any 
site to the presence or absence of a fermion at that site. We introduce a 
fermion annihilation operator $\psi_i$ at each site, and write the spin at 
the site as
\bea
{\hat S}_i^z &=& {\hat \psi}_i^{\dagger} {\hat \psi}_i -1/2 = {\hat n}_i - 1/2
\nonumber\\
{\hat S}_i^- &=& (-1)^i ~{\hat \psi}_i ~e^{i\pi \sum_j {\hat n}_j} ~,
\label{jorwig}
\eea 
where the sum runs from one boundary of the chain up to the $(i-1)^{\rm th}$ 
site (we assume open boundary conditions here for convenience), $n_i =0$ or 
$1$ is the fermion occupation number at site $i$, and the expression for 
${\hat S}_i^+$ is obtained by taking the hermitian conjugate of ${\hat 
S}_i^-$, The string factor in the definition of ${\hat S}_i^-$ is added in 
order to ensure the correct statistics for different sites; the fermion 
operators at different sites anticommute, whereas the spin operators commute. 

We now find that 
\beq
{\hat H} ~=~ -~ ~\sum_i ~[~\frac{J}{2} ~({\hat \psi}_i^{\dagger} {\hat 
\psi}_{i+1} +h.c.) ~+~ J_z ~({\hat n}_i-1/2)({\hat n}_{i+1}-1/2) ~-~ h ~(
{\hat n}_i -1/2) ~]~.
\eeq
We see that the spin-flip operators ${\hat S}_i^{\pm}$ lead to hopping terms 
in the fermion Hamiltonian, whereas the ${\hat S}^z_i {\hat S}^z_{i+1}$ 
interaction term leads to an interaction between fermions on adjacent sites.

Let us first consider the noninteracting case given by $J_z =0$.
By Fourier transforming the fermions, ${\hat \psi}_k = \sum_j {\hat \psi}_j
e^{-ikja}/\sqrt{N}$, where $a$ is the lattice
spacing and the momentum $k$ lies in the first Brillouin 
zone $-\pi /a < k \le \pi /a$, we find that the Hamiltonian is given by 
\beq
{\hat H} ~=~ \sum_k ~\omega_k ~{\hat \psi}_k^{\dagger} {\hat \psi}_k ~,
\eeq 
where 
\beq
\omega_k ~=~ -J ~\cos (ka) ~-~ h~.
\eeq
The noninteracting ground state is the one in which all the single-particle
states with $\omega_k <0$ are occupied, and all the states with 
$\omega_k >0$ are empty. If we set the magnetic field $h=0$, the 
magnetization per site $m \equiv \sum_i S_i^z /N$ will be zero in the
ground state; equivalently, in the fermionic language, the ground state is 
precisely half-filled. Thus, for $m=0$, the Fermi points ($\omega_k =0$) lie 
at $ka = \pm \pi/2 \equiv k_F a$. Let us now add the magnetic field term. In 
the fermionic language, this is equivalent to adding a chemical potential 
term (which couples to ${\hat n}_i$ or ${\hat S}^z_i$). In that case, the 
ground state no longer has $m=0$ and the fermion model is no longer 
half-filled. The Fermi points are then given by $\pm k_F$, where 
\beq
k_F a ~=~ \pi ~( m ~+~ \frac{1}{2} ) ~.
\eeq
It turns out that this relation between $k_F$ (which governs the oscillations 
in the correlation functions as discussed below) and the magnetization $m$ 
continues to hold even if we turn on the interaction $J_z$, although the 
simple picture of the ground state (with states filled below some energy and 
empty above some energy) no longer holds in that case.

In the linearized approximation, the modes near the two Fermi points have 
the velocities $\partial \omega_k / \partial k = \pm v$, where
$v$ is some function of $J$, $J_z$ and $h$. Next, we introduce the slowly 
varying fermionic fields ${\hat \psi}_R$ and ${\hat \psi}_L$ 
as indicated above; these are functions of a coordinate $x$ which must be an
integer multiple of $a$. Finally, we bosonize these fields. The spin fields 
can be written in terms of either the fermionic or the bosonic fields. For 
instance, ${\hat S}^z$ is given by the fermion density as in Eq. (\ref{jorwig})
which then has a bosonized form given in Eq. (\ref{den}). Similarly, 
\bea
{\hat S}^+(x) ~&=&~ (-1)^{x/a} ~[e^{-ik_Fx/a} {\hat \psi}_R^\dagger (x)+ 
e^{ik_Fx/a} {\hat \psi}_L^\dagger (x)] ~\times \nonumber \\
&& ~~~~~~~~~~~~ [e^{i\pi \int_{-\infty}^{x} dx^{\prime} ({\hat 
\psi}^{\dagger}(x^{\prime} ) {\hat \psi} (x^{\prime} ) + 1/2a)} + h.c. ]~,
\eea
where $(-1)^{x/a} = \pm 1$ since $x/a$ is an integer. This can now be written
entirely in the bosonic language; the term in the exponential is given by
\beq
\int_{-\infty}^x dx^{\prime} ~{\hat \psi}^{\dagger} (x^{\prime}) {\hat \psi} 
(x^{\prime} )= -~{1\over \sqrt{\pi}} \int_{-\infty}^x dx^{\prime} ~
\frac{\partial {\hat \phi}}{\partial x^{\prime}} ~=~ -~{1\over \sqrt{\pi}} ~[
{\hat \phi}_R (x) + {\hat \phi}_L(x)]~,
\eeq
where we have ignored the contribution from the lower limit at $x^{\prime} = 
-\infty$.

We can now use these bosonic expressions to compute the two-spin, equal-time
correlation functions $G^{ab}(x) \equiv <0|T {\hat S}^a (x) {\hat S}^b 
(0)|0>$. We find that
\bea
G^{zz} (x) & = & m^2 ~+~ \frac{c_1}{x^2} ~+~ 
c_2 ~\frac{\cos (2k_F x)}{x^{2K}} ~, \nonumber \\
G^{+-}(x) + G^{-+}(x) & = & c_3 ~\frac{(-1)^{x/a}}{x^{1/2K}} ~+~
c_4 ~\frac{(-1)^{x/a}~\cos (2k_F x)}{x^{2K+1/2K}} ~,
\eea 
where $c_1, ..., c_4$ are some constants. The parameters $K$ and $v$ are 
functions of $J_z /J$ and $h/J$; the exact dependence can be found from the 
web site given in Ref. \cite{cabr2}. For $h=0$,
$K$ is given by the analytical expression 
\beq
{1\over K} ~=~ 1 ~+~ {2\over \pi} \sin^{-1}({J_z\over J}) ~.
\eeq
Note that at the $SU(2)$ invariant point $J_z =J$ and $h=0$, we have $K=1/2$, 
and the two correlations $G^{zz}$ and $G^{+-}$ have the same forms.

In addition to providing a convenient way of computing correlation functions, 
bosonization also allows us to study the effects of small perturbations.
For instance, a physically important perturbation is a dimerizing term
\beq
V ~=~ \delta ~\sum_i ~(-1)^{i} ~[~ \frac{J}{2} ~({\hat S}_i^+ {\hat S}_{i+1}^-
+ {\hat S}_i^- {\hat S}_{i+1}^+) ~+~ J_z {\hat S}_i^z {\hat S}_{i+1}^z ~] ~,
\eeq
where $\delta$ is the strength of the perturbation. Upon bosonizing, we find
that the scaling dimension of this term is $K$. Hence it is relevant if
$K < 2$; in that case, it produces an energy gap in the system which scales
with $\delta$ as
\beq
\Delta E ~\sim ~\delta^{1/(2-K)} ~.
\eeq
This kind of phenomenon occurs in spin-Peierls systems such as $\rm CuGeO_3$;
below a transition temperature $T_{sp}$, they go into a dimerized phase which
has a gap \cite{bouc}.

\section{Density Matrix Renormalization Group Method}

The method which held the promise of overcoming the difficulty of exploding 
dimensionalities is the renormalization group (RG) technique, in which one 
systematically throws out the degrees of freedom of a many-body system. While 
this technique found dramatic success in the Kondo problem \cite{wilson}, its 
straightforward extension to interacting lattice models was quite inaccurate
\cite{oldrg}.

In early 1992, the key problems associated with the failure of the old RG 
method were identified and a different renormalization procedure based on the 
eigenvalues of the many-body density matrix of proper subsystems was
developed \cite{white1,white2}. This method has come to be known as the 
density matrix renormalization group (DMRG) method and has found dramatic 
success in solving quasi-one-dimensional many-body Hamiltonians.

In a real-space RG approach, one begins by subdividing the total system into 
several blocks $A_n$ and proceeds to iteratively build effective blocks so 
that at each iteration, each effective block represents two or more blocks of 
the previous iteration, without increasing the Fock space dimensionality of 
the blocks from what existed at the previous iteration. Usually, one starts 
with each $A_n$ consisting of a single site. Since the Hilbert space grows 
exponentially with the increase in system size, one truncates the number 
of states kept at each iteration. 

The main reason for the failure of the old RG methods is the choice of the 
states retained at each stage of the iteration \cite{white1}. White 
\cite{white2},
recognized that the weakness of the old RG procedure was in the truncation of 
the Fock space of a block based on the eigenvalues of the block Hamiltonian 
being renormalized. He replaced this choice by introducing a completely 
different truncation scheme than the one that was used in the old quantum RG 
procedures. The choice is the eigenvalues of the reduced density matrix of 
the block constructed from the desired state of the full Hamiltonian. The 
truncated Fock space is now spanned by the $m$ eigenvectors of the reduced 
density matrix of order $l \times l$ ($m \le l$) corresponding to the $m$ 
{\it highest} eigenvalues of the reduced many-body density matrix. The reason 
for choosing the eigenvalues of the reduced density matrix as a criterion for 
implementing a cut-off is that, the larger the density matrix eigenvalue, the 
larger is the weight of the eigenstate of the density matrix in the 
expectation value of any property of the system. This result becomes evident 
when all the dynamical operators are expressed as matrices in the basis of 
the eigenvectors of the density matrix. The expectation value of any 
operator $\hat A$ is simply
\beq 
< {\hat A} > ~=~ \sum_i A_{ii} \rho_i ~/ ~\sum_i \rho_i ~, 
\eeq
where $\rho_i$ is the density matrix eigenvalue. The larger the value of
a particular $\rho_i$, the larger is its contribution to the expectation 
value, for a physically reasonable spread in the diagonal matrix elements
$A_{ii}$.

The many-body density matrix of a part of the system can be easily constructed
as follows. Let us begin with given state $|\psi>_{S}$ of $S$, which is called
the {\it universe} or {\it superblock}, consisting of the system (which we 
call a block) $A$ and its {\it environment} $A^{\prime}$. Let us assume that 
the Fock space of $A$ and $A^{\prime}$ are known, and can be labeled as 
$|i>_A$ and $|j>_{A^{\prime}}$ respectively. The representation of $|\psi>_S$ 
in the product basis of $i_A$ and $i_{A^{\prime}}$ can be written as
\beq
|\psi>_{S} ~=~ \sum_{ij} \psi_{ij} |i>_A \times |j>_{A^{\prime}} ~,
\eeq
where we assume the coefficients $\psi_{ij}$ to be real, without loss of 
generality. Then the reduced many-body density matrix for block $A$, is 
defined as
\beq
{\bf\rho}_{kl} ~=~ \sum_{j} \psi_{kj} \psi_{lj}~.
\eeq
The eigenvalue $\rho_i$ of the density matrix ${\bf\rho}$ gives the
probability of finding the corresponding eigenstate $|\mu_i>_A$ in the 
projection of $|\psi>_{S}$ on block $A$. It therefore follows that the 
eigenvectors with highest eigenvalues of the density matrix of $A$, are the 
optimal or most probable states to be retained while the system is augmented.

In the early literature in quantum chemistry, the eigenvectors corresponding
to large eigenvalues of one-particle density matrices were employed as the
orbital basis for carrying out a configuration interaction (CI) calculation.
The eigenvectors of the density matrix were called the `natural' orbitals,
and it was observed that the CI procedure converged rapidly when the `natural'
orbitals were employed in setting up the Slater determinants \cite{mcweeny}.

The DMRG scheme differs from the `natural' orbital scheme in two important
respects: (i) the reduced density matrices are many-body density matrices,
and (ii) the size of the system in terms of the number of sites being studied
at each iteration is usually augmented by two sites. However, the Hamiltonian
matrix that one encounters from iteration to iteration, remains roughly of 
the same order while the matrix elements keep changing (renormalized). In 
this sense, the procedure can be called a renormalization procedure. The 
coupling constants (the Hamiltonian matrix elements) keep changing while the 
system size increases, as in the RG procedure carried out within a blocking 
technique. 

\subsection{Implementation of the DMRG method}

We now describe the procedure to carry out the computations. One starts 
the computation with a small size system, $2n$, which can be exactly solved, 
$1 \le n \le 4$, depending on the degree of freedom at each site. By exact 
diagonalization, one gets the desired eigenstate of that system. The density
matrices of the left and right blocks, each consisting of $n$ sites (in 
principle it is not necessary to have the same number of sites for the two 
blocks, although in practice this is what is most generally used), are obtained
from the desired eigenstate. The density matrices are diagonalized and at the 
first iteration usually all the density matrix eigenvectors (DMEV) are 
retained. The Hamiltonian matrix of the left and right blocks (denoted by $A$ 
and $A^{\prime}$) obtained in any convenient basis are transformed into the 
density matrix eigenvector basis. So also are the matrices corresponding to 
the relevant site operators in both blocks. Now, the iterative procedure 
proceeds as follows.

\begin{enumerate}

\item Construct a superblock $S= A \bullet \bullet A^{\prime}$, consisting
of the block $A$, two additional sites $\sigma$ and $\sigma^{\prime}$ and 
the block $A^{\prime}$. Thus, at the first iteration, the system $S$ has 
$n+1+1+n=2n+2$ sites.

\item Set up the matrices for the total Hamiltonian of the superblock
$S$ in the direct product basis of the DMEV of the blocks $A$ and $A^{\prime}$ 
and the Fock space states of the new sites. Considering that the new sites
are spin-$S$ sites with $(2S+1)$ Fock states each, the order of the total 
Hamiltonian matrix will be $m^2(2S+1)^2 \times m^2(2S+1)^2$, where $m$ 
is the dimension of the block DMEV basis. 

\item Diagonalize the Hamiltonian of the superblock $S=2n+2$ to find the 
desired eigenstate $|\psi>$. Using the state $|\psi>$, evaluate all properties
of the superblock of interest.

\item Construct the reduced many-body density matrix, ${\bf\rho}$, for the 
new block $A \bullet$. If the system does not possess reflection symmetry, 
construct the density matrix, ${\bf\rho^{\prime}}$, for the new right block 
$\bullet A^{\prime}$ as well.

\item Diagonalize the density matrix, ${\bf\rho}$, and if necessary 
${\bf\rho^\prime}$. Usually, the density matrix is block-diagonal in the 
$z$-component of the total spin of the block, and it becomes computationally
efficient to exploit such quantum numbers. Construct a nonsquare matrix 
${\bf O}$, with $m$ columns, each column being an eigenvector of the density 
matrix corresponding to one of the $m$ largest eigenvalues. The number of rows 
in the matrix ${\bf O}$ corresponds to the order of the density matrix.

\item Construct the matrices corresponding to the Hamiltonian, $H_{A 
\bullet}$, of the new left block $A \bullet$, and the site spin operators 
(${\hat S}^z$ and ${\hat S}^+$) of all the necessary sites. The ${\hat S}^-$
operators are simply the adjoints of the ${\hat S}^+$ operators.

\item Renormalize all the matrices corresponding to the block and site 
operators by using the RG transformation matrix ${\bf O}$, {\it e. g.}, 
${\tilde H}_{A \bullet} = O H_{A \bullet} O^{\dag}$. The resulting 
renormalized matrices are of order $m \times m$ and the procedure amounts to 
a simultaneous change of basis and a truncation.

\item Replace the block $A$ by ${A \bullet}$. If the system does not possess 
reflection symmetry, replace $A^{\prime}$ by ${\tilde A^{\prime} \bullet}$.

\item Go to step 1.

\end{enumerate}

\noindent Using the block-diagonal nature of the density matrix, besides 
reducing the requirement in CPU time, also allows one to label the DMEV by the
appropriate $z$-component of the total spin of the block ($M_{s,A}$). The 
Fock space of the individual sites that are added at each iteration are 
eigenstates of the site spin and number operators. This allows us to
target a definite projected spin ($M_s$) state of the total system.

We now briefly describe the mathematical notations that we have used
so far for various states. A state of $A \bullet$ is given by the tensor 
product of a state of $A$ with quantum number $q$ and an index $i$, and a 
state $\sigma$ of the additional site. Thus, 
\beq
|q, i, \sigma >_{A \bullet} ~=~ |q, i >_{A} \times |\sigma >~.
\eeq

\noindent A state of a superblock $S = A \bullet \bullet A^{\prime}$ is 
given by
\beq
|q_A, \mu, \sigma; q_{A^{\prime}}, \nu, \tau > ~=~ |q_A,\mu, \sigma >_{A 
\bullet} \times |q_{A^{\prime}}, \nu, \tau >_{\bullet A^{\prime}} ~.
\eeq

\noindent The eigenstate of the Hamiltonian of the super-block can be 
written as
\beq
|\psi>_{S} ~=~ \sum_{q_A, q_{A^{\prime}}, \mu, \nu, \sigma, \tau} \psi^{q_A, 
q_{A^{\prime}}, \sigma, \tau}_{\mu, \nu} |q_A, \mu, \sigma; q_{A^{\prime}}, 
\nu, \tau >_{S} ~.
\eeq

\noindent The density matrix for $A \bullet$ then will have a block 
structure and can be expressed as
\beq
\rho^{q_A, \sigma}_{\mu, \nu} ~=~ \sum_{q_{A^{\prime}}, \mu^{\prime}, \tau}
\psi^{q_A, q_{A^{\prime}}, \sigma, \tau}_{\mu, \mu^{\prime}}
\psi^{q_A, q_{A^{\prime}}, \sigma, \tau}_{\nu, \mu^{\prime}} ~.
\eeq

\noindent The above algorithm is called the infinite lattice DMRG algorithm 
because this procedure is best suited for the system in the thermodynamic 
limit, {\it i.e.}, when the properties of the system are extrapolated to the 
infinite system size limit. 

\subsection{Finite size DMRG algorithm}

If we are interested in accurate properties of the system at a required size, 
then it is possible to improve upon the accuracies obtainable from the 
infinite DMRG procedure. This involves recognizing that the reduced 
many-body density matrices at each iteration correspond to a different system 
size. For example, when we carry out the DMRG procedure to obtain the 
properties of a system of $2M$ sites, at an iteration corresponding to $2p$ 
sites ($n \le p \le M$), the reduced density matrix we construct is 
that of a block of $p$ sites in a system 
of $2p$ sites. However, if our interest is in the $2M$-site system, we should
employ the density matrix of the block of $p$ sites in a $2M$-site system. It 
is possible to construct, iteratively, the $p$-site reduced density matrix of 
the $2M$-site system. This is achieved by the so called finite-size algorithm 
\cite{white2}. This method provides highly accurate solutions even when the 
states of the full Hamiltonian have inhomogeneous (symmetry breaking) 
properties.

To obtain the $2M$-site result, we should perform the infinite lattice 
algorithm up to $l=(M-1)$ sites first storing all operators in each iteration.
Now the algorithm for finite lattices with reflection symmetry (left block = 
right block), proceeds as follows. 

\begin{enumerate}

\item On reaching a system size of $2M$ sites, obtain the density matrix of 
the block of $M$ sites.

\item Use the density matrix of $M$ sites on the left and that of $(M-2)$ 
sites on the right, add two new sites as in the infinite DMRG procedure and 
obtain the desired eigenstate of the $2M$ system. 

\item Now obtain the reduced density matrix of the $(M+1)$ sites from the
eigenstate of the previous iteration obtained in the direct product basis of 
the DMEVs of the $M$-site, $(M-2)$-site density matrices, and the Fock space 
states of the individual sites.

\item Go back to step 2, replacing $M$ by $(M+1)$ and $(M-2)$ by $(M-3)$ and
iterate until a single site results on the right and $(2M-3)$ sites result 
on the left.

\item Since the system has reflection symmetry, use the density matrix of the 
$(2M-3)$ sites on the right and construct the $2M$ system as built-up 
from three individual sites on the left and $(2M-3)$ sites on the right. 
Obtain the desired eigenstate of the $2M$ system in this basis.

\item Now obtain the new $2$-site density matrix on the left and $(2M-4)$
site density matrix on the right. Replace the single-site on the left 
by two sites and $(2M-3)$ sites on the right by $(2M-4)$ sites in step 5.

\item Repeat steps 5 and 6 until $(M-1)$ sites are obtained both on the left 
and right. The properties of the $2M$ system obtained from the eigenstates at 
this stage corresponds to the first iteration of the finite-size algorithm. 
We can now go back to step 1 and carry through the steps to obtain properties 
at later iterations of the finite-size DMRG algorithm.

\end{enumerate}

In systems without reflection symmetry, the DMEVs of the right and
left parts are not identical even if the sizes of the reduced systems are 
the same. The finite-DMRG algorithm in this case involves first constructing
the density matrices of the left part for sizes greater than $M$ and on 
reaching the density matrix of $(2M-3)$ sites, reducing the size of the 
left-part and increasing that of the right, from one site to $(M-1)$. This
will result in the refined density matrices of both the right and the left
block of the total system, for block sizes of $(M-1)$. At this stage, we 
can compute all the properties and continue the reverse sweep until the right 
block is of size $(2M-3)$ and the left block is of size $1$. The forward sweep 
that follows will increase the block size on the left and decrease that of the 
right. We would have completed the second iteration when the two block sizes 
are equal. The forward and reverse sweeps can be continued until we reach the 
desired convergence in the properties of the whole system.

\subsection{Calculation of properties in the DMRG basis}

At the end of each iteration, one can calculate the properties of the targeted 
state \cite{mx}. The reduced many-body density matrix computed at each 
iteration can be used to calculate the static expectation values of any site 
operator or their products. Care should be taken to use the density matrices 
appropriate to the iteration. The expectation value of a site property 
corresponding to the operator ${\hat A}_i$ can be written as
\beq
< {\hat A}_i > ~=~ {\rm Tr} ~({\bf \rho} {\bf A}_i) ~.
\eeq
${\bf \rho}$ is the density matrix of the block in which the site $i$ is 
situated and ${\bf A}_i$ is the matrix of the renormalized site operator 
at site $i$. For calculating correlation functions, one can use a similar
equation. The correlation function between two site operators belonging to 
separate blocks can be written as
\beq
< {\hat A}_i {\hat A}_j > ~=~ {\rm Tr} ({\bf \rho}{\bf A}_i {\bf A}_j) ~.
\eeq
However, the accuracy of this procedure turns out be very poor if 
the sites $i$ and $j$ belong to the same block \cite{white2}. The reason is 
that a feature implicit in the above procedure is the resolution of identity 
by expansion in terms of the complete basis. Unfortunately, the basis in which 
the site operators are represented is incomplete and such an expansion is 
therefore error prone. To circumvent this difficulty, it has been suggested
\cite{white2} that, one obtains the matrix representation of the products of 
the site operators from the first occurrence of the product pair $<ij>$ and, 
by renormalizing the product operator ${\hat A}_i {\hat A}_j$, at every 
subsequent iteration until the end of the RG procedure. Then, the correlation 
function between ${\hat A}_i$ and ${\hat A}_j$ (where $i$ and $j$ belong to 
the same block) can be evaluated as
\beq
< {\hat A}_i {\hat A}_j > ~=~ {\rm Tr} (\rho {\bf A}_i {\bf A}_j)~.
\eeq
This procedure is found to be more accurate in most cases.

\subsection{Remarks on the applications of DMRG}

The DMRG method is currently the most accurate method for large quantum 
lattice models in one dimension. It can be applied to interacting bosonic, 
fermionic or spin models as well as to models which have interactions amongst 
them. The overall accuracy of the DMRG method is exceptionally high for 
one-dimensional systems with only nearest neighbor interactions. For a 
spin-$1/2$ chain where exact Bethe {\it ansatz} ground-state energy is 
available, the DMRG ground state energy per site in units of the exchange 
constant $J$, is found to be accurate to seven decimal places with a cut-off 
$m=100$ \cite{white2}. The method is found to be almost as accurate for the 
one-dimensional Hubbard model, 
where again it is possible to compare the DMRG results with exact results 
obtained from the exact analytic Bethe {\it ansatz} solution \cite{white3}. 

Since higher dimensionality is equivalent to longer-range interactions within 
one dimension, the model also restricts the range of interactions in one 
dimension. It has been noted that the number of DMEVs that should be retained
in a calculation on higher dimensional systems, for accuracies comparable to 
accuracy in one dimension, scales exponentially with dimensionality. Thus, to 
obtain accuracy comparable to that obtained in a chain of $L$-sites for a 
cut-off $m$, in a $L \times L$ square lattice, the number of DMEVs needed to 
be retained for the corresponding 2-dimensional lattice is $\approx m^2$.

Extending the range of interactions to next-nearest neighbors 
does not significantly deteriorate the accuracy \cite{skp1}. However, 
inclusion of cyclic boundary conditions reduces the accuracy of the method 
significantly, although in one-dimension, the DMRG method still would 
outperform any other method for the same system size. In the DMRG procedure, 
the most accurate quantity computed is the total energy. In dealing with 
other quantities such as correlation functions, caution must be exercised 
in interpreting the results.

The density matrix eigenvalues sum to unity and the {\it truncation error}, 
which is defined as the sum of the density matrix eigenvalues corresponding to 
the discarded DMEV, gives a qualitative estimate as to the accuracy of the
calculation as well as providing a framework for the extrapolation to the
$m \rightarrow \infty$ limit. The accuracy of the results obtained in this way 
is unprecedented \cite{white4,white5}. The accuracy of the ground state energy 
per site for the spin-$1$ chain is limited by the precession of machine 
arithmetic, {\it i.e.}, $e_{0} = 1.401484038971(4)$. Similarly, the accuracy 
persists even while calculating for the Haldane gap, {\it e.g.}, the gap is
evaluated to be $0.41050(2)$.

Another aspect of the DMRG technique worth noting is that the method is best 
suited for targeting one eigenstate at a time. However, it is possible to 
obtain reasonable results for a set of states by using an average many-body 
reduced density matrix constructed as a weighted sum of the density matrices 
corresponding to each of the states in question. One way of constructing the 
average density matrix is by using a statistical weight for the chosen set 
of states; the averaged density matrix in this instance is given by
\beq
\rho_{\beta ;kl} ~=~ \sum_{ij} ~\psi_{i;kj} ~\psi_{i;lj} ~\exp [-\beta 
\epsilon_i ] ~/ ~\sum_i ~\exp [-\beta \epsilon_i] ~,
\eeq
where $\beta=1/k_B T$ with $k_B$ and $T$ being the Boltzmann constant and 
temperature respectively. One can thus extend the DMRG method to finite 
temperatures.

Finite size algorithms have been used extensively to study the edge states and 
systems with impurities, where substantial improvement of the accuracy is 
needed to characterize the various properties of a finite system. The DMRG 
method has been applied to diverse problems in magnetism: study of spin chains 
with $s > 1/2$ \cite{suzu}, chains with dimerization and/or frustration
\cite{kato,skp1,scho,burs}, coupled spin chains \cite{skp1,chen,naru}, to 
list a few. Highly accurate studies of the structure factor and string order 
parameter (topological long range order) \cite{white4} as well as edge states 
in Haldane phase systems \cite{qin} have been performed. Dynamical properties 
for both spin and fermionic systems with DMRG have also been reported within 
the Maximum Entropy Method \cite{pang} as well as the continued fraction 
\cite{hallberg} and correction 
vector \cite{skp2} approaches. Also, DMRG has been 
successfully formulated to obtain low-temperature thermodynamic properties for
various spin systems \cite{caron1,skp3}, and the solution of models of spin 
chains dynamically coupled to dispersionless phonons \cite{caron2}. 
Additionally, Nishino and Okunishi have derived two reformulations of DMRG, 
namely, the product wave function renormalization group (PWFRG) \cite{nish1}, 
and the corner transfer matrix renormalization group (CTMRG) \cite{nish2} 
methods. These methods offer the means of calculating dynamical correlation 
functions in spin chains as well as highly accurate results for the 
$2$-dimensional Ising model at criticality. 

\section{\bf Frustrated and Dimerized Spin Chains}

It is well known that the one-dimensional XY chain can be mapped on to a
one-dimensional noninteracting spinless fermion model. The isotropic spin 
chain will then map on to a chain of interacting spinless fermions. According 
to Peierls' theorem, a partly filled one-dimensional band of noninteracting
fermions is unstable with respect to a lattice distortion that results in an 
insulating ground state. It has been shown that introduction of interactions 
in the Peierls system leads to a stronger instability. The mapping between
the Heisenberg spin chains with equal nearest neighbor exchange interactions
(uniform spin chain) and the spinless fermion model suggests that such a spin 
chain is also unstable with respect to a lattice distortion leading to 
alternately strong and weak nearest neighbor exchange constants, {\it i.e.},
a dimerized 
spin chain. What is of importance is the fact that such a dimerization is 
unconditional: no matter how strong the lattice is, the lattice dimerizes, 
since the exchange energy gained due to dimerization always exceeds the 
strain energy. This result follows from the fact that the gain in exchange 
energy varies as $\delta^2 {\rm ln} \delta$ while the strain energy loss 
varies as $\delta ^2$, where $\delta$ is the magnitude of dimerization that 
leads to the nearest neighbor exchange constants alternating as 
$J(1\pm \delta)$. 

In recent years, many systems which closely approximate the one-dimensional 
spin chain have been synthesized. What has been observed in these experimental
systems is that besides the nearest neighbor antiferromagnetic exchange, there
also exists a second neighbor exchange $J_2$ of the same sign and comparable 
magnitude. Such a second neighbor interaction has the effect of frustrating 
the spin alignment favored by the nearest neighbor interaction. Therefore, a 
realistic study of these systems require modeling them using both dimerization 
and frustration. Theoretically, spin chains with only frustration ($J-J_2$ 
model) were studied by Majumdar and Ghosh. Interestingly, they showed that for 
$J_2= J/2$, the ground state is doubly degenerate and is spanned by the two 
possible Kekule structures (Fig. 11). It is quite gratifying to note that a 
century after the Kekule structure for benzene was proposed, there actually 
exists a Hamiltonian for which Kekule structure happens to be the ground state!

While most of the discussion above is restricted to spin-$1/2$ chains, there
has been considerable interest in the higher-spin chains following the
conjecture of Haldane which predicts that for uniform spin chains, the 
excitation spectrum of integer spin chains is qualitatively different from
that of half-odd-integer spin chains. The latter have a gapless excitation
spectrum while the excitation spectrum of the former are gapped. The synthesis
and study of integer spin chains have indeed confirmed this conjecture.

Notwithstanding many interesting exact analytic solutions for spin chains,
there still exist a large number of situations for which such solutions
have been elusive. The exact solutions are basically confined to the uniform
Heisenberg model and the frustrated and dimerized model along the line 
$2J_2~+~\delta~=~1$ in the $J_2-\delta$ plane, with $J = 1$. Reliable
numerical study of these models therefore requires developing techniques 
which are highly accurate so that the results of large finite systems can
be scaled or extrapolated to the thermodynamic limit. As has already been
discussed, the DMRG technique is ideally suited due to its high accuracy
for quasi-one-dimensional systems.

The Hamiltonian for the frustrated and dimerized spin chain is given in Eq. 
(\ref{ham1}) and is schematically shown in Fig. 8. A few low-lying states in 
a sector with a given value of the total spin component, $M_S$ are obtained 
in representative points in the $J_2 - \delta$ plane, using the DMRG method. 
The ground state is always the first (lowest energy) state in the $M_S = 0$ 
sector. The accuracy of the DMRG method depends crucially on the number of 
eigenstates of the density matrix, $m$, which are retained. Working with 
$m=100$ to $120$ over the entire $J_2 - \delta$ plane gives accurate results. 
This can be verified by comparing the DMRG results for these $m$ values with 
exact numerical diagonalizations of chains with up to $16$ sites for spin-$1$ 
systems \cite{tone2} and $22$ sites for spin-$1/2$ systems \cite{rama}. The 
chain lengths studied vary from $150$ sites for $J_2 > 0$ to $200$ sites for 
$J_2 =0$. The DMRG results are also tracked as a function of $N$, the chain 
length, to verify that convergence is reached well before $150$ sites in all 
cases. The numerical results are much better convergent for open chains than 
for periodic chains, a feature generic to the DMRG technique 
\cite{white2,white5}. 

The quantum phase diagrams obtained for a spin-$1/2$ chain is shown in Fig. 
12. The system is gapless on the line $A$ running from $J_2
=0$ to $J_{2c} = 0.241$ for $\delta =0$, and is gapped everywhere else in the 
$J_2- \delta$ plane. There is a disorder line $B$ given by $2J_2 +\delta =1$; 
the peak in the structure factor $S(q)$ is at $q_{max} =\pi$ to the left of 
$B$ (region I), decreases from $\pi$ to $\pi /2$ as we go from $B$ up to the 
line $C$ (region II), and is at $q_{max} =\pi /2$ to the right of $C$ (region
III). This is in agreement with the results obtained in section 4 using the 
NLSM approach. The correlation length $\xi$ goes through a minimum on line $B$.

In the spin-$1$ case (Fig. 10), the phase diagram is more 
complex. There is a solid line marked $A$ which runs from $(0,0.25)$ to about 
$(0.22 \pm 0.02, 0.20 \pm 0.02)$ shown by a cross. To within numerical 
accuracy, the gap is zero on this line and the correlation length $\xi$ is as 
large as the system size $N$. The rest of the `phase' diagram is gapped. 
However the gapped portion 
can be divided into different regions characterized by other interesting 
features. On the dotted lines marked $B$, the gap is finite. Although $\xi$ 
goes through a maximum when we cross $B$ in going from region II to region I 
or from region III to region IV, its value is much smaller than $N$. There is 
a dashed line $C$ extending from $(0.65,0.05)$ to about $(0.73,0)$ on which 
the gap appears to be zero (to numerical accuracy), and $\xi$ is very large 
but not as large as $N$. The straight line $D$ satisfying $2 J_2 + \delta 
= 1$ extends from $(0,1)$ to about $(0.432, 0.136)$. Regions II and III are 
separated by line $E$ which goes down to about $(0.39,0)$. Across $D$ and $E$, 
the peak in the structure factor decreases from $\pi$ (Neel) in regions I and 
II to less than $\pi$ (spiral) in regions III and IV. 
In regions II and III, the ground state for an {\it open} chain has a 
four-fold degeneracy (consisting of states with $S=0$ and $S=1$), 
whereas it is nondegenerate in regions I and IV with $S=0$. The regions II and
III, where the ground state is four-fold degenerate for an open chain, can be 
identified with the Haldane phase; the regions I and IV correspond to a 
non-Haldane singlet phase. The lines $B$, $D$ and $E$ meet in a small region 
V where the ground state of the system is numerically very difficult to find.
Note that the numerically zero gap at $(0.73,0)$ is unexpected from either
bosonic mean-field theory \cite{rao3} or the NLSM approach discussed earlier.

For the spin-$1$ system, there is a striking similarity between the 
ground state properties of the dimerized and frustrated model (\ref{ham1}) 
as a function of $J_2$ (with $\delta =0$) and the biquadratic model,
\beq
{\hat H} ~=~ \sum_i ~[ ~{\hat {\bf S}}_i \cdot {\hat {\bf S}}_{i+1} ~+~ 
\beta ~(~ {\hat {\bf S}}_i \cdot {\hat {\bf S}}_{i+1} ~)^2 ~]~,
\label{ham2}
\eeq
as a function of (positive) $\beta$ \cite{scho}. 
For $J_2 < 0.39$ and $\beta < 1/3$, both models are in the Neel phase and
are gapped. For $J_2 > 0.39$ and $\beta > 1/3$, the two models are in the
spiral phase and are generally gapped; however, model (\ref{ham1}) is
`gapless' for $J_2 = 0.73$ while model ({\ref{ham2}) is gapless for $\beta =
1$. Qualitatively, the cross-over from the Neel to the spiral phase (but {\it 
not} the gaplessness at a particular value of $J_2$ or $\beta$) can be 
understood through the following classical argument. Let us set the 
magnitudes of the spins equal to $1$ and define the angle between spins
${\bf S}_i$ and ${\bf S}_{i+n}$ to be $n \theta$. The angle $\theta$ can 
be obtained by minimizing $\cos \theta + J_2 \cos 2 \theta$ in (\ref{ham1}),
and $\cos \theta + \beta \cos^2 \theta$ in (\ref{ham2}). This gives us
a Neel phase ($\theta = \pi$) if $J_2 \le 1/4$ and $\beta \le 1/2$ in the
two models, and a spiral phase for larger values of $J_2$ and $\beta$
with $\theta = \cos^{-1} (-1/4 J_2)$ and $\theta = \cos^{-1} (-1/2 \beta)$
respectively. The actual crossover points from Neel to spiral is different for
spin-$1$ than these classical values.
In the classical limit $S \rightarrow \infty$, the ground state of the model 
is in the Neel phase for $4J_2 < 1 - \delta^2$, in a spiral phase for 
$1 - \delta^2 < 4J_2 < (1 - \delta^2)/\delta$ and in the colinear phase 
for $(1 - \delta^2)/\delta < 4J_2$ (Fig. 9). 

As can be seen from Fig. 8, $\delta =1$ results in two coupled spin chains 
wherein the interchain coupling is $2$ and the intrachain coupling is $J_2$. 
By using DMRG, one can study the dependence of the gap $\Delta$ and the 
two-spin correlation function $C(r)$ on the interchain coupling $J$. In Fig. 
13 is plotted $\Delta$ vs $J$ for both spin-$1/2$ and spin-$1$
systems. For spin-$1/2$, the system is gapped for any nonzero value of the 
interchain coupling $J$, although the gap vanishes as $J \rightarrow 0$. 
The gap increases and correspondingly the correlation length decreases with 
increasing $J$. In the case of coupled spin-$1$ chains, one finds the somewhat 
surprising result that both the gap and the correlation length $\xi$ are 
fairly large for moderate values of $J$. Note that the variation of the gap 
with $J$ for spin-$1$ (shown as circles) is much less than that for spin-$1/2$ 
(crosses). 

The NLSMs derived in section 4 can be expected to be accurate only for large 
values of the spin $S$. It is interesting to note that the numerically 
obtained `phase' boundary between the Neel and spiral phases
for spin-$1$ is closer to the the classical ($S \rightarrow 
\infty$) boundary $4J_2 = 1 - \delta^2$ than for spin-$1/2$. For instance, the
crossover from Neel to spiral occurs, for $\delta =0$, at $J_2 = 0.5$ for 
spin-$1/2$, at $0.39$ for spin-$1$, and at $0.25$ classically.

To conclude this section, we have studied a two-parameter `phase' diagram for 
the ground state of isotropic antiferromagnetic spin-$1/2$ and spin-$1$ 
chains. The spin-$1$ diagram is considerably more complex than the 
corresponding spin-$1/2$ chain, with surprising features like a `gapless' 
point inside the spiral `phase'; this point could be close to a critical 
point discussed earlier in the literature \cite{affl2,suth}. It would be 
interesting to establish this more definitively. Our results show that 
frustrated spin chains with small values of $S$ exhibit features not 
anticipated from large $S$ field theories.

\section{Alternating $(S_1,S_2)$ ferrimagnetic spin chains}

Ferrimagnets belong to a class of magnets which show spontaneous
magnetization below a certain critical temperature. There have been
a number of experimental efforts to synthesize molecular materials 
showing spontaneous magnetization at low temperatures \cite{stein76,kahn1}. 
These are quasi-one-dimensional bimetallic molecular magnets in which each 
unit cell contains two spins with different spin values, with
the general formula $\rm ACu(pbaOH)(H_2O)_3.2H_2O$, where $\rm pbaOH$ is
2-hydroxo-1,3-propylenebis(oxamato) and ${\rm A}= {\rm Mn}$, $\rm Fe$, $\rm 
Co$, $\rm Ni$; they
belong to the alternating or mixed spin chain family \cite{kahn1,kahn2}. These
alternating spin compounds have been seen to exhibit ferrimagnetic behavior.
It has been possible to vary the spin at each site from low values where
quantum effects dominate to large values which are almost classical.

The thermodynamic behavior of these alternating
spin compounds is very interesting \cite{kahn2,kahn3}. In very low magnetic 
fields, these systems show one-dimensional ferrimagnetic behavior. The 
$\chi T$ vs $T$ (where $\chi$ is the magnetic susceptibility and $T$ the
temperature) plots show a rounded minimum. As the temperature is increased,
$\chi T$ decreases sharply, goes through a minimum before increasing 
gradually. The temperature at which this minimum occurs differs from
system to system and depends on the site spins of the chain. The variation of 
the field induced magnetization with temperature is also interesting as the 
ground state is a magnetic state. These exciting observations have 
motivated us to study the ferrimagnetic systems with arbitrary spins $s_1$ 
and $s_2$ alternating from site to site. It would also be of interest to know 
the thermodynamic properties of these systems with varying $s_1$ and $s_2$.

\subsection{Ground state and excitation spectrum}

We start our discussion with the Hamiltonian for a chain with spins $s_1$ and
$s_2$ on alternating sites (with $s_1 > s_2$, without loss of generality).
\beq
{\hat H} ~=~ J ~\sum_{n} ~[~ (1+\delta) ~{\hat {\bf S}}_{1,n} \cdot {\hat 
{\bf S}}_{2,n} ~+~ (1-\delta) ~{\hat {\bf S}}_{2,n} \cdot {\hat {\bf 
S}}_{1,n+1} ~]~,
\label{ham3}
\eeq
where the total number of sites is $2N$ and the sum is over the
total number of unit cells $N$. ${\hat {\bf S}}_{i,n}$ corresponds to the 
spin operator for the site spin $s_i$ in the $n$-th unit cell. The 
exchange integral $J$ is taken to be positive for all our calculations;
$\delta$ is the dimerization parameter and it lies in the range $\{ 0,1 \}$.

Before describing our numerical results, we briefly summarize the results
of a spin wave analysis for the purposes of comparison \cite{skp3}. We will 
first state the results for $\delta =0$. According to spin wave theory, the 
ground state has total spin $S_G =N(s_1 -s_2)$. Let us define a function
\beq
\omega(k) ~=~ J ~\sqrt{(s_1 -s_2)^2 + 4 s_1 s_2 \sin^2 (k/2)} ~,
\eeq
where $k$ denotes the wave number. Then the ground state energy per site
is given by
\beq
\epsilon_0 ~=~ \frac{E_0}{2N} ~=~ - ~Js_1 s_2 ~+~ \frac{1}{2} ~\int_0^{\pi} ~
\frac{dk}{\pi} ~[~ -~J (s_1 +s_2 ) ~+~ \omega(k) ~]~.
\eeq
The lowest branch of excitations is to states with spin $S=S_G -1$, with 
the dispersion
\beq
\omega_{1}(k) ~=~ J ~(-s_1 + s_2) ~+~ \omega(k) ~;
\eeq
the gap vanishes at $k=0$. There is a gapped branch of excitations to
states with spin $S=S_G +1$, with the dispersion
\beq
\omega_{2}(k) ~=~ J ~(s_1 - s_2) ~+~ \omega(k) ~;
\eeq
the minimum gap occurs at $k=0$ and is given by $\Delta = 2J (s_1 - s_2)$.
In the ground state with $S^z = S_G$, the sublattice magnetizations are given 
by the expectation values
\bea
<{\hat S}^z_{1,n}> &=& (~s_1 +\frac{1}{2}~) ~-~ \frac{1}{2} ~\int_0^{\pi} ~
\frac{dk}{\pi} ~\frac{J(s_1+s_2)}{\omega (k)} ~, \nonumber \\
<{\hat S}^z_{2,n}> &=& s_1 ~-~ s_2 ~-~ <{\hat S}^z_{1,n}> ~.
\eea
The various two-spin correlation functions decay exponentially with distance;
the inverse correlation length is given by $\xi^{-1} = \ln (s_1 /s_2)$. The 
results with dimerization ($\delta > 0$) are very similar. In fact, within 
spin wave theory, the minimum gap $\Delta$ to states with spin $S=S_G +1$ is 
independent of $\delta$.

We now use the powerful DMRG method to study the system defined by Eq. 
(\ref{ham3}) both with and without dimerization, $\delta \ne 0$ and $\delta =
0$ respectively. We have considered alternating spin-$3/2$/spin-$1$ 
(hereafter designated as ($3/2,1$)), spin-$3/2$/spin-$1/2$ (to be called 
($3/2,1/2$)) and spin-$1,1/2$ (called ($1,1/2$)) chains with
open boundary condition for the Hamiltonian (\ref{ham3}). We compute the 
ground state properties for these three systems by studying chains with 
$80$ to $100$ sites. The number of dominant density matrix eigenstates,
$m$, that we have retained at each DMRG iteration also varies between 
$80$ to $100$. With the increase of the Fock space
dimensionality of the site spins, we increase $m$ to obtain more accurate
results. We follow the usual steps for the "infinite system" DMRG method 
discussed above \cite{white2,skp1,hall}, except that the 
alternating chains studied here are not symmetric between the left and right 
halves; hence the density matrices for these two halves have to be separately
constructed at every iteration of the calculations. We have also verified 
the convergence of our results by 
varying the values of $m$ and the system size. 

The ground states of all the systems lie in the $S^z =N(s_1 -s_2)$ sector,
as verified from extensive checks carried out by
obtaining the low-energy eigenstates in different $S^z$ sectors of a 
$20$-site chain. A state corresponding to the lowest energy in 
$S^z =N(s_1 -s_2)$ is found in all subspaces with $\vert S^z \vert \le 
N(s_1 -s_2)$, and is absent in subspaces with $\vert S^z \vert > N(s_1 -s_2)$. 
This shows that the spin in the ground state is $S_G =N(s_1 -s_2)$. (Actually, 
the lowest energy states in the different $S^z$ sectors are found to be 
degenerate only up to $10^{-5} J$. Such small errors are negligible 
for studying thermodynamics at temperatures larger than, say, $10^{-2} J$). 

In Fig. 14, we show the expectation value of site spin operator 
${\hat S}^z_{i,n}$ ( spin density) at all the sites for the ($3/2,1$), 
($3/2,1/2$) and ($1,1/2$) chains. The spin densities 
are uniform on each of the sublattices in the chain for all the three systems. 
For the ($3/2,1$) chain, the spin density at a spin-$3/2$ is $1.14427$ 
(the classical value is $3/2$), while, at a spin-$1$ site it is $-0.64427$ 
(classical value $1$). For the ($3/2,1/2$)
case, the spin density at a spin-$3/2$ site is $1.35742$ and at a 
spin-$1/2$ site it is $-0.35742$. For the ($1,1/2$) case, the 
value at a spin-$1$ site is $0.79248$ and at a spin-$1/2$ site it is 
$-0.29248$. These can be compared with the spin wave values of $1.040$ and 
$-0.540$; $1.314$ and $-0.314$; and $0.695$ and $-0.195$ for the spin-$s_1$ 
and spin-$s_2$ sites of the ($3/2,1$); ($3/2,1/2$); and ($1,1/2$) systems 
respectively. We note that the spin wave analysis overestimates the quantum 
fluctuations in case of systems with small site spin values. We also notice 
that there is a greater quantum fluctuation when the difference
in site spin $|s_1 -s_2|$ is larger. This is also seen in spin-wave theory. 
The spin density distribution in an alternating ($s_1,s_2$) chain is more 
similar to that of a ferromagnetic chain rather than an antiferromagnet, 
with the net spin of each unit cell perfectly aligned (but with small quantum
fluctuations on the individual sublattices). In a ferromagnetic ground 
state, the spin density at each site has the classical value appropriate
to the site spin, whereas for an antiferromagnet, this averages out to zero 
at each site as the ground state is nonmagnetic. From this viewpoint, the 
ferrimagnet is similar to a ferromagnet and is quite unlike an 
antiferromagnet. The spin wave analysis also yields the same physical picture.

Because of the alternation of spin-$s_1$ and spin-$s_2$ sites along the 
chain, one has to distinguish between three different types of pair 
correlations, namely, $< {\hat S}^z_{1,0} {\hat S}^z_{1,n}>$, $< {\hat 
S}^z_{2,0} {\hat S}^z_{2,n}>$ and $< {\hat S}^z_{1,0} {\hat S}^z_{2,n}>$.
We calculate all the three correlation functions with the mean values
subtracted out, since the mean values are nonzero in all these three
systems unlike in pure antiferromagnetic spin chains. In the DMRG procedure,
we have computed these correlation functions from the sites inserted at the 
last iteration, to minimize numerical errors. In Fig. 15, we plot the two-spin
correlation functions in the ground state as a function of the distance
between the spins for an open chain of $100$ sites for all three cases. All 
three correlation functions decay rapidly with distance for each of the three 
systems. From the figure it is clear that, except for the $< {\hat S}^z_{1,0} 
{\hat S}^z_{2,n}>$ correlation, all other correlations are almost zero even 
for the shortest possible distances. The $< {\hat S}^z_{1,0} 
{\hat S}^z_{2,n}>$ correlation has an appreciable value [$-0.2$ for 
($3/2,1$), $-0.07$ for ($3/2,1/2$) and $-0.094$ for ($1,1/2$)] only for the 
nearest neighbors. This rapid decay of the correlation functions makes it 
difficult to find the exact correlation length $\xi$ for a lattice model, 
although it is clear that $\xi$ is very small (less than one unit cell) for 
the ($3/2,1/2$) and ($1,1/2$) cases, and a little 
greater ($1<\xi <2$) for the ($3/2,1$) system. Spin wave theory gives 
$\xi =2.47$ for ($3/2,1$), $\xi =0.91$ for ($3/2,1/2$), and $\xi =1.44$ for 
($1,1/2$) cases. (We should remark 
here that our $\xi$ is not to be confused with the conventional definition of 
the correlation length; the latter is actually infinite in these systems 
due to the long-range ferrimagnetic order).

The lowest spin excitation of all the three chains is to a state with $S=S_G 
-1$. To study this state, we target the $2^{\rm nd}$ state in the $S^z =S_G -
1$ sector of the chain. To confirm that this state is a $S=S_G -1$ state, we 
have computed the $2^{\rm nd}$ state in $S^z =0$ sector and find that
it also has the same energy. However, the corresponding state is
absent in $S^z$ sectors with $\vert S^z \vert > S_G -1 $. Besides, from exact 
diagonalization of all the states of all the $s_1 -s_2$ alternating spin
chains with $8$ sites, we find that the energy orderings of the states is 
such that the lowest excitation is to a state with spin $S=S_G -1$. We have 
obtained the excitation gaps for all the three alternating spin chains in the
limit of infinite chain length by extrapolating from the plot of spin gap
vs the inverse of the chain length (Fig. 16). We find that 
this excitation is gapless in the infinite chain limit for all three cases.

To characterize the lowest spin excitations completely, we also have computed
the energy of the $S=S_G +1$ state by targeting the lowest state in the 
$S^z=S_G +1$ sector. In Fig. 17,
we plot the excitation gaps to the $S=S_G +1$ state from the ground state
for all three systems as a function of the inverse of the chain length. 
The gap saturates to a finite value of $(1.0221 \pm 0.0001)J$ for the 
($3/2,1$) case, $(1.8558 \pm 0.0001)J$ for ($3/2,1/2$), and $(1.2795 \pm 
0.0001)J$ for ($1,1/2$). It appears that the gap is also higher when the 
difference in site spins $|s_1 -s_2|$, is larger. The site spin densities 
expectation values computed in this state for all three cases are found to 
be uniform ({\it i.e.}, independent of the site) on each of the sublattices.
This leads us to believe that this excitation cannot be characterized 
as the states of a magnon confined in a box, as has been observed for a 
spin-$1$ chain in the Haldane phase \cite{white4}. 

We have also studied the spin excitations in the dimerized alternating
($s_1,s_2$) chains, defined in Eq. (\ref{ham3}).
We calculate the lowest spin excitation to the $S=S_G -1$ state
from the ground state. We find that the $S=S_G -1$ state is gapless from 
the ground state for all values of $\delta$. This result agrees with the spin 
wave analysis of the general ($s_1,s_2$) chain. The systems remain gapless 
even while dimerized unlike the pure antiferromagnetic dimerized spin chains. 
There is a smooth increase of the spin excitation gap from the ground state 
to the $S=S_G +1$ state with increasing $\delta$ for all three systems 
studied here. We have plotted this gap vs $\delta$ in Fig. 
18. The gap shows almost a linear behavior 
as a function of $\delta$, with an exponent of $1.0 \pm 0.01$ for all three 
systems. This seems to be an interesting feature of all ferrimagnets.
The spin wave analysis however shows that this excitation gap is independent
of $\delta$ for the general ($s_1,s_2$) chain. The similar behaviors of these 
three alternating spin systems suggest that a ferrimagnet can be considered as
a ferromagnet with small quantum fluctuations. 

\subsection{Low-temperature thermodynamic properties}

We have varied the size of the system from $8$ to $20$ sites to calculate 
the thermodynamic properties. We impose periodic boundary conditions to 
minimize finite size effects with ${\hat {\bf S}}_{1,N+1}$ = ${\hat {\bf 
S}}_{1,1}$, so that the number of sites equals the number of bonds.
We set up the Hamiltonian matrices in the DMRG basis for all allowed 
$S^z$ sectors for a ring of $2N$ sites. We can diagonalize these matrices
completely to obtain all the eigenvalues in each of the $S^z$ sectors. 
As the number of DMRG basis states increases rapidly with 
increasing $m$, we retain a smaller number of dominant density matrix
eigenvectors in the DMRG procedure, {\it i.e.}, $50 \le m \le 65$, 
depending on the $S^z$ sector as well as the size of the system. We have 
checked the dependence of properties (with $m$ in the range $50 \le m \le 65$) 
for the system sizes we have studied ($8 \le 2N \le 20$), and have confirmed 
that the properties do not vary significantly for the temperatures at which 
they are computed; this is true for all the three systems. 

 It may appear surprising that the DMRG technique which essentially targets a 
single state, usually the lowest energy state in a chosen $S^z$ sector, should
provide accurate thermodynamic properties since these properties are governed 
by the energy level spacings and not by the absolute energy of the ground 
state. However, there are two reasons why the DMRG procedure yields reasonable
thermodynamic properties at low temperatures. Firstly,
the projection of the low-lying excited state eigenfunctions on the DMRG space
which contains the ground state is substantial; hence these excited 
states are well described in the chosen DMRG space. Secondly, the low-lying 
excitations of the full system are often the lowest energy states in different
sectors in the DMRG procedure; hence their energies are quite accurate 
even on an absolute scale.

The canonical partition function $Z$ for the $2N$ site ring can be written as
\beq
Z ~=~ \sum_j ~e^{-\beta ( E_j - B(M)_j )} ~,
\label{thermo1}
\eeq
where the sum is over all the DMRG energy levels of the
$2N$ site system in all the $S^z$ sectors. $E_j$ and $(M)_j$ denote the 
energy and the $\hat z$-component of the total spin of the state $j$, and 
$B$ is the strength of the magnetic field in units of $1/g\mu_B$ ($g$ is the 
gyromagnetic ratio and $\mu_B$ is the Bohr magneton) along the ${\hat z}$ 
direction. The field induced magnetization $< M >$ is defined as
\beq
< M > ~=~ \frac{\sum_j ~(M)_j ~e^{-\beta ( E_j - B(M)_j )}}{Z} ~.
\label{thermo2}
\eeq
The magnetic susceptibility $\chi$ is related to the fluctuation in 
magnetization
\beq
\chi ~=~ \beta ~[~ < M^{2} > - < M >^{2} ~]~.
\label{thermo3}
\eeq
Similarly, the specific heat $C_V$ is related to the fluctuation in the energy
and can be written as
\beq
\frac{C_V}{k_B} ~=~ \beta^2 ~[~ < E^{2} > - < E >^{2} ~]~.
\label{thermo4}
\eeq

In the discussion to follow, we present results on the $20$-site ring although 
all calculations have been carried out for system sizes from $8$ to $20$ sites.
This is because the qualitative behavior of the properties we have studied
are similar for all the ring sizes in this range for all three systems.

The dependence of magnetization on temperature for different magnetic field 
strengths are shown in Fig. 19 for all three systems. At low 
magnetic fields, the magnetization shows a sharp decrease at 
low temperatures and shows paramagnetic behavior at high temperatures. 
As the field strength is increased, the magnetization shows a slower decrease 
with temperature, and for high field strengths the magnetization shows 
a broad maximum. This behavior can be understood 
from the type of spin excitations present in these systems. The lowest
energy excitation at low magnetic fields is to a state with spin $s$
less than $S_G$. Therefore, the magnetization initially decreases at
low temperatures. As the field strength is increased, the gap to 
spin states with $S > S_G$ decreases as the Zeeman coupling to these
states is stronger than to the states with $S \le S_G$. The critical 
field strengths at which the magnetization increases with temperature
varies from system to system since this corresponds to the lowest spin gap
of the corresponding system. The behavior
of the system at even stronger fields turns out to be remarkable.
The magnetization in the ground state ($T=0$) shows an abrupt increase
signalling that the ground state at this field strength has $S^z >S_G$.
The temperature dependence of the magnetization shows a broad 
maximum indicating the presence of states with even higher spin values
lying above the ground state in the presence of this strong field. 
In all three cases, the ground state at very high field strengths
should be ferromagnetic. For the systems at such high fields,
the magnetization decreases slowly with increase of temperature as no 
other higher spin states lie above the ground state. While we have not
studied such high field behaviors, we find that the field strength 
corresponding to switching the spin of the ground state $s_{G}$ to $s_{G}+1$ 
is higher for ($3/2,1/2$) system compared to ($3/2,1$)
and ($1,1/2$) systems. The switching field appears to
depend on the value of $|s_1 -s_2 |$. We see in Fig. 19
that in the ($3/2,1$) and ($1,1/2$) cases, the ground state
has switched to the higher spin state at the highest magnetic field strength
we have studied but in the ($3/2,1/2$) case, the ground state has not 
switched even at the field strength indicating that the excitation gap for 
this system is larger than the other two. For the ($3/2,1/2$) case, the 
same situation should occur at very high magnetic fields. Thus, we predict
that the highest $S^z$ is attained in the ground state at high
magnetic field and this field strength increases with increase in
the site spin difference $|s_1 - s_2|$.

The dependence of $\chi T/2N$ on temperature for different field strengths
are shown in Fig. 20 for all three systems. For zero field, the zero 
temperature value of $\chi T$ is infinite in the thermodynamic limit; for 
finite rings it is finite and equal to the average of the square of the 
magnetization in the ground state. For the ferrimagnetic ground state $\chi 
T/2N$, as $T \rightarrow 0$, is given by $S_G (S_G +1)/6N$. As the 
temperature increases, this product decreases and shows a minimum
before increasing again. For the three systems studied here, the minimum
occurs at different temperatures depending on the system. 
For the ($3/2,1$) alternating spin system, it is at $k_B T=(0.8 \pm 
0.1)J$, while for the ($3/2,1/2$) and ($1,1/2$) cases, 
it occurs at $k_B T=(1.0 \pm 0.1)J$ and $k_B T=(0.5 \pm 0.1)J$ respectively. 
The minimum occurs due to the states with $S^z < S_G$ getting populated at 
low temperatures. In the infinite chain limit, these states turn out
to be the gapless excitations of the system. The subsequent
increase in the product $\chi T$ is due to the higher energy-higher spin 
states being accessed with further increase in temperature.
This increase is slow in ($3/2,1/2$) case, as in this 
system very high spin states are not accessible within the chosen temperature 
range. Experimentally, it has been found in the bimetallic chain compounds
that the temperature at which the minimum occurs in the $\chi T$ 
product depends upon the magnitude of the spins $s_1$ and $s_2$ \cite{kahn3}.
The $Ni^{II}-Cu^{II}$ bimetallic chain shows a minimum in $\chi T/2N$ at a 
temperature corresponding to $55$ ${\rm cm}^{-1}$ (80K);
an independent estimate of the exchange constant in this system is
$100$ ${\rm cm}^{-1}$ \cite{kahn4}. This is in very good agreement with the
minimum theoretically found at temperature $(0.5 \pm 0.1)J$ for the 
($1,1/2$) case. The minimum in $\chi T/2N$ vanishes at 
$B=0.1 J/g\mu_B$ which corresponds to about $10 T$ for all three systems. It
would be interesting to study the magnetic susceptibility of these
systems experimentally under such high fields. The low-temperature 
zero-field behavior of our systems can be compared with the one-dimensional 
ferromagnet. In the latter, the spin wave analysis shows that the $\chi T$ 
product increases as $1/T$ at low temperatures \cite{taka2}. 

In finite but weak fields, the behavior of $\chi T$ is different. 
The magnetic field opens up a gap and $\chi T$ falls exponentially 
to zero for temperatures less than the gap in the applied field for all
three systems. Even in this case a minimum is found at the same temperature 
as in the zero-field case for the corresponding system, for the same reason 
as discussed in the zero field case.

In stronger magnetic fields, the behavior of $\chi T$ 
from zero temperature up to $k_B T=J_{min}$ ($J_{min}$ is the temperature
at which the minimum in $\chi T$ is observed) is qualitatively different. The 
minimum in this case vanishes for all three systems. In these field strengths, 
the states with higher $S^z$ values are accessed even below $k_B T=J_{min}$.
The dependence of $\chi T$ above $k_B T=J_{min}$ at all field strengths is 
the same in all three systems.
In even stronger magnetic fields, the initial sharp increase is suppressed. 
At very low temperature, the product $\chi T$ is nearly zero and increases 
almost linearly with $T$ over the temperature range we have studied. This 
can be attributed to a switch in the ground state at this field strength. 
The very high temperature behavior of $\chi T$ should be independent of 
field strength and should saturate to the Curie law value corresponding to 
the mean of magnetic moments due to spin-$s_1$ and spin-$s_2$.

The temperature dependence of specific heat also shows a marked dependence on 
the magnetic field at strong fields. This dependence is shown in Fig. 
21 for various field strengths for all the three systems. In zero and weak 
magnetic fields, the specific heat shows a broad maximum at different 
temperatures which are specific to the system. Interestingly, the temperature
at which the specific heat shows a maximum closely corresponds to the 
temperature where the low-field $\chi T$ has a minimum for the corresponding
system. For a strong magnetic field ($B \sim J$), there is a dramatic
increase in the peak height at about the same temperature corresponding to 
the specific system, although the qualitative dependence is still the same 
as at low magnetic fields in all three cases. This phenomena indicates that 
the higher energy high-spin states are brought to within $k_B T$ of the 
ground state at this magnetic field strength for all three cases.

Studies of thermodynamic properties of the dimerized alternating spin chains 
in these three cases show qualitatively similar trends to that of the 
corresponding uniform systems; this follows from the fact that the low-energy 
spectrum does not change qualitatively upon dimerization.

\section{Magnetization Properties of a Spin Ladder}

As mentioned earlier, a quantum spin system can sometimes exhibit 
magnetization plateaus as a function of an applied magnetic field [37-45].
In this section, we 
will use the finite system DMRG method to study the magnetic properties of a 
three-legged spin-$1/2$ ladder \cite{tand}. We consider the Hamiltonian 
\beq
{\hat H} ~=~ \jp ~\sum_a ~\sum_n ~ {\hat {\bf S}}_{a,n} \cdot {\hat {\bf 
S}}_{a+1,n} ~+~ J ~ \sum_{a=1}^3 ~\sum_n ~{\hat {\bf S}}_{a,n} \cdot {\hat 
{\bf S}}_{a,n+1} ~ -~ h ~\sum_{a=1}^3 ~\sum_n ~{\hat S}_{a,n}^z ~,
\label{ham4}
\eeq
where $a$ denotes the chain index, $n$ denotes the rung index, $h$ denotes 
the magnetic field (we have absorbed the gyromagnetic ratio $g$ and the Bohr 
magneton $\mu_B$ in the definition of $h$), and $J, \jp > 0$. It is convenient
to scale out the parameter $J$, and quote all results in terms of the two 
dimensionless quantities $\jp /J$ and $h/J$. If the length of each chain 
is $L$, the total number of sites is $N = 3L$. Since the total ${\hat S}^z$ 
is a good quantum number, it is more convenient to do the numerical 
computations {\it without} including the magnetic-field term in (\ref{ham4}), 
and then to add the effect of the field at the end of the computation. 
For the ground state properties, we have only considered an open boundary 
condition (OBC) in the rung direction, namely, the summation over $a$ in the 
first term of (\ref{ham4}) runs over $1,2$. However, for low-temperature 
properties, we have studied both OBC, as well as a periodic boundary condition 
(PBC) in the rung direction in which we sum over $a=1,2,3$ in the first term. 

We have done DMRG calculations (using the finite system algorithm 
\cite{white2}) with open boundary conditions in the chain direction. We have 
gone up to $120$ sites, {\it i.e.}, a chain length of $40$. 
The number of dominant density matrix eigenstates, corresponding to
the $m$ largest eigenvalues of the density matrix, that we retained 
at each DMRG iteration was $m =80$. In fact, we varied the value of $m$ 
from $60$ to $100$, and found that $m =80$ gives satisfactory results in 
terms of agreement with exact diagonalization for small systems and good 
numerical convergence for large systems. 
For inputting the values of the couplings into the numerical 
programmes, it is more convenient to think of the system as a single chain 
(rather than as three chains) with the Hamiltonian
\beq
{\hat H} ~=~ \frac{2}{3} ~\jp ~\sum_i ~[~ 1 ~-~ \cos ~(~\frac{2\pi i}{3}~)~]~ 
{\hat {\bf S}}_i \cdot {\hat {\bf S}}_{i+1} ~+~ J ~\sum_i ~{\hat {\bf S}}_i 
\cdot {\hat {\bf S}}_{i+3} ~.
\eeq
The system is grown by adding two new sites at each iteration.
Note that our method of construction ensures that we obtain the 
three-chain ladder structure after every third iteration when the total 
number of sites becomes a multiple of $6$. At various system sizes, starting
from $48$ sites and going up to $120$ sites in multiples of $6$ sites, we
computed the energies after doing three finite system iterations; we found 
that the energy converges very well after three iterations. The energy data 
is used in Figs. 22 and 23 below. After reaching 
$120$ sites, we computed the spin correlations after doing three finite system
iterations. This data is used in Figs. 24 and 25.
All our numerical results quoted below are for $J / \jp = 1/3$. We chose this 
particular value of the ratio because there is a particularly broad 
magnetization plateau at $m_s =1/2$ which can be easily found numerically.

We now describe the various ground-state properties we have found with OBC 
along the rungs. We looked for magnetization plateaus at 
$m_s =0$, $1/2$ and $1$. For a system 
with $N$ sites, a given value of magnetization per rung, $m_s$, corresponds 
to a sector with total $S^z$ equal to $M=m_s N/3$. Using the infinite
system algorithm, we found the lowest energies $E_0 (S^z,N)$ in the three 
sectors $S^z = M+1, M$ and $M-1$. Then we examined the 
three plots of $E_0 /NJ$ vs $1/N$ and extrapolated the
results up to the thermodynamic limit $N \rightarrow \infty$. We fitted these
plots with the formula $E_0 /NJ = e_i + a_i /N + b_i/N^2$, where the label 
$i=1,2,3$ denotes the $S^z$ sectors $M+1, M$ and $M-1$. 
In the thermodynamic limit, the values of the three intercepts $e_i$ should 
match since those are just the energy per site for the three states whose 
$S^z$'s differ by only $1$. However, the three slopes $a_i$ are not equal 
in general. We now show that there is a magnetization plateau if $a_1 + a_3 
- 2 a_2$ has a nonzero value. Since the three energies $E_0$ are computed 
without including the magnetic field term, the upper critical field $h_{c+}$ 
where the states with $S^z =M+1$ and $M$ become degenerate is given by
\beq
h_{c+} (N) ~=~ E_0 (M+1, N) ~-~ E_0 (M, N) ~.
\eeq 
Similarly, the lower critical field $h_{c-}$ where the states with $S^z =M$ 
and $M-1$ become degenerate is given by
\beq
h_{c-} (N) ~=~ E_0 (M, N) ~-~ E_0 (M-1, N) ~.
\eeq 
We therefore have a finite interval $\Delta h (N) = h_{c+} (N) - h_{c-} (N)$ 
in which the lowest energy state with $S^z = M$ is the ground state of the 
system with $N$ sites in the presence of a field $h$. If this interval 
has a nonzero limit as $N \rightarrow \infty$, we have a magnetization 
plateau. Thus, in the thermodynamic limit, the plateau width
$\Delta h /J$ is equal to $a_1 + a_3 - 2 a_2$. 

We will now quote our numerical results for $J / \jp =1/3$. For a rung 
magnetization of $m_s =1/2$, {\it i.e.}, $M=N/6$, we found the three slopes
$a_i$ to be equal to $3.77, -0.02$ and $-1.93$; see Fig. 22. 
This gives the upper and lower critical fields to be
\bea
h_{c+}/J ~=~ a_1 ~-~ a_2 ~&=&~ 3.79 ~, ~~~~~~~~ h_{c-}/J ~=~ a_2 ~-~ a_3 ~=~ 
1.91 ~, \nonumber \\
\Delta h/J ~&=&~ (h_{c+} ~-~ h_{c-})/J ~=~ 1.88 ~.
\label{plat1}
\eea
This is a sizeable plateau width. For a rung magnetization of $m_s =1$,
we found the $a_i$ to be equal to $4.97, -0.24$ and $-5.43$.
Thus the upper and lower critical fields are 
\beq
h_{c+}/J ~=~ 5.21 ~, ~~~~~~~~ h_{c-}/J ~=~ 5.19 ~, ~~~~~~~~ \Delta h/J ~=~ 
0.02 ~.
\label{plat2}
\eeq
Finally, for a rung magnetization of $m_s =0$, we need the energies of
states with $M=0$ and $M= \pm 1$. Since the last two states must have the
same energy, we have $a_1 = a_3$ and it is sufficient to plot only $E_0 (0,N)$
and $E_0 (1,N)$ vs $1/N$. We found $a_1$ and $a_2$ to be equal to $0.39$
and $0.34$. This gives the upper and lower fields to be
\beq
h_{c+}/J ~=~ 0.05 ~, ~~~~~~~~ h_{c-}/J ~=~ -0.05 ~, ~~~~~~~~ \Delta h/J ~=~ 
0.10 ~.
\label{plat3}
\eeq
The plateau widths given in (\ref{plat2}) and (\ref{plat3}) are rather small. 
In Fig. 23, we indicate the plateau widths $\Delta h (N) /J$ 
as a function of $1/N$ for $m_s =1/2$, $0$ and $1$. 

Next, we computed various two-spin correlations for the $120$-site system.
These are denoted by $\langle {\hat S}_{a,l}^z {\hat S}_{b,n}^z 
\rangle$ and $\langle {\hat S}_{a,l}^+ {\hat S}_{b,n}^- \rangle$. For the $zz$
correlations, it is convenient to subtract the product of the two separate
spin densities. At $m_s =1/2$, we found that all these correlations decay 
very rapidly to zero as the rung separation $\vert l - n \vert$ grows. In 
fact, the fall offs were so fast that we were 
unable to compute sensible correlation lengths. The correlation lengths 
are of the order of one or two rungs as can be seen in Fig. 24.

On the other hand, for the state at $m_s =0$, we found that all the 
correlations decay quite slowly. The decays are consistent with power law 
fall offs of the form $A (-1)^{\vert l-n \vert} / \vert l-n \vert^{\eta}$. It
is difficult to find $\eta$ very accurately since the maximum value of $\vert 
l - n \vert$ is only $20$; this is because we fixed one site to be in the 
middle of the chain (to minimize edge effects), and the maximum chain length 
is $40$ for our DMRG calculations. For $m_s =0$, the exponent $\eta$ for 
all the correlations was found to be around $1$. There was no difference in 
the behaviors of the $zz$ and $+-$ correlations since this was an isotropic 
system; $m_s =0$ is the ground state if the magnetic field is zero.

For the state at $m_s =1$ (which is the ground state only for a substantial
value of the magnetic field), we found that the $+-$ correlations again
decay quite slowly consistent with a power law. The exponents $\eta$ for 
the different $+-$ correlations varied from $0.61$ to $0.70$ with an 
average value of $0.66$; see Fig. 25 for an example. 

We now describe some low-temperature thermodynamic properties of the
three-chain system obtained using DMRG. Although DMRG is normally expected to 
be most accurate for targeting the lowest states in different $S^z$ sectors,
earlier studies of mixed spin chains have shown that DMRG is quite reliable
for computing low-temperature properties also \cite{skp3}. There are two 
reasons for this; the low-lying excited states generally have a large 
projection onto the space of DMRG states which contains the ground state, and
the low-lying excitations in one sector are usually the lowest states in 
nearby $S^z$ sectors. 

We first checked that for systems with $12$ sites, the results obtained using 
DMRG agree well with those obtained by exact diagonalization. We then used 
DMRG to study the magnetization, susceptibility and specific heat of $36$-site
systems with both OBC and PBC along the rungs. We computed these quantities 
using the techniques described in Eqs. (\ref{thermo1}-\ref{thermo4}).
The plots of magnetization vs magnetic field for various temperatures
are shown in Fig. 26 for OBC along the rungs. 
The temperature $T$ is measured in units of $J/k_B$. We see that the 
plateau at $m_s =1/2$ disappears quite rapidly as we increase the temperature.
The plateau has almost disappeared at $T=0.4$ which is substantially lower 
than the width $\Delta h/J = 1.88$. The magnetic susceptibility 
is (exponentially) small at low temperatures in the region of the plateau 
because the magnetic excitations there are separated from the ground state 
by a gap. Similar results for the magnetization and susceptibility are 
found for the case of PBC along the rungs.

However, the specific heats at the $m_s =1/2$ plateau demonstrate an 
interesting difference between OBC and PBC along the rungs. While it is very 
small at low temperatures for OBC (see Fig. 27), it is not small 
for PBC, although it shows a plateau in the same range of magnetic fields as 
the magnetization itself. This observation strongly suggests that the 
system with PBC along the rungs has {\it nonmagnetic} excitations which
do not contribute to the magnetization or susceptibility, but do contribute
to the specific heat. Fig. 28 gives a more direct comparison
between OBC and PBC along the rungs. Although these nonmagnetic excitations 
were studied by previous authors \cite{cabr1,cabr2,kawa}, we believe that our
specific heat plots prove their existence most physically. To show these 
excitations even more explicitly, we present in Fig. 29 all the energy 
levels for a $12$-site chain in the sector $S^z =2$ ({\it i.e.}, $m_s =1/2$) 
using exact diagonalization. It is clear that the ground state is well 
separated from the excited states for OBC, but it is at the bottom of a
band of excitations for PBC; these excitations are nonmagnetic since
they have the same value of $S^z$ as the ground state. 

We summarize our results for a three-chain spin-$1/2$ ladder with a large 
ratio of interchain coupling to intrachain coupling. There is a wide plateau
with rung magnetization given by $m_s =1/2$ for both OBC and PBC along the 
rungs. For the case of OBC, the two-spin correlations are extremely 
short-ranged, and the magnetic susceptibility and specific heat are very 
small at low temperature in the plateau. All these are consistent with the 
large magnetic gap. At other values of $m$, the two-spin correlations fall 
off as power laws. For the case of PBC, the magnetic susceptibility is again 
very small at low temperature in the plateau. However the specific heat goes 
to zero much more slowly which dramatically shows the presence of 
nonmagnetic excitations. 

To summarize this review, we have discussed a variety of numerical and 
analytical methods for studying spin clusters and quasi-one-dimensional spin 
systems. The methods discussed for spin clusters are directly relevant to 
areas of current interest such as quantum tunneling in the presence of 
time-dependent magnetic fields [17-20,95].
One-dimensional spin systems sometimes
have unusual properties (such as a disordered ground state with an excitation
gap above it) which are not observed in similar models in higher dimensions. 
The techniques described here, particularly the DMRG method, are well-suited 
for studying extended systems in one dimension.

\vskip .5 true cm
 
\leftline{\bf Acknowledgments}

We thank H. R. Krishnamurthy, Sumathi Rao, R. Chitra, Y. Anusooya, Kunj Tandon,
Siddhartha Lal, C. Raghu and Indranil Rudra for several collaborations over 
the years. The work on mixed spin systems was motivated by discussions with 
late Professor Olivier Kahn.

\vskip .5 true cm

\newpage
\clearpage

\noindent
Table 1: Low-Lying states of $\rm Mn_{12}Ac$, relative to the ground state
for the parameter in question. Entries in parenthesis in
cases A, B and C correspond to the effective Hamiltonian results of Sessoli
{\it et al.} \cite{gat1}. Case D corresponds to the parameters suggested by
Chudnovsky \cite{chud}. The parameters corresponding to different cases are:
case (A) $J_1$=225K, $J_2$=90K, $J_3$=90K, $J_4$=0K; case (B) $J_1$=225K,
$J_2$=90K, $J_3$=93.8K, $J_4$=0K; case (C) $J_1$=225K, $J_2$=90K, $J_3$=86.2K,
$J_4$=0K; case (D) $J_1$=215K, $J_2$=85K, $J_3$=-85K, $J_4$=-45K; case (E)
$J_1$=215K, $J_2$=85K, $J_3$=85K, $J_4$=-64.5K. All the energies are in K.

\begin{center}
\begin{tabular}{|c|c|c|c|c|}
\hline
\hspace{0cm} {\bf Case A} \hspace{0cm} & \hspace{0cm} {\bf Case B}
\hspace{0cm} & \hspace{0cm} {\bf Case C} \hspace{0cm} &
\hspace{0cm} {\bf Case D} \hspace{0cm} & \hspace{0cm} {\bf Case E}
\hspace{0cm} \\ \hline
state~~~S~~E(K) & state~~~S~~E(K) & state~~~S~~E(K) &
state~~~S~~E(K) & state~~~S~~E(K) \\ \hline

$^e{B}$~~~~0~~0.0 & $^e{B}$~~~~0~~0.0 & $^e{B}$~~~~6~~0.0 & $^e{A}$~~~~10~~
0.0 & $^e{A}$~~~~10~~ 0.0 \\
(8) &(0) & (10) & & \\ \hline

$^o{E}$~~~~1~~10.8 & $^o{E}$ ~~~~1~~16.2 & $^o{E}$ ~~~~1~~15.5 & 
$^o{E}$ ~~~~9~~223 & $^o{E}$ ~~~~9~~35.1 \\

~~~~~~(9)~(6.4) &~~~~~~(8)~(1.4) &~~~~~~(8)~(2.7) & & \\

~~~~~~(10)~(6.4) & & & & \\ \hline

$^o{B}$ ~~~~1~~19.8 & $^o{B}$ ~~~~1~~20.0 & $^o{B}$ ~~~~1~~19.6 & 
$^o{B}$ ~~~~9~~421.2 & $^e{B}$ ~~~~8~~62.1 \\

~~~~~~(0)~(6.8) & &~~~~~~(9)~(5.0) & & \\ \hline

$^e{A}$ ~~~~2~~24.7 & $^e{A}$ ~~~~ 2~~30.5 & $^e{A}$ ~~~~2~~23.8 & 
$^o{B}$ ~~~~9~~~425.1 & $^o{E}$ ~~~~7~~~82.4 \\ \hline

$^o{E}$ ~~~~3~~39.0 & $^e{B}$ ~~~~4~~58.4 & $^o{E}$ ~~~~1~~~28.8 & 
$^e{B}$ ~~~~8~~439.5 & $^e{A}$ ~~~~6~~~99.7 \\ \hline

$^e{E}$ ~~~~2~~49.9 & $^e{E}$ ~~~~2~~60.9 & $^e{B}$ ~~~~6~~53.6 & 
$^e{B}$ ~~~~8~~443.7 & $^e{B}$ ~~~~0~~102.0 \\ \hline

$^e{B}$ ~~~~4~~57.1 & $^o{A}$ ~~~~3~~64.3 & $^e{B}$ ~~~~6~~54.4 & 
$^e{B}$ ~~~~8~~458.1 & $^e{A}$ ~~~~2~~121.0 \\ \hline

$^e{B}$ ~~~~8~~57.8 & $^e{E}$ ~~~~2~~80.0 & $^e{B}$ ~~~~8~~57.2 & 
$^o{A}$ ~~~11~~573.4 & $^0{B}$ ~~~~1~~133.3 \\ \hline

$^e{B}$ ~~~~2~~57.8 & $^o{A}$ ~~~~3~~88.1 & $^e{E}$ ~~~~2~~63.0 & 
$^o{E}$ ~~~~9~~583.8 & $^e{E}$ ~~~~2~~177.1 \\ \hline

$^o{B}$ ~~~~3~~78.4 & $^e{A}$ ~~~~6~~88.3 & $^o{A}$ ~~~~3~~77.0 & 
$^e{E}$ ~~~~8~~632.8 & $^o{A}$ ~~~~3~~211.3 \\ \hline

$^o{B}$ ~~~~3~~86.8 & $^o{B}$ ~~~~3~~112.8 & $^o{B}$ ~~~~3~~85.3 & 
$^o{A}$ ~~~~9~~ 640.5 & $^o{A}$ ~~~~3~~220.8 \\ \hline

$^e{A}$ ~~~~6~~105.7 & $^o{B}$ ~~~~5~~114.6 & $^e{E}$ ~~~~2~~86.1 & 
$^e{E}$ ~~~~8~~658.3 & $^e{E}$ ~~~~4~~249.9 \\ \hline

$^o{B}$ ~~~~3~~113.4 & $^o{B}$ ~~~~5~~158.4 & $^e{A}$ ~~~~6~~97.1 & 
$^e{A}$ ~~~~8~~767.1 & $^0{B}$ ~~~~5~~278.5 \\ \hline

$^e{E}$ ~~~~4~~117.3 & $^o{A}$ ~~~~1~~165.2 & $^e{A}$ ~~~~6~~98.2 & 
$^e{B}$ ~~~~8~~807.6 & $^o{A}$ ~~~~7~~332.1 \\ \hline

$^o{B}$ ~~~~5~~154.2 & $^o{A}$~~~~ 1~~181.6 & $^o{B}$ ~~~~3~~112.2 & 
$^e{A}$~~~~ 8~~815.8 & $^o{A}$~~~~ 7~~340.8 \\ \hline
\end{tabular}
\end{center}

\newpage
\clearpage

Table 2 : Energies (in unit of K) of few low lying states in
$\rm Fe_8$. The exchange constants corresponding to the various cases
are: case (1) J$_1$ = 150K, J$_2$ = 25K, J$_3$ = 30K , J$_4$ = 50K;
case (2) J$_1$ = 180K, J$_2$ = 153K, J$_3$ = 22.5K , J$_4$ = 52.5K;
case (3) J$_1$ = 195K, J$_2$ = 30K, J$_3$ = 52.5K , J$_4$ = 22.5K.
All the energies are in K.

\begin{center}
\begin{tabular}{|c|c|c|}
\hline
~~~ {\bf Case 1}~~~~ & ~~~ {\bf Case 2}~~~~ & ~~~ {\bf Case 3} ~~~~ \\ \hline

~~~S~~~~~~E(K)~~~ & ~~~S~~~~~~E(K)~~~ & ~~~S~~~~~~E(K)~~~ \\ \hline

~~10~~~~~~0.0~~~ & ~10~~~~~~~0.0~~~ & ~10~~~~~~~0.0~~~ \\ \hline

~~~9~~~~~~13.1~~~ & ~~9~~~~~~~3.4~~~ & ~~~9~~~~~~39.6 ~~~ \\ \hline

~~~8~~~~~~27.3~~~ & ~~~8~~~~~~10.2~~~ & ~~~9~~~~~~54.2~~~ \\ \hline

~~~9~~~~~~41.7~~~ & ~~~~7~~~~~~20.1 ~~~ & ~~~9~~~~~~62.4~~~ \\ \hline
\end{tabular}
\end{center}

\newpage
\clearpage

\leftline{\large {\bf Figure Captions}}
\vspace*{1cm}

\begin{enumerate}

\item A schematic representation of the VB diagrams for eight spin-$1/2$
objects. States $|1>$ and $|2>$ are legal VB diagrams and $|3>$ is an
illegal VB diagram.

\item VB diagram for spin-$1/2$ objects with total spin 1/2 ($|1>$) and
total spin 1 ($|2>$). Their bit representations ($0$ and $1$) as well as the 
unique integer $I_k$ representing them are shown. $P$ and $P^{\prime}$ are 
the phantom sites. $|3>$ is a singlet VB diagram
corresponding to two spin-$1$, a spin-$5/2$ and a spin-$3/2$ object.

\item Effect of operation by the operator (${\hat {\bf S}}_i \cdot {\hat {\bf 
S}}_j - 1/4$) on a state with a singlet line between sites $i$ and $j$ and on 
a state with sites $i$ and $j$ singlet paired with two different sites 
$i^\prime$ and $j^\prime$. 

\item A schematic diagram of the exchange interactions between the Mn ions in 
the $\rm Mn_{12}Ac$ molecule.

\item Spin densities in the ground state (S=10, $\rm M_s$=10) of $\rm Mn_{12}
Ac$ for parameter values $J_1$=215K, $J_2$=85K, $J_3$=85K and $J_4$=-64.5K. 

\item A schematic diagram of the exchange interactions between the Fe ions in 
the $\rm Fe_8$ molecule.

\item Spin density in the ground state (S=10,$ \rm M_s$=10) of $\rm Fe_8$ for 
three different parameter values: (a) J$_1$ = 150K, J$_2$ = 25K, 
J$_3$ = 30K, J$_4$ = 50K, (b) J$_1$ = 180K, J$_2$ = 153K, J$_3$ = 22.5K, J$_4$
= 52.5K, and (c) J$_1$ = 195K, J$_2$ = 30K, J$_3$ = 52.5K, J$_4$ = 22.5K.

\item Schematic picture of the frustrated and dimerized spin chain.

\item Classical phase diagram of the spin chain in the $J_2 - \delta$ plane.

\item `Phase' diagram for the spin-$1$ chain in the $J_2 - \delta$ plane.

\item Doubly degenerate ground states (a) and (b) of the $J-J_2$ chain (see 
Fig. 8 for $\delta=0$) at $J_2=J/2$. The solid line between sites $i$ and $j$
represents a singlet, $[|\uparrow_i \downarrow_j> - |\downarrow_i 
\uparrow_j>]/{\sqrt 2}$.

\item `Phase' diagram for the spin-$1/2$ chain in the $J_2 - \delta$ plane. 

\item Gap $\Delta$ vs $J$ for coupled spin chains ($\delta = 1$). 
Spin-$1/2$ and spin-$1$ data are indicated by crosses and circles respectively.

\item Expectation values of the $z$-components of the two spins vs the unit 
cell index $n$ for an alternating spin chain. The upper and lower points are 
for the spin-$s_1$ and the spin-$s_2$ sites respectively.

\item Subtracted two-spin correlation functions as a function of distance 
between the two spins. (a) spin-$s_1$ spin-$s_1$ correlations, (b) spin-$s_2$ 
spin-$s_2$ correlations, and (c) spin-$s_1$ spin-$s_2$ correlations. In each 
figure, squares correspond to ($3/2,1$), circles to ($3/2,1/2$) and triangles 
to ($1,1/2$) systems.

\item Energy difference (units of $J$) between the ground state and the lowest
energy state with spin $S=S_G -1$ as a function of inverse system size. $S_G$ 
is the total spin of the ground state.

\item Excitation gap (units of $J$) from the ground state (spin $S=S_G$) to 
the state with spin $S=S_G +1$ as a function of the inverse system size. 

\item Excitation gap (units of $J$) to the state with spin $S=S_G +1$ from 
the ground state ($S=S_G$) as a function of $\delta$ for the dimerized 
alternating chain. The exponent is $1.0 \pm 0.01$ for all three systems.

\item Plot of magnetization per site as a function of temperature 
$T$ for four different values of the magnetic fields $B$. Squares are for $B=
0.1 J/g\mu_B$, circles for $B=0.5 J/g\mu_B$, triangles for $B=J/g\mu_B$ 
and diamonds for $B=2 J/g\mu_B$.

\item $\chi T$ ( defined in the text) per site as a function of 
temperature $T$ for various magnetic fields $B$. Zero field results are shown 
by squares, $B=0.01 J/g\mu_B$ by circles, $B=0.1 J/g\mu_B$ by triangles 
and $B=J/g\mu_B$ by diamonds.

\item Specific heat per site as a function of temperature $T$ for four 
different values of magnetic fields $B$. Zero field data are shown by squares,
$B=0.01 J/g\mu_B$ by circles, $B=0.1 J/g\mu_B$ by triangles and $B=J/g
\mu_B$ by diamonds.

\item The energy/site in units of $J$ vs $1/N$ at the $m_s =1/2$ plateau, for
the three-chain ladder with $J/ \jp =1/3$. The curves indicate quadratic fits 
for (a) $E_0 (M+1,N)$, (b) $E_0 (M,N)$, and (c) $E_0 (M-1,N)$.

\item Plateau widths vs $1/N$ for (a) $m_s =1/2$, (b) $m_s =0$, and (c) $m_s 
=1$. 

\item Correlation function $<{\hat S}^+_{2,l}{\hat S}^-_{2,n}>$ at the $m_s 
=1/2$ plateau for $J/\jp =1/3$.

\item Correlation functions $<{\hat S}^+_{1,l} {\hat S}^-_{2,n}>$ in the 
$m_s =1$ state for $J/\jp =1/3$.

\item Magnetization vs magnetic field at six different temperatures, for a 
$36$-site system with OBC along the rungs and $J/ \jp =1/3$.

\item Specific heat in units of $k_B$ vs magnetic field at six different 
temperatures, for a $36$-site system with OBC along the rungs and $J/ \jp 
=1/3$.

\item Comparisons of specific heat and susceptibility vs the magnetic field
for a $36$-site systems with OBC and PBC along the rungs.

\item Comparison of the energy spectra in units of $J$ of the $12$-site
system with OBC and PBC along the rungs. The energies in the $S^z = 2$ sector 
are shown for $J/ \jp =1/3$.

\end{enumerate}

\newpage
\clearpage

\begin{figure}
\begin{center}
\epsfig{figure=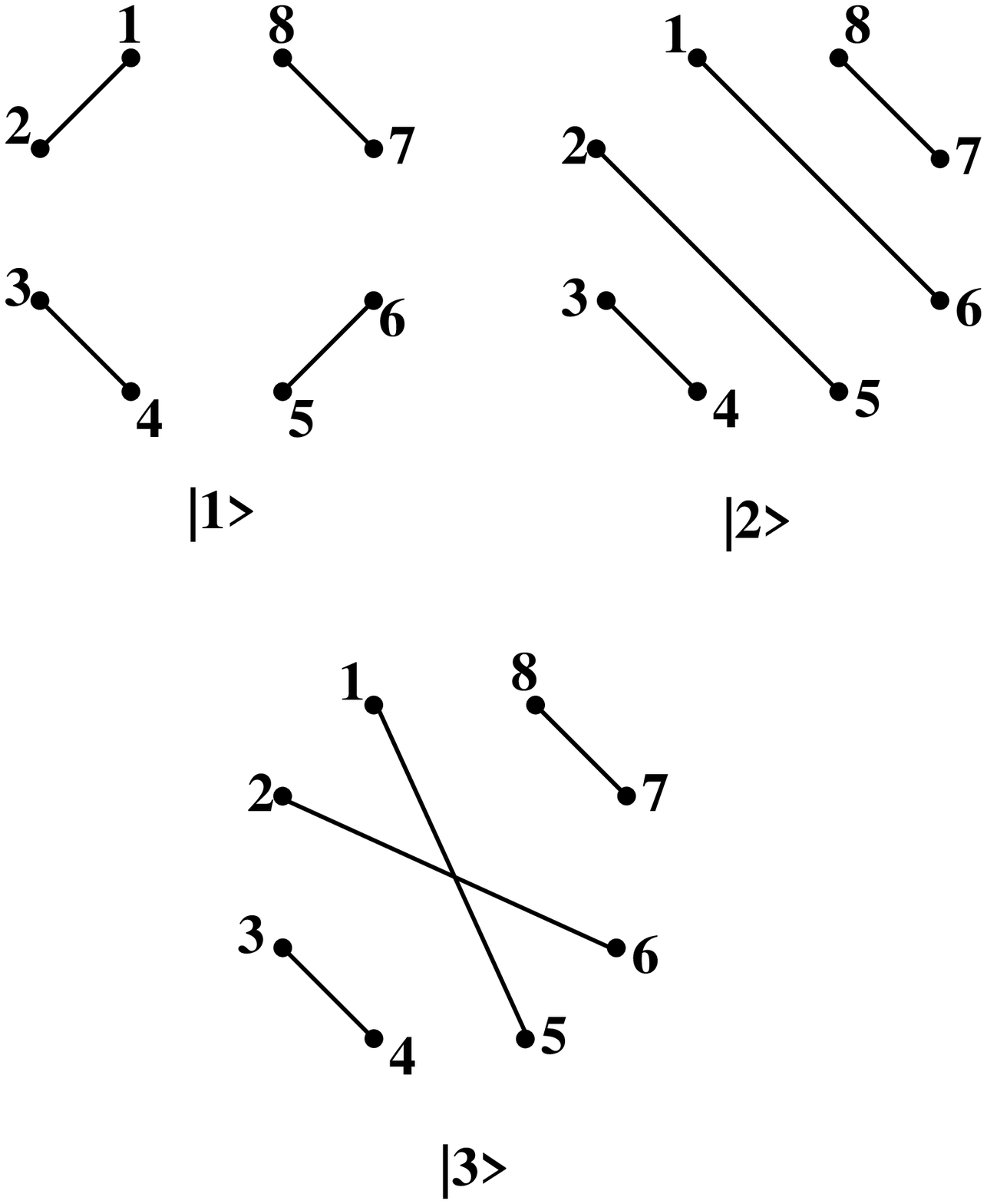,height=16cm}
\end{center}
\vspace*{0.6cm}
\centerline{Fig. 1}
\label{1vb1}
\end{figure}

\newpage
\clearpage

\begin{figure}
\begin{center}
\epsfig{figure=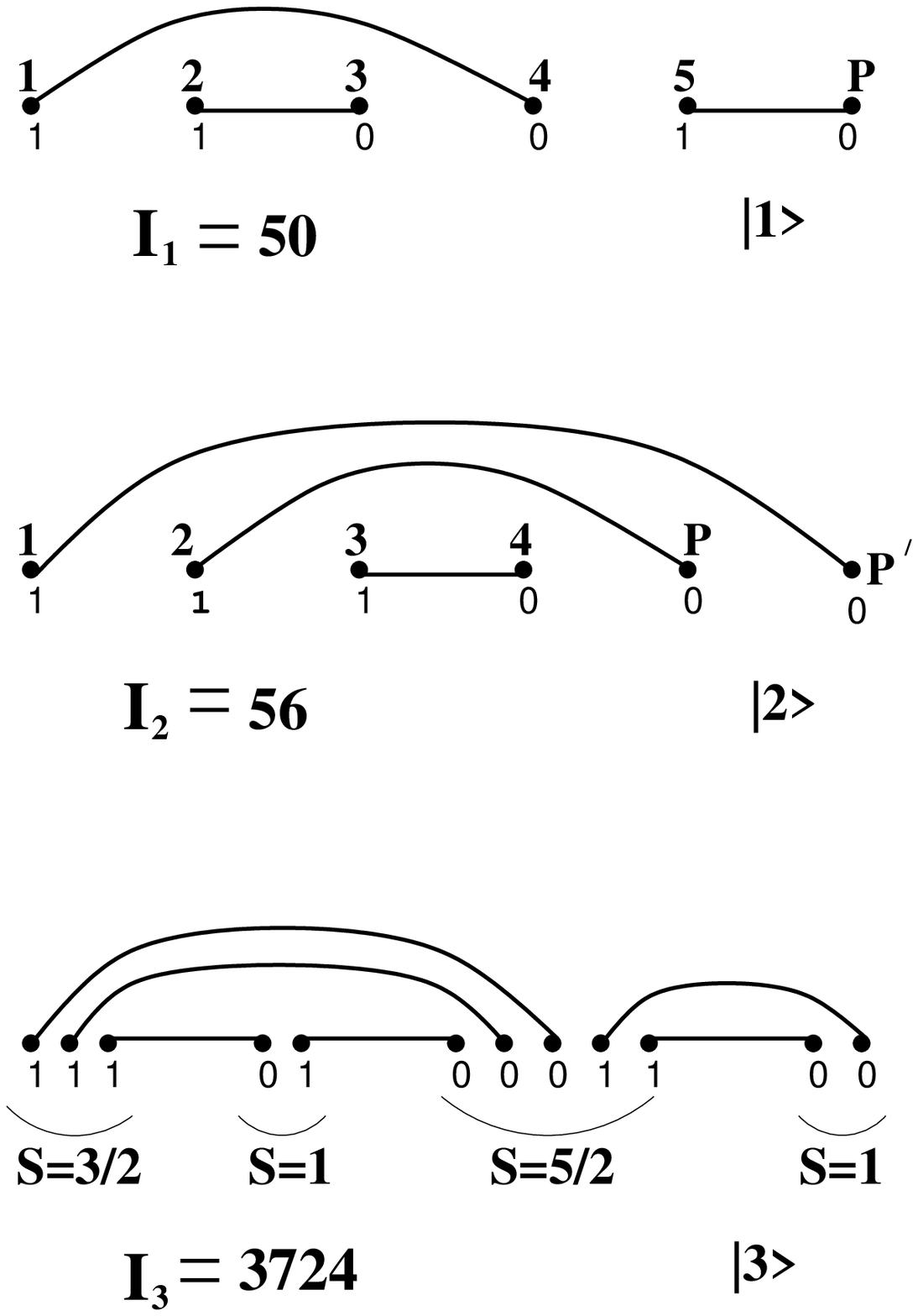,height=16cm}
\end{center}
\vspace*{0.6cm}
\centerline{Fig. 2}
\label{2vb2}
\end{figure}

\newpage
\clearpage

\begin{figure}
\begin{center}
\epsfig{figure=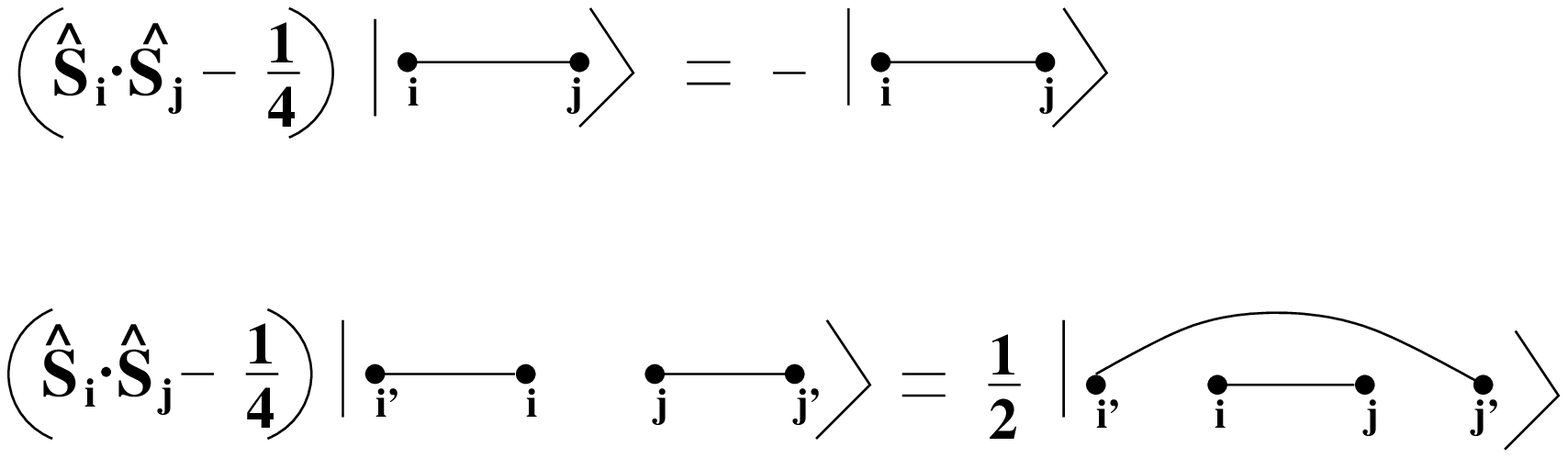,width=16cm}
\end{center}
\vspace*{1.0cm}
\centerline{Fig. 3}
\label{3vb3}
\end{figure}

\newpage
\clearpage

\begin{figure}
\begin{center}
\epsfig{figure=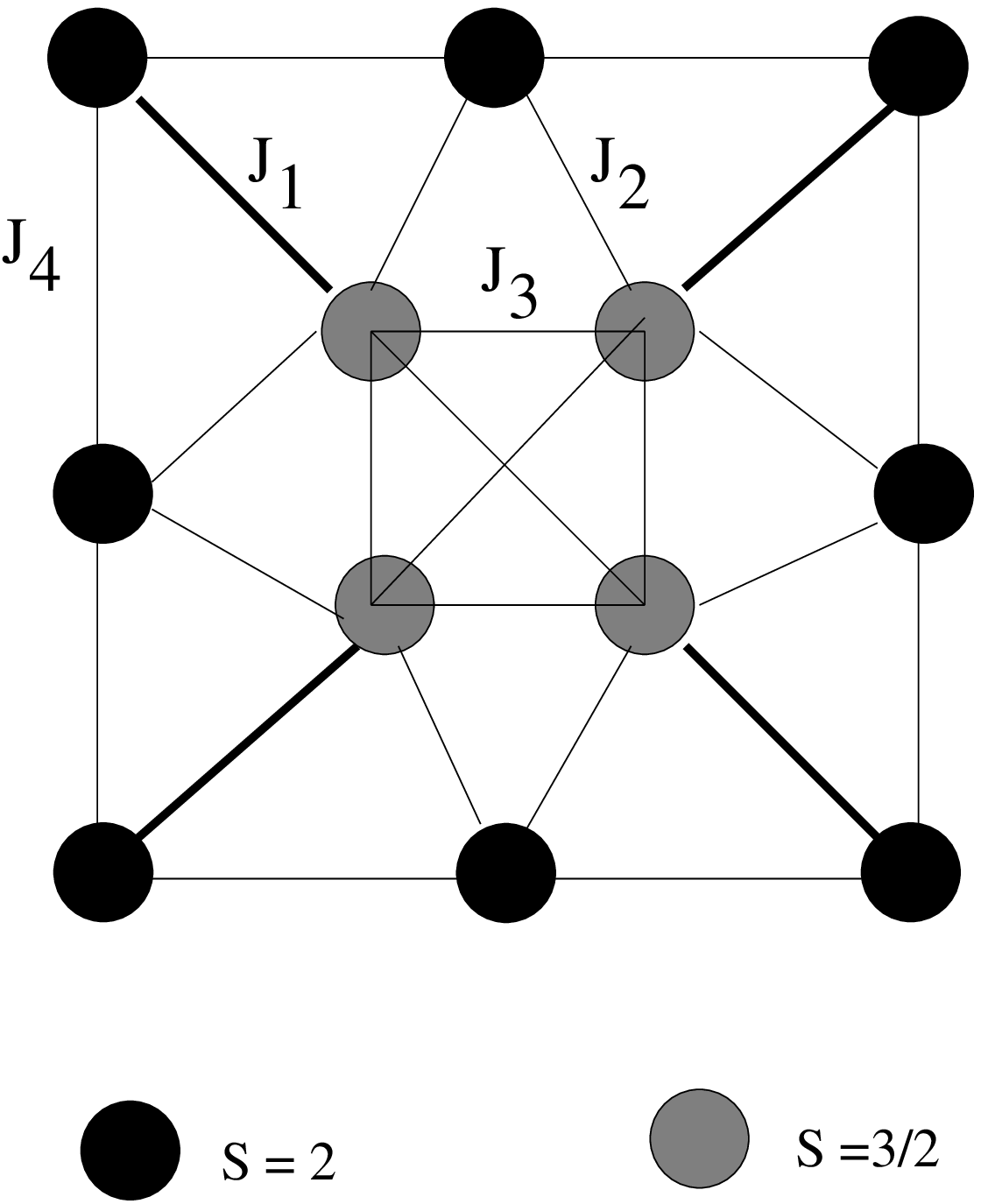}
\end{center}
\vspace*{1.0cm}
\centerline{Fig. 4}
\label{4mn12ac}
\end{figure}

\newpage
\clearpage

\begin{figure}
\begin{center}
\epsfig{figure=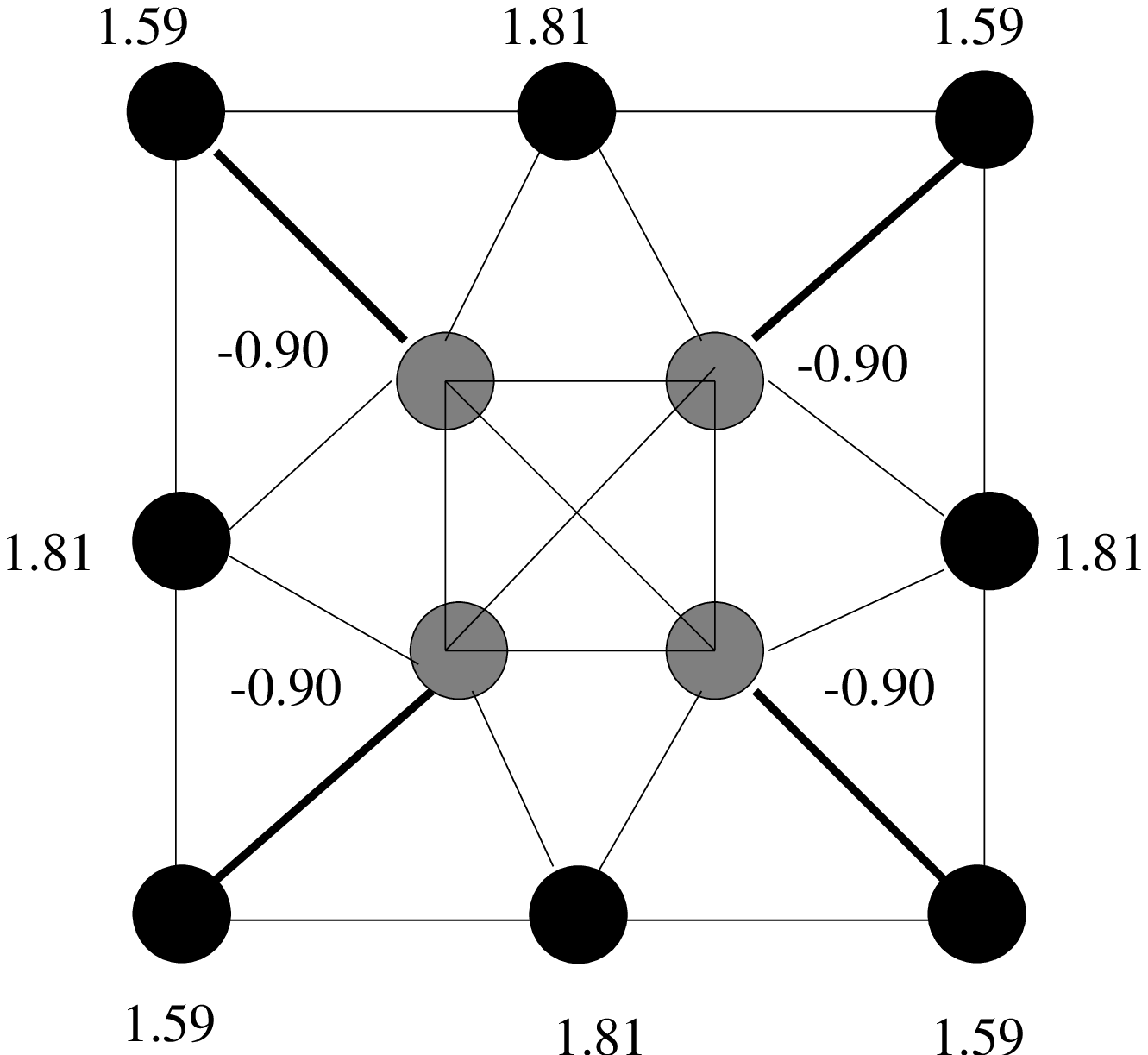}
\end{center}
\vspace*{1.0cm}
\centerline{Fig. 5}
\label{5mn12spn}
\end{figure}

\newpage
\clearpage

\begin{figure}
\begin{center}
\epsfig{figure=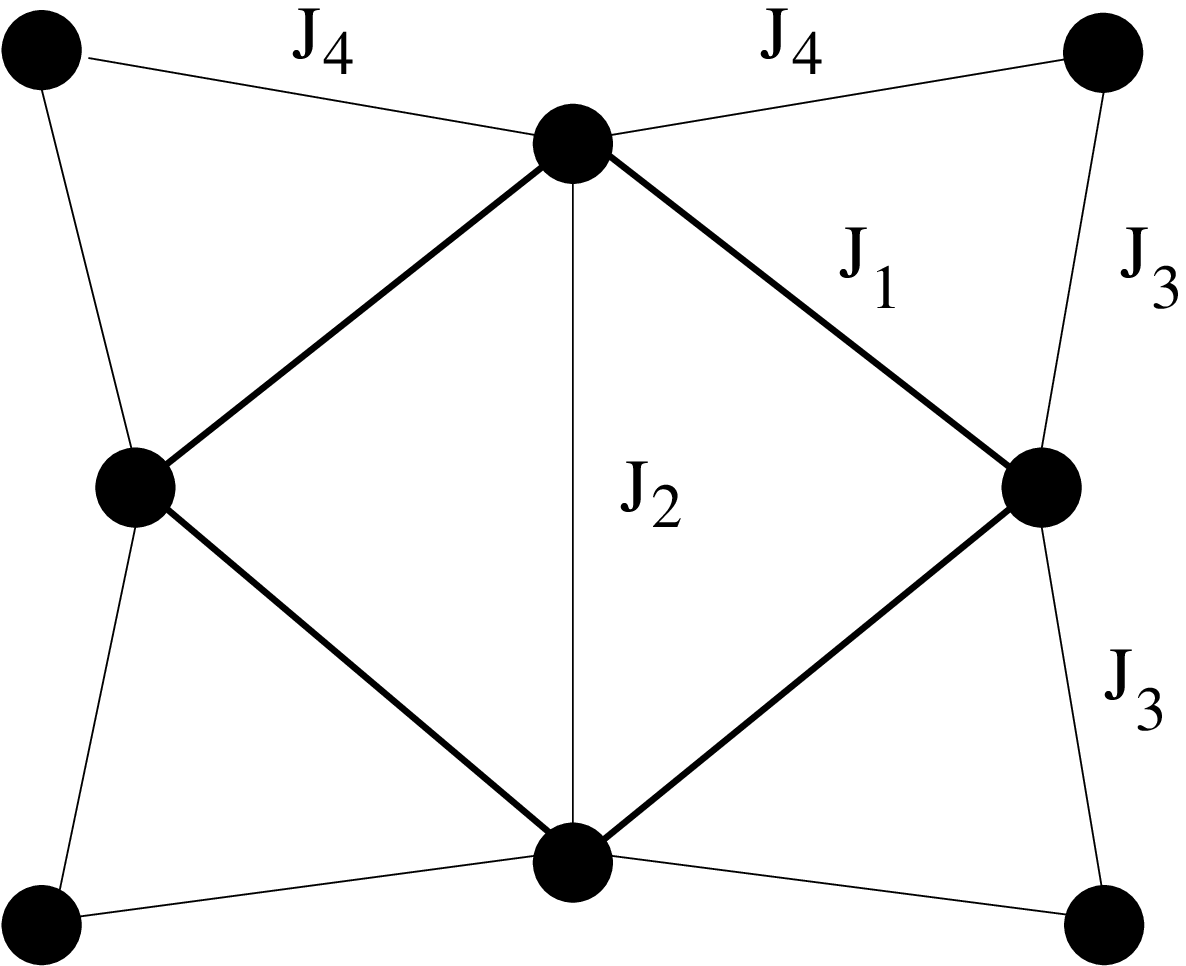}
\end{center}
\vspace*{1.0cm}
\centerline{Fig. 6}
\label{6fe8}
\end{figure}

\newpage
\clearpage

\begin{figure}
\begin{center}
\epsfig{figure=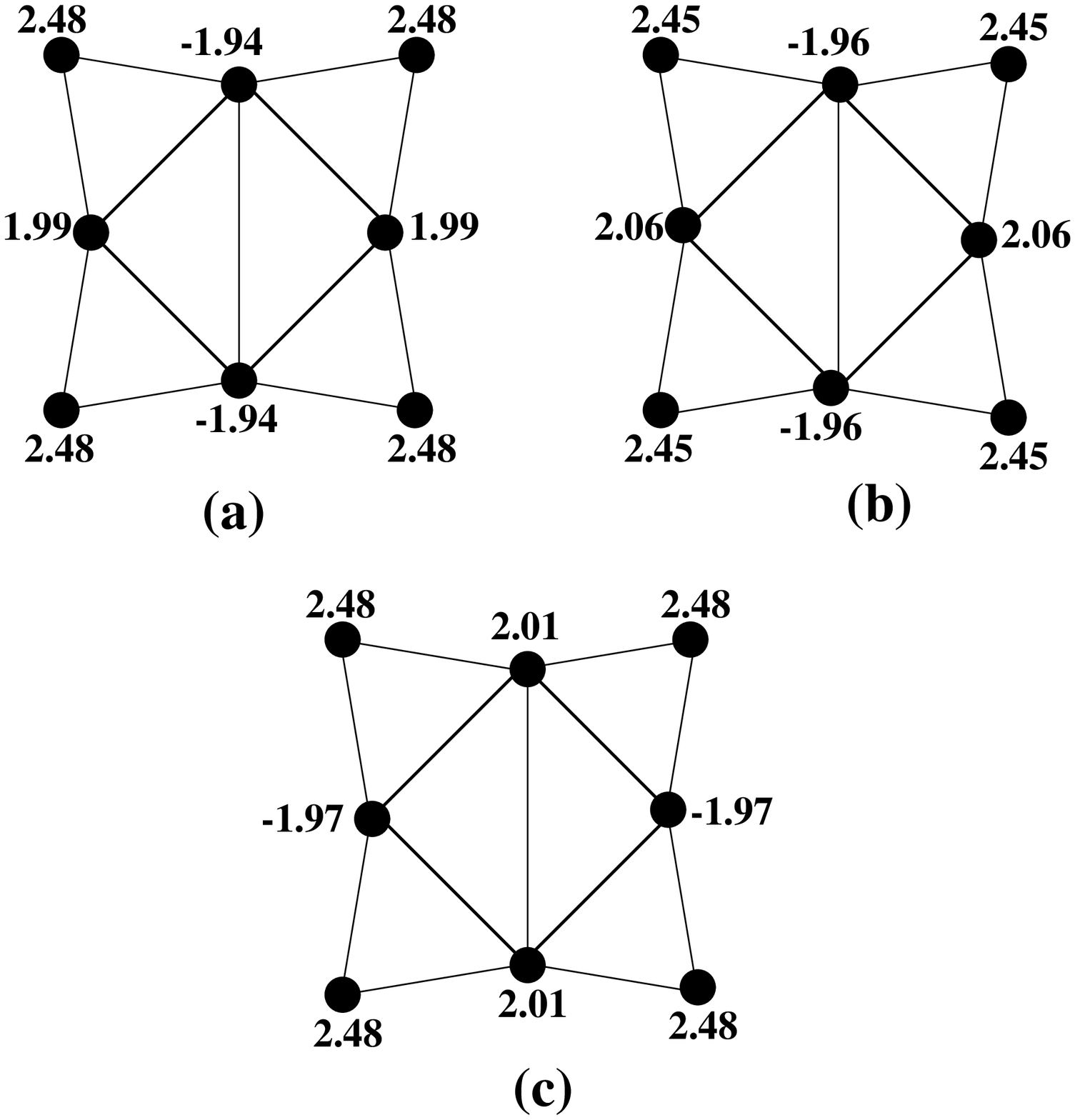,width=16cm}
\end{center}
\vspace*{0.4cm}
\centerline{Fig. 7 }
\label{7fe8spn}
\end{figure}

\newpage
\clearpage

\begin{figure}
\begin{center}
\epsfig{figure=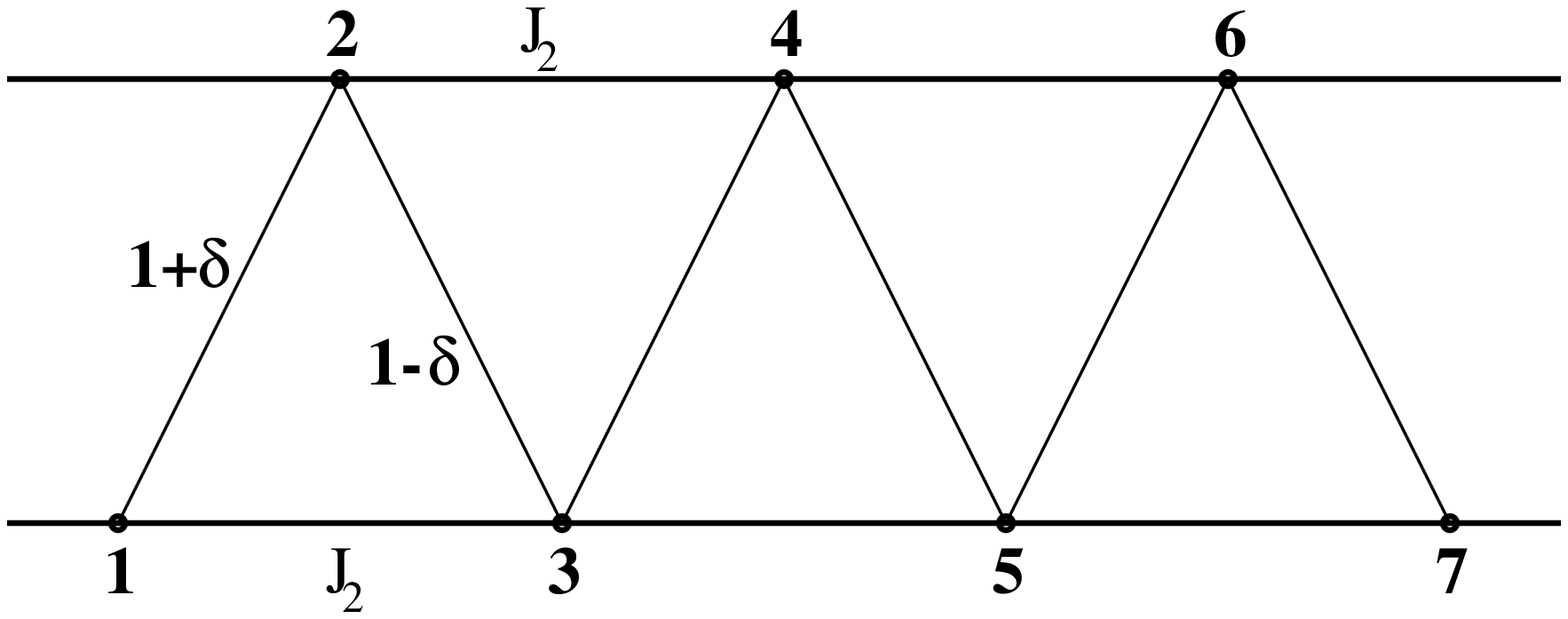,angle=270,width=16cm}
\end{center}
\vspace*{0.8cm}
\centerline{Fig. 8}
\label{11j1j2delta}
\end{figure}

\newpage
\clearpage

\begin{figure}
\begin{center}
\epsfig{figure=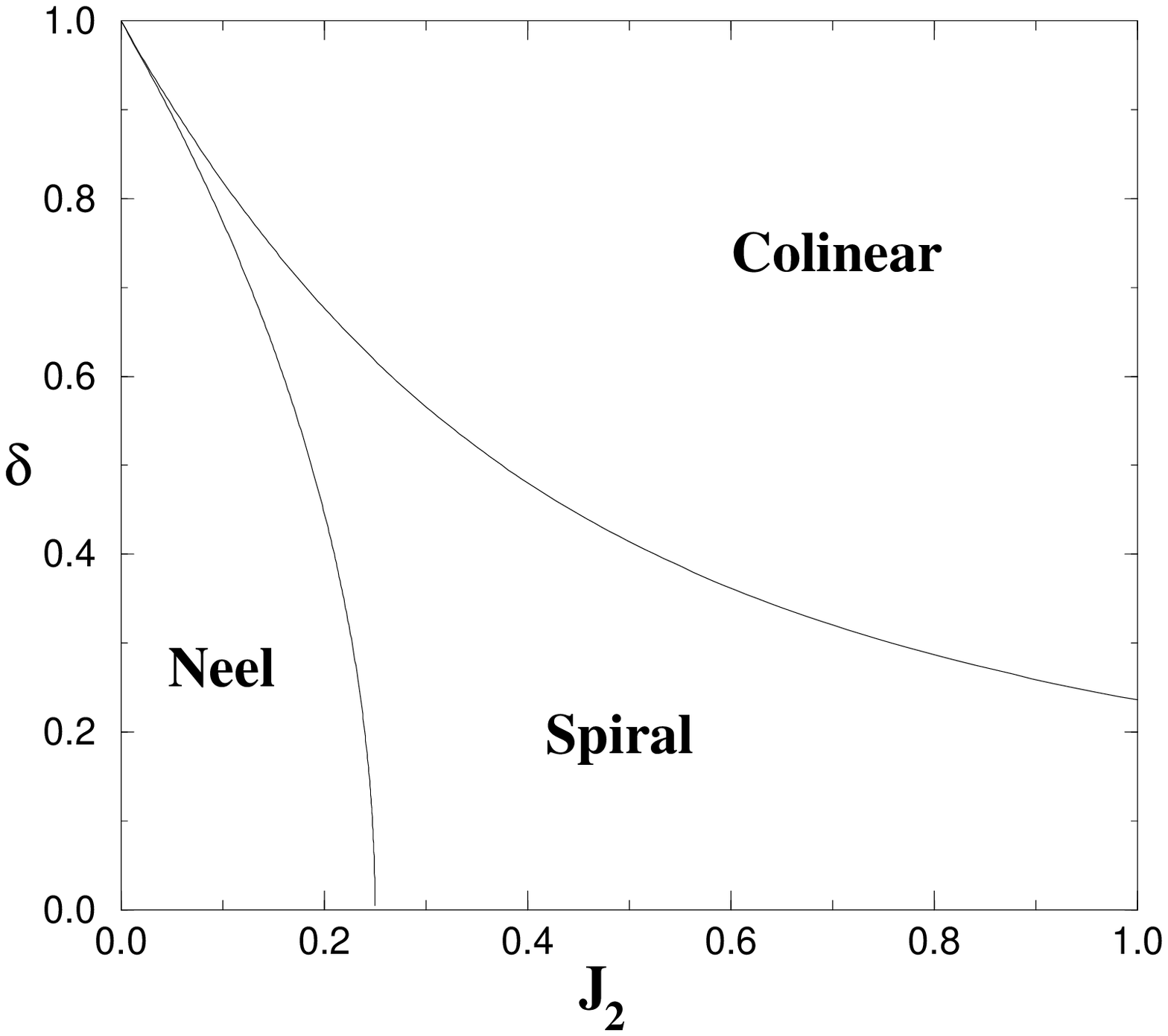,width=16cm}
\end{center}
\centerline{Fig. 9}
\label{8classical}
\end{figure}

\newpage
\clearpage

\begin{figure}
\begin{center}
\epsfig{figure=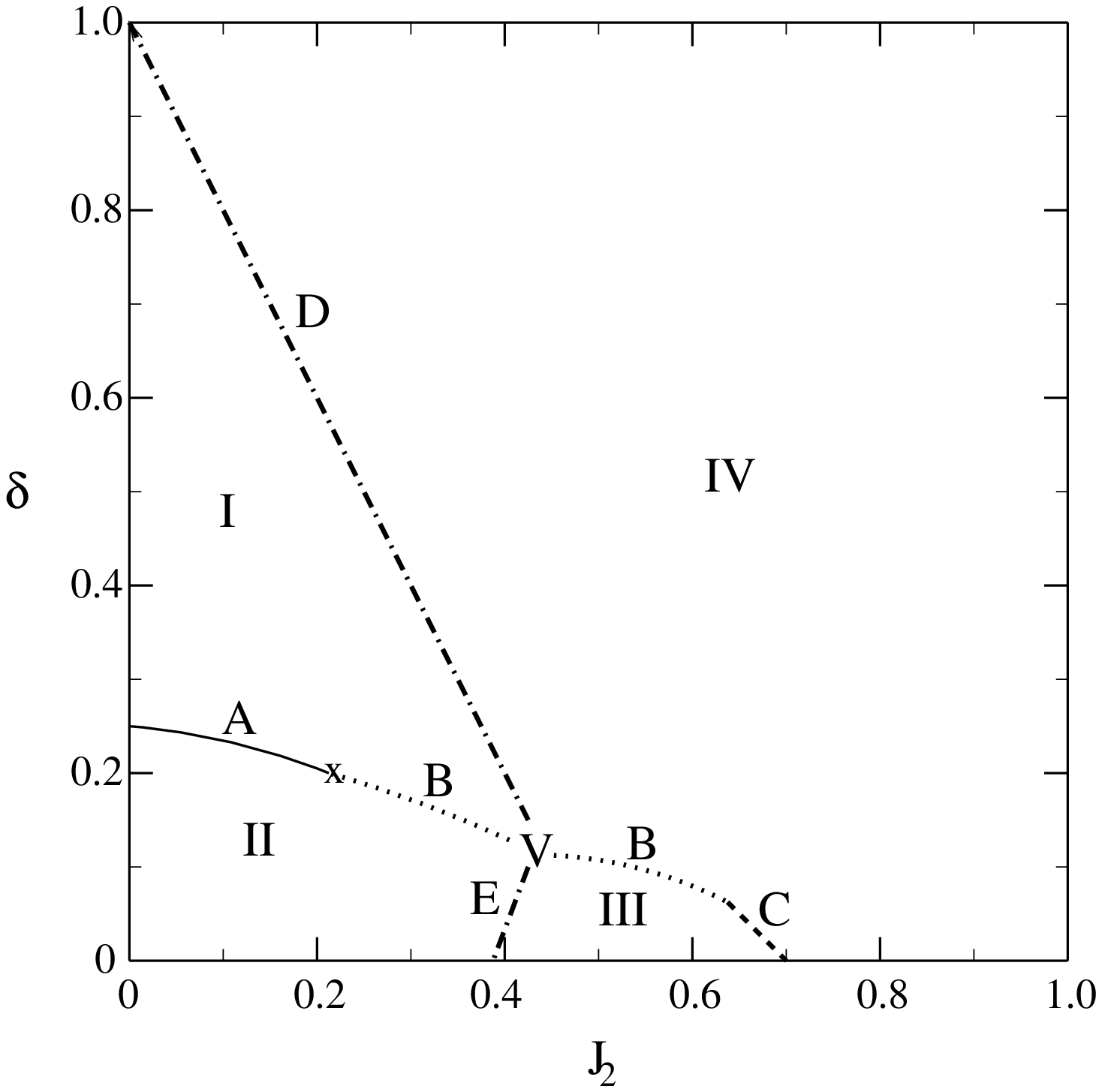}
\end{center}
\vspace*{0.2cm}
\centerline{Fig. 10}
\label{9spin1phase}
\end{figure}

\newpage
\clearpage

\begin{figure}
\begin{center}
\epsfig{figure=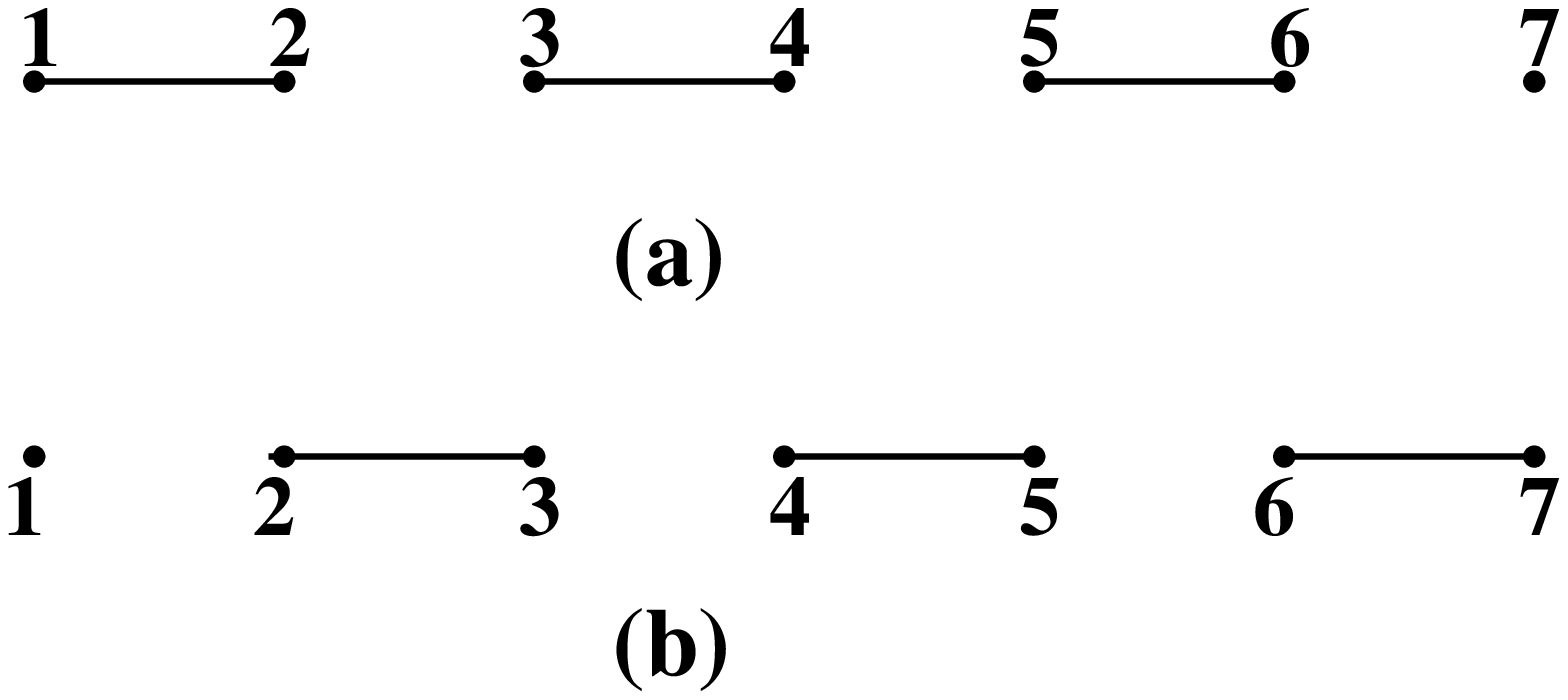}
\end{center}
\vspace*{1.2cm}
\centerline{Fig. 11}
\label{10fdsc1}
\end{figure}

\newpage
\clearpage

\begin{figure}
\begin{center}
\epsfig{figure=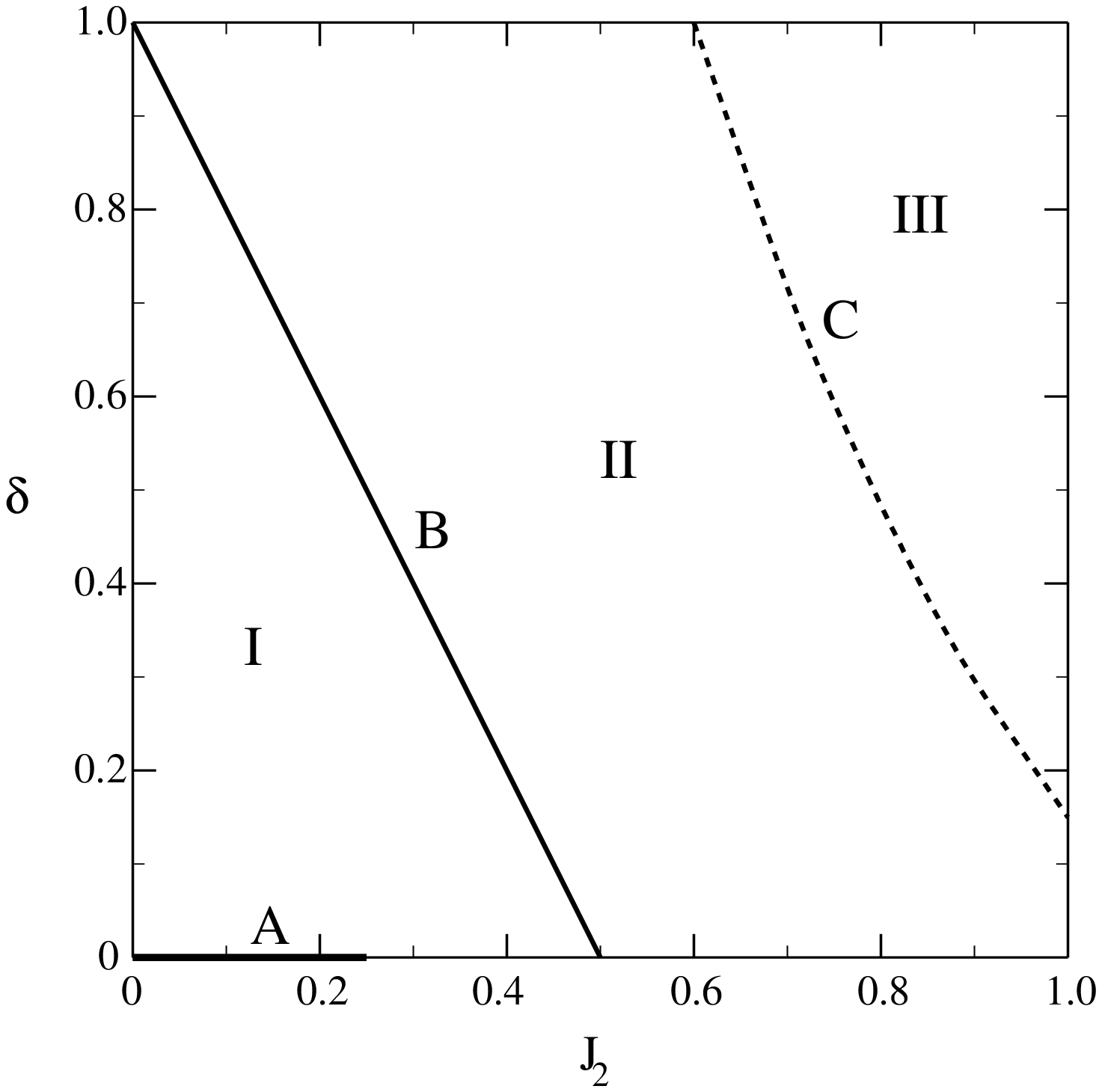}
\end{center}
\vspace*{0.4cm}
\centerline{Fig. 12}
\label{12spinhalfphase}
\end{figure}

\newpage
\clearpage

\begin{figure}
\begin{center}
\epsfig{figure=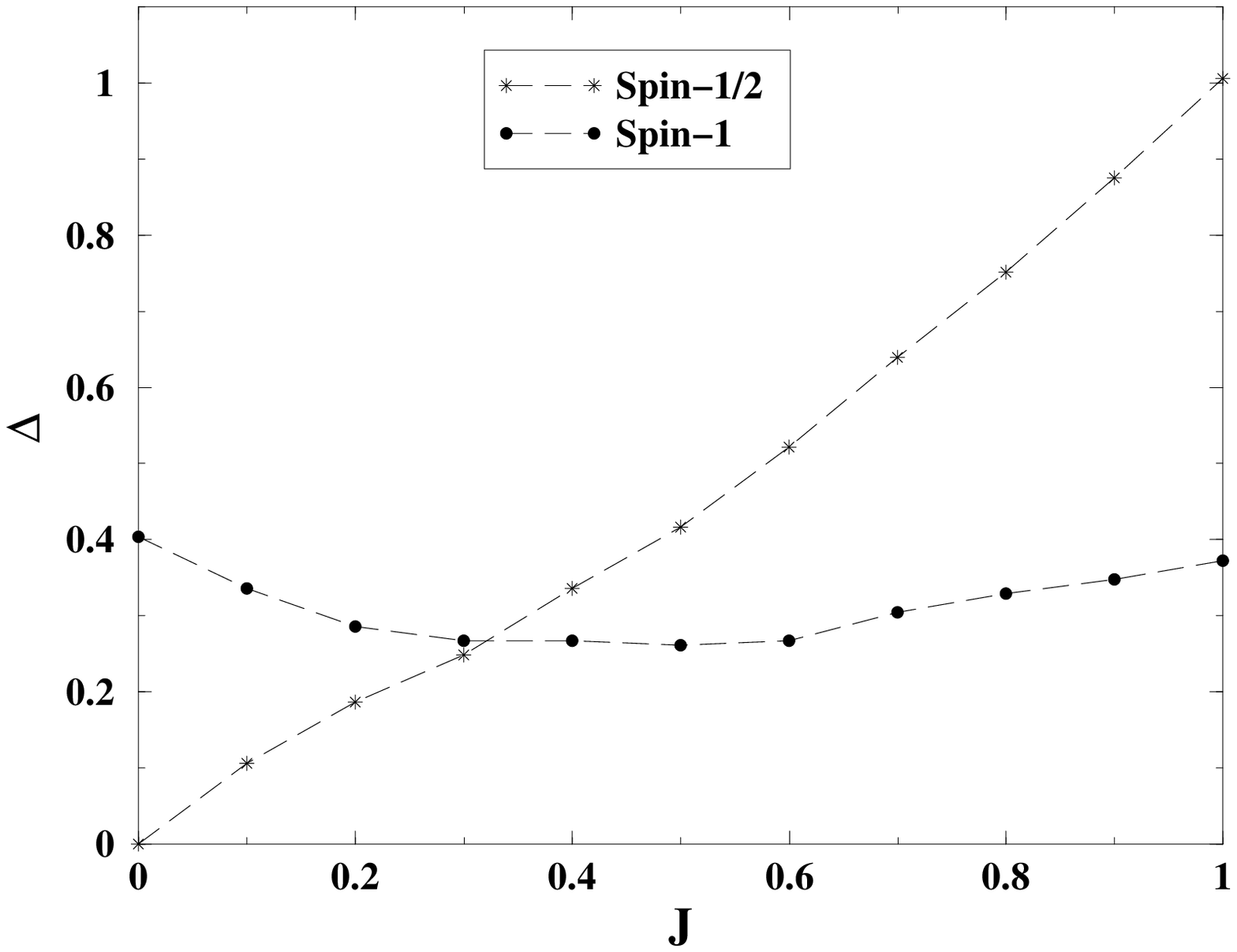}
\end{center}
\vspace*{0.2cm}
\centerline{Fig. 13}
\label{13coupled}
\end{figure}

\newpage
\clearpage

\begin{figure}
\begin{center}
\epsfig{figure=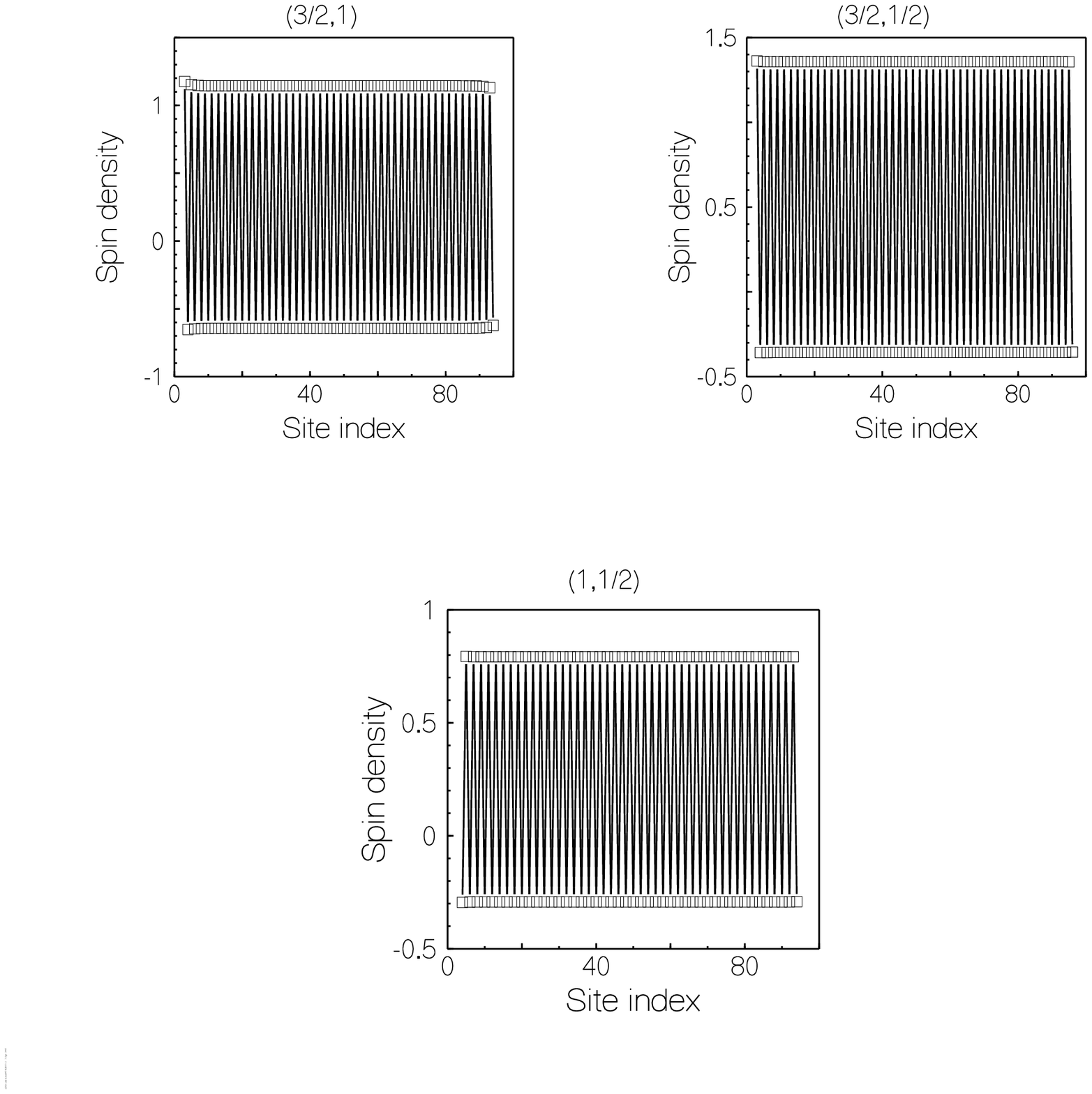,width=18cm}
\end{center}
\vspace*{-0.6cm}
\centerline{Fig. 14}
\label{14s1s2spn}
\end{figure}

\newpage
\clearpage

\begin{figure}
\begin{center}
\epsfig{figure=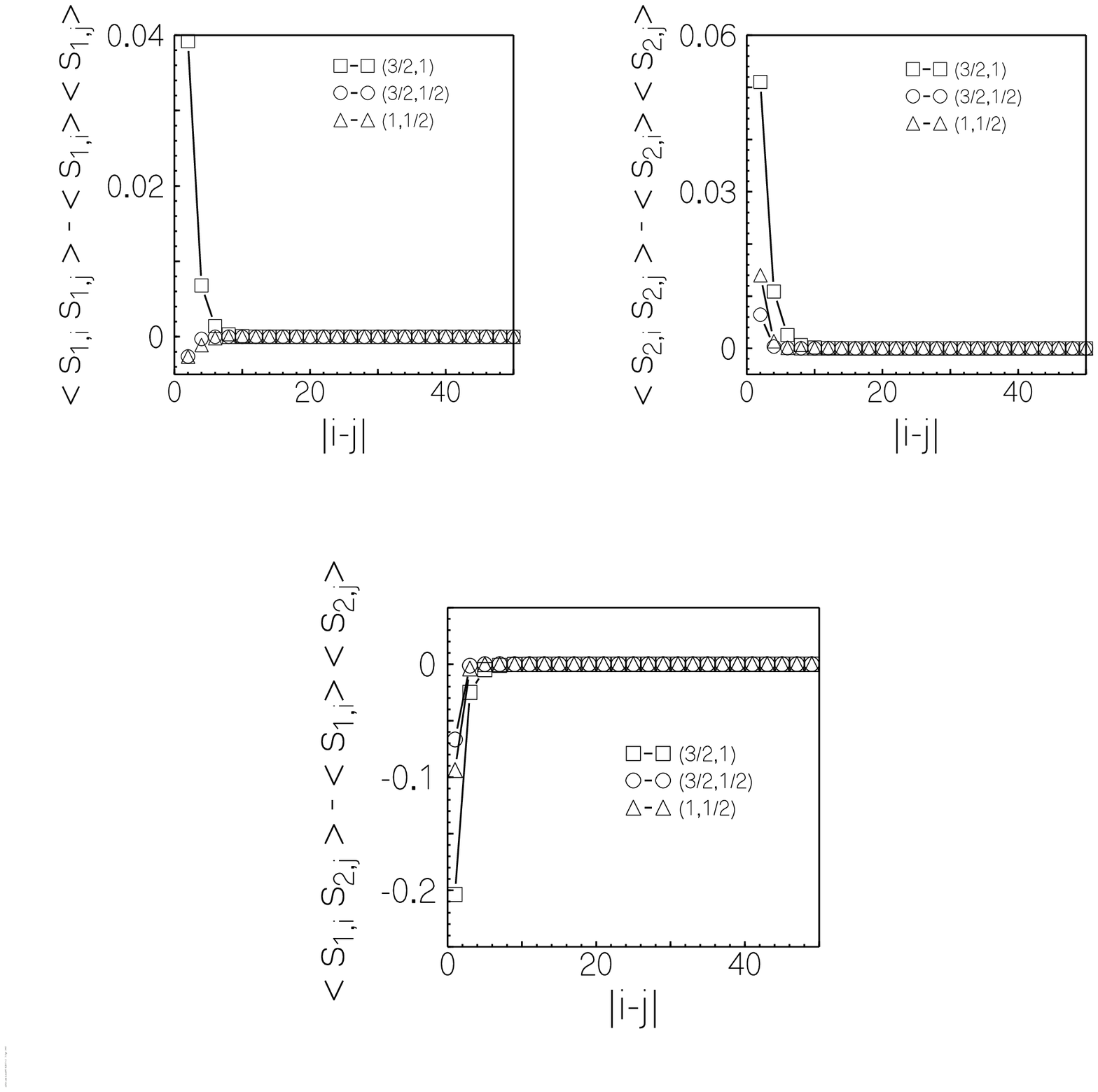,width=18cm}
\end{center}
\vspace*{-0.6cm}
\centerline{Fig. 15}
\label{15s1s2corr}
\end{figure}

\newpage
\clearpage

\begin{figure}
\begin{center}
\epsfig{figure=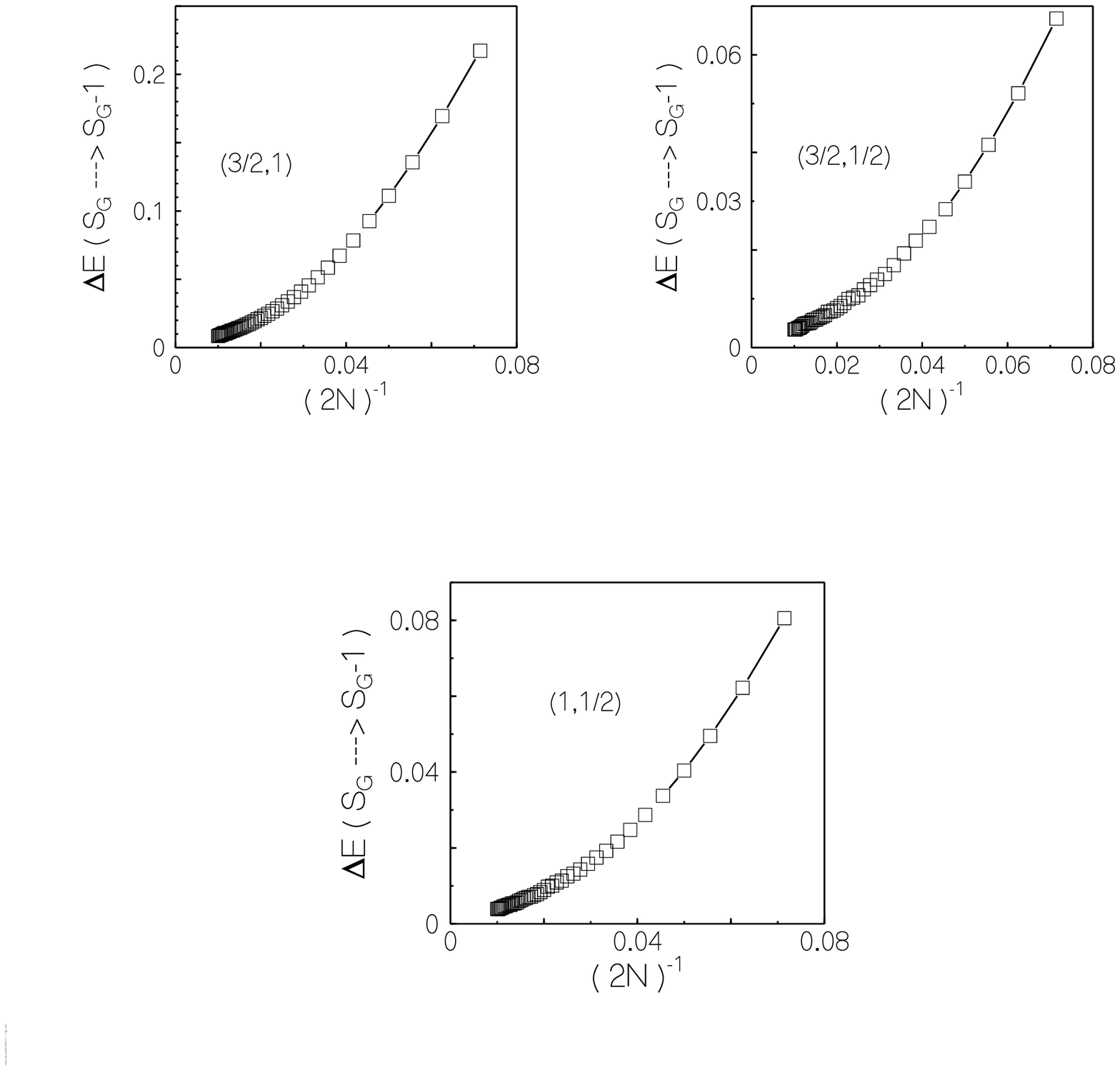,width=18cm}
\end{center}
\vspace*{-0.6cm}
\centerline{Fig. 16}
\label{16s1s2gapless}
\end{figure}

\newpage
\clearpage

\begin{figure}
\begin{center}
\epsfig{figure=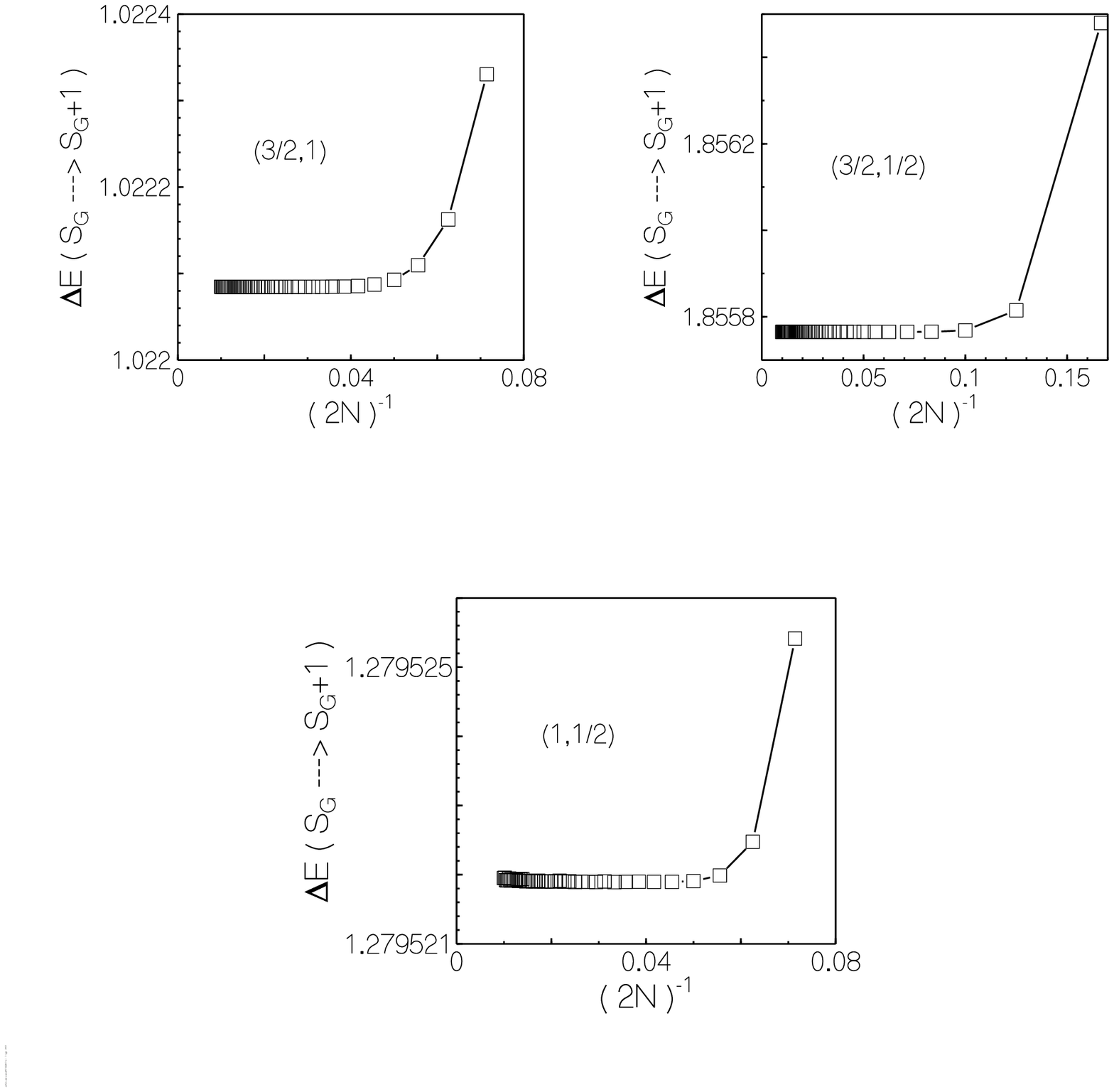,width=18cm}
\end{center}
\vspace*{-0.6cm}
\centerline{Fig. 17}
\label{17s1s2gapped}
\end{figure}

\newpage
\clearpage

\begin{figure}
\begin{center}
\epsfig{figure=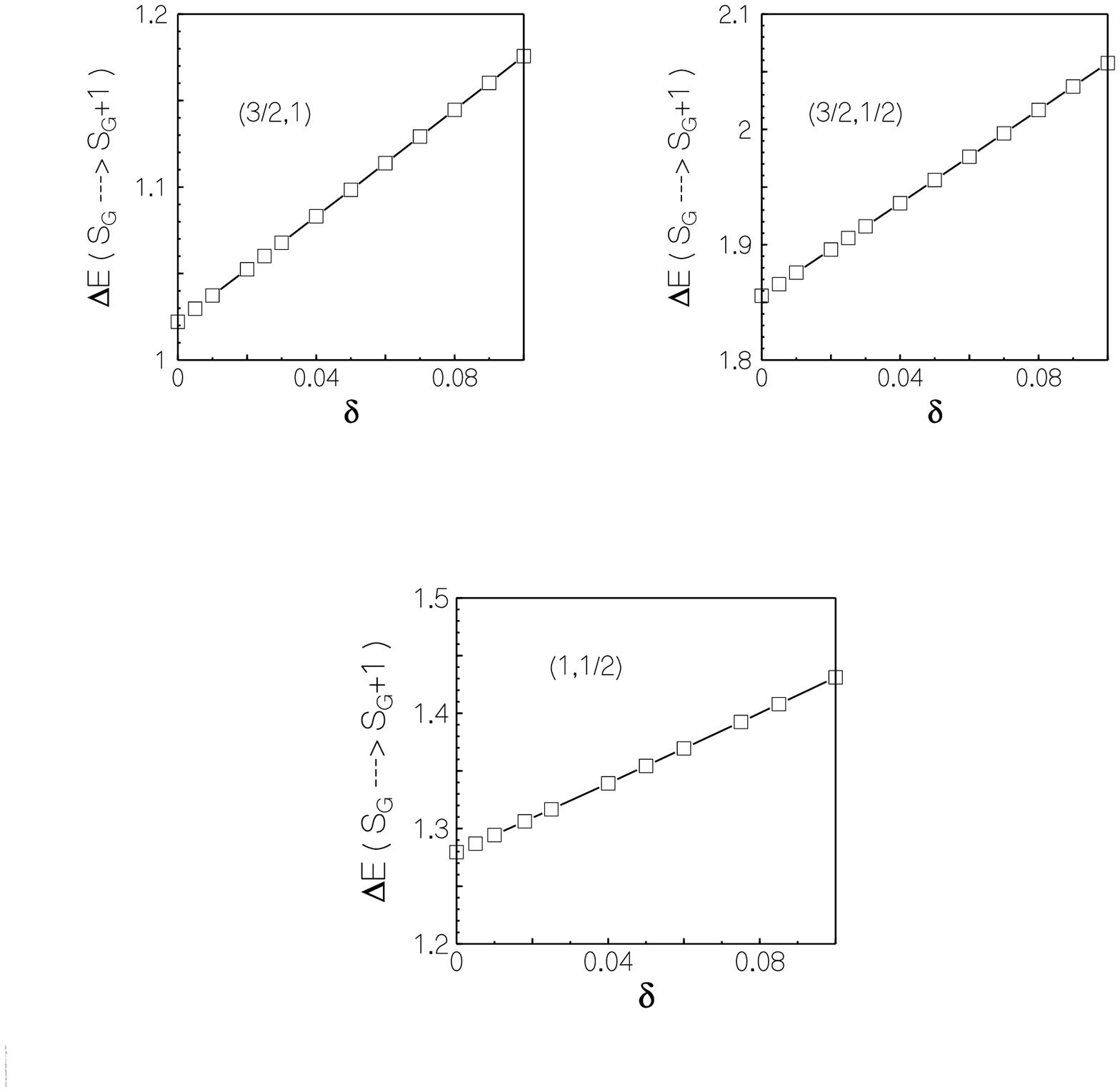,width=18cm}
\end{center}
\vspace*{-0.6cm}
\centerline{Fig. 18}
\label{18s1s2gapdelta}
\end{figure}

\newpage
\clearpage

\begin{figure}
\begin{center}
\epsfig{figure=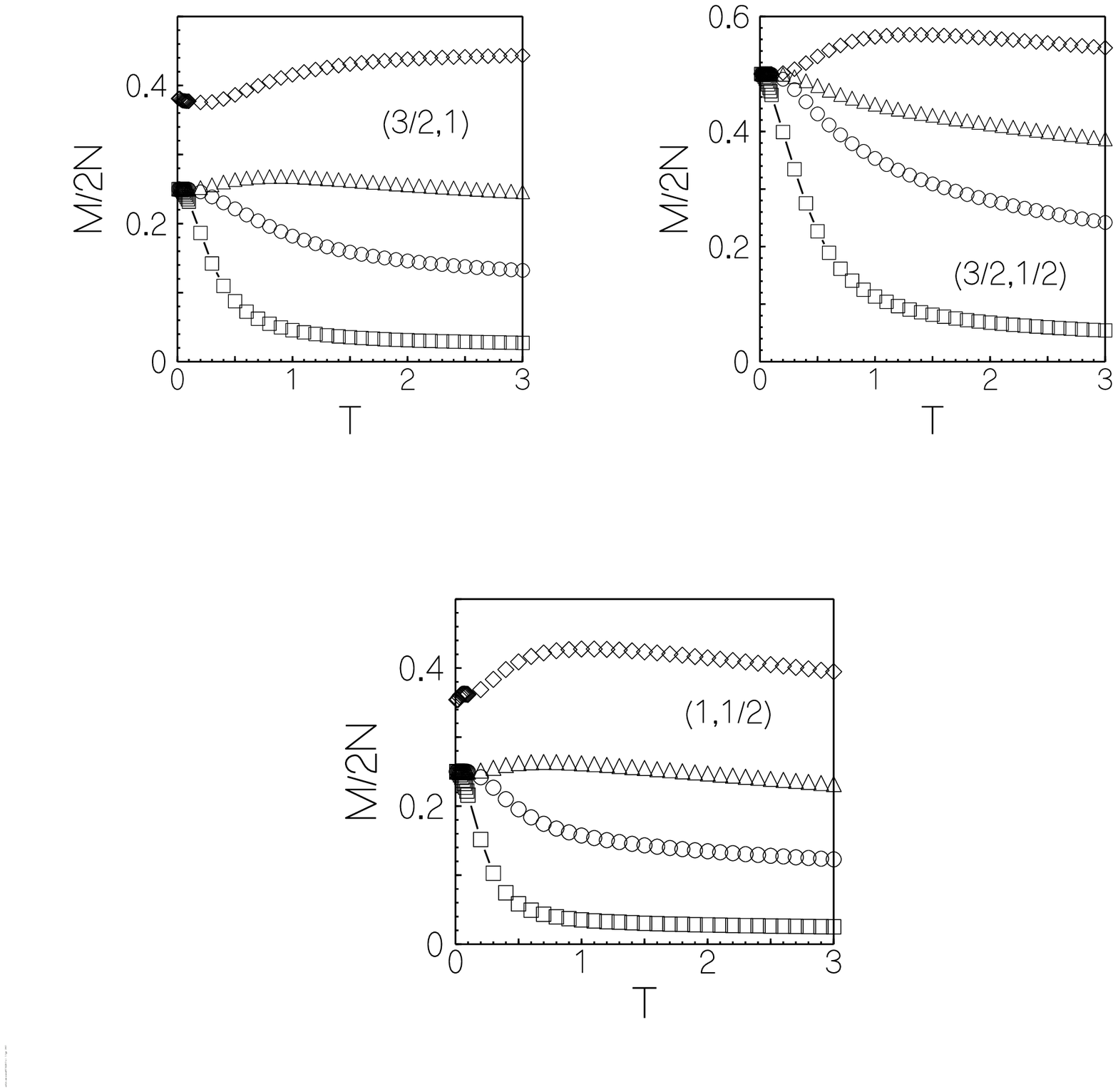,width=18cm}
\end{center}
\vspace*{-0.6cm}
\centerline{Fig. 19}
\label{19s1s2mvst}
\end{figure}

\newpage
\clearpage

\begin{figure}
\begin{center}
\epsfig{figure=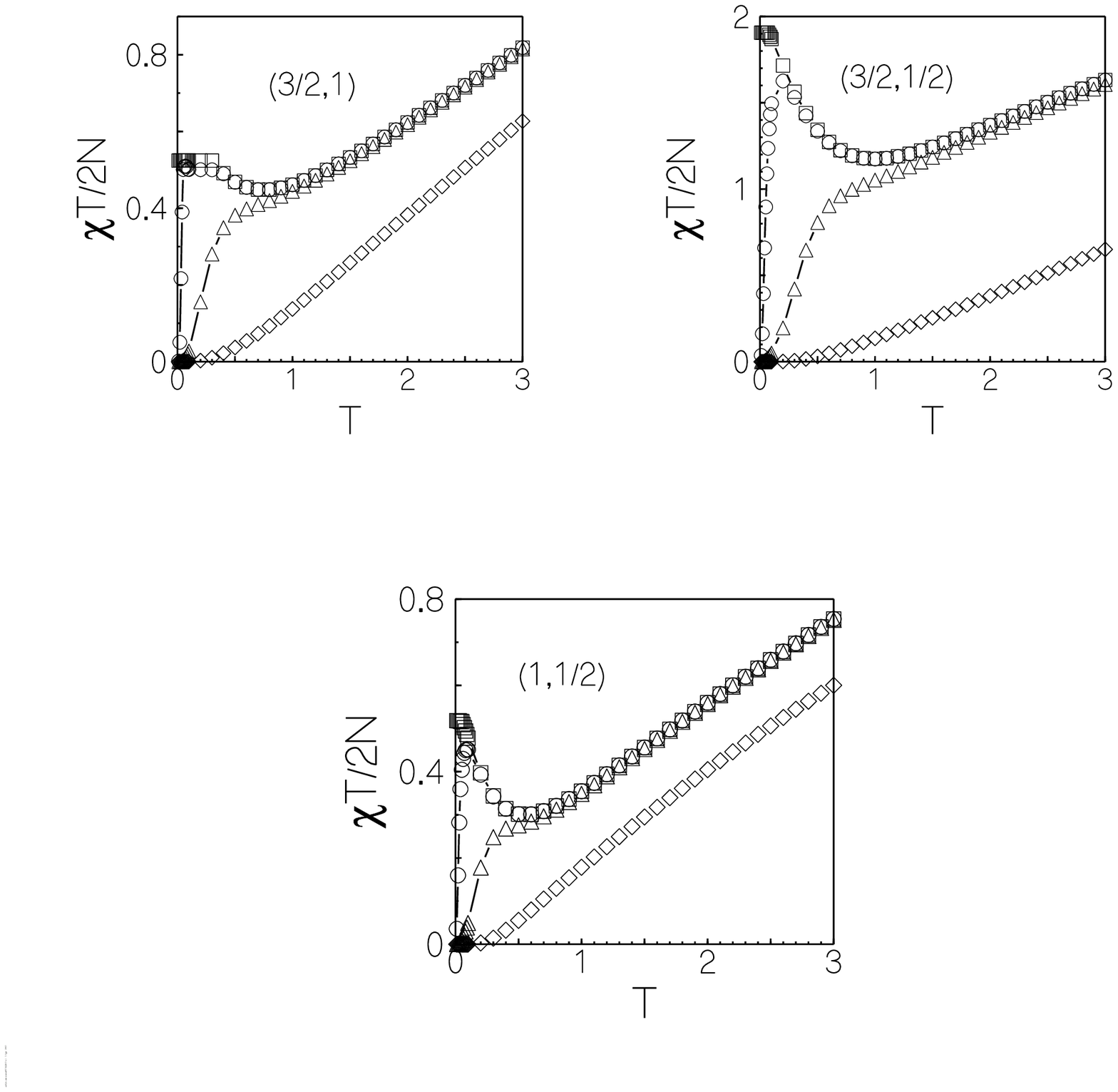,width=18cm}
\end{center}
\vspace*{-0.6cm}
\centerline{Fig. 20}
\label{20s1s2chi}
\end{figure}

\newpage
\clearpage

\begin{figure}
\begin{center}
\epsfig{figure=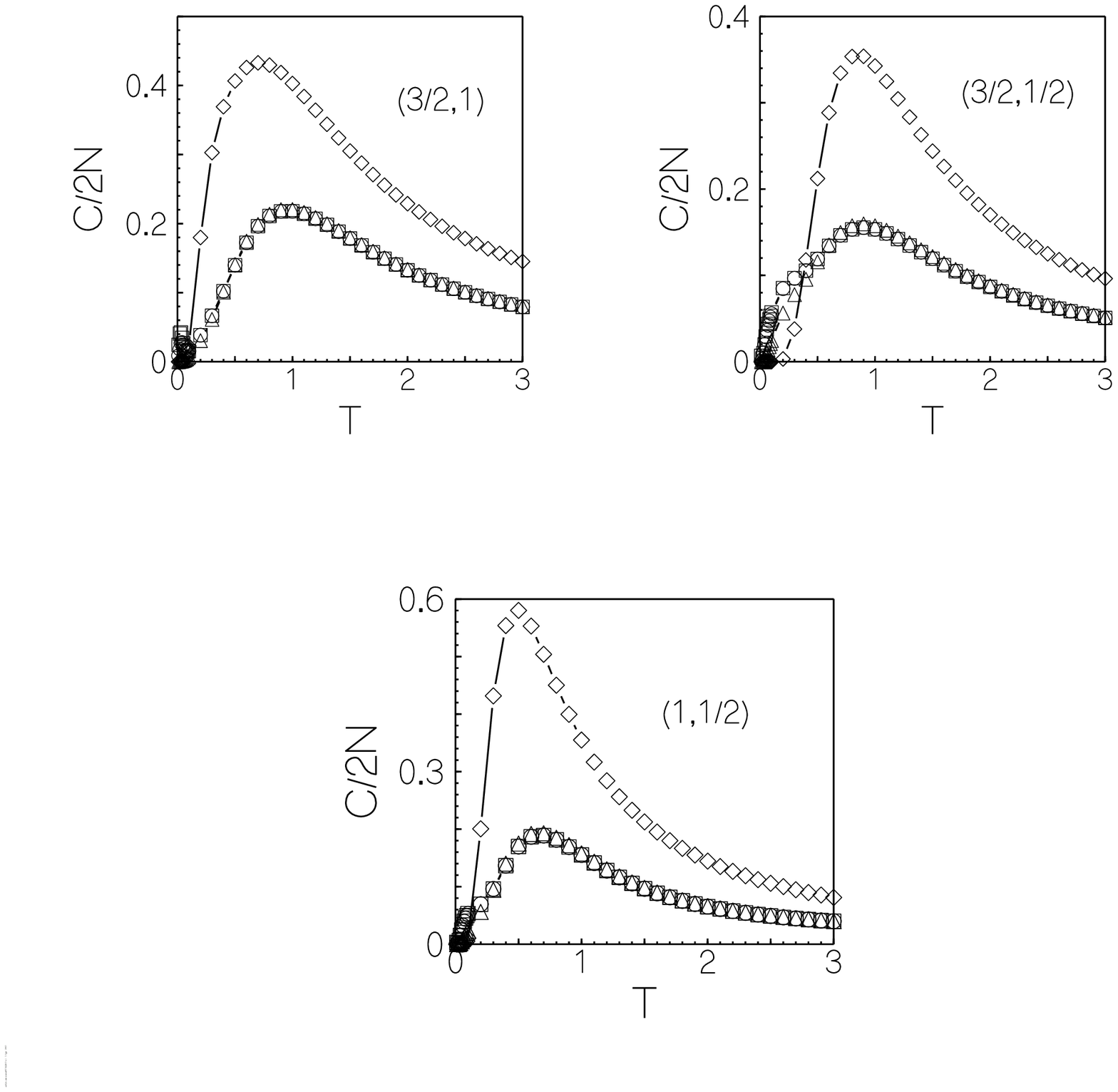,width=18cm}
\end{center}
\vspace*{-0.6cm}
\centerline{Fig. 21}
\label{21s1s2cvst}
\end{figure}

\newpage
\clearpage

\begin{figure}
\begin{center}
\epsfig{figure=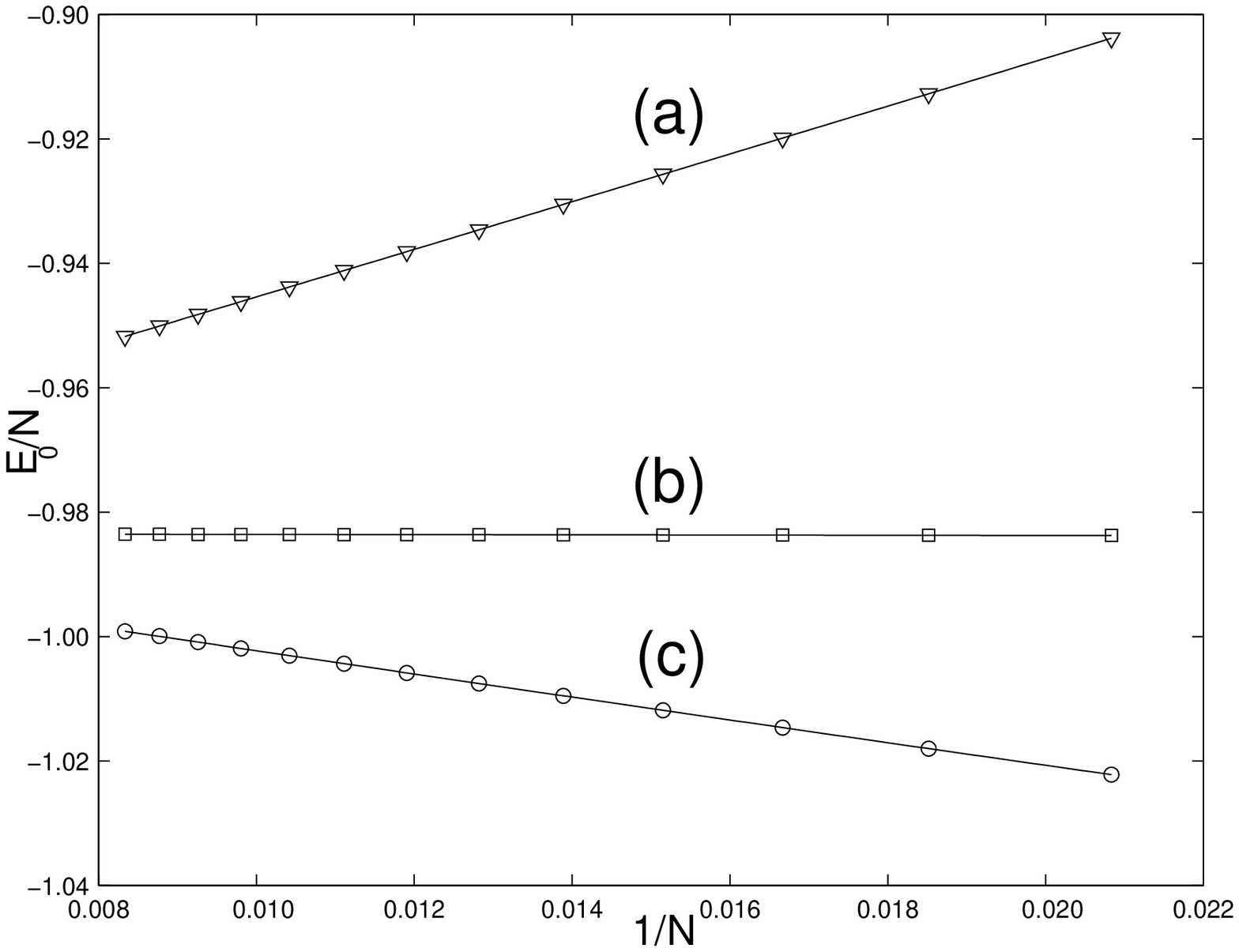,width=14cm}
\end{center}
\vspace*{1.0cm}
\centerline{Fig. 22}
\label{22quadfit}
\end{figure}

\newpage
\clearpage

\begin{figure}
\begin{center}
\epsfig{figure=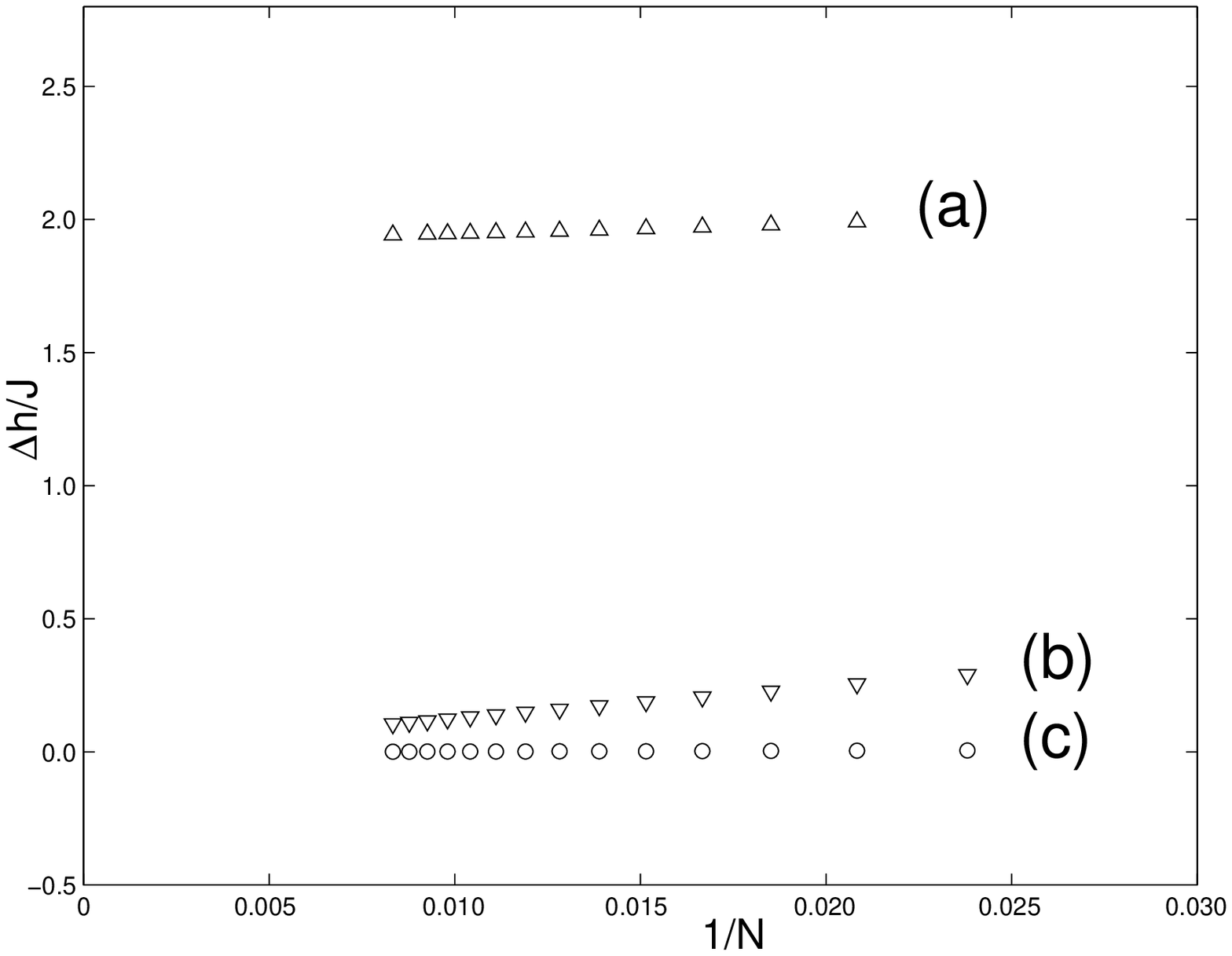,width=14cm}
\end{center}
\vspace*{1.0cm}
\centerline{Fig. 23}
\label{23platgap}
\end{figure}

\newpage
\clearpage

\begin{figure}
\begin{center}
\epsfig{figure=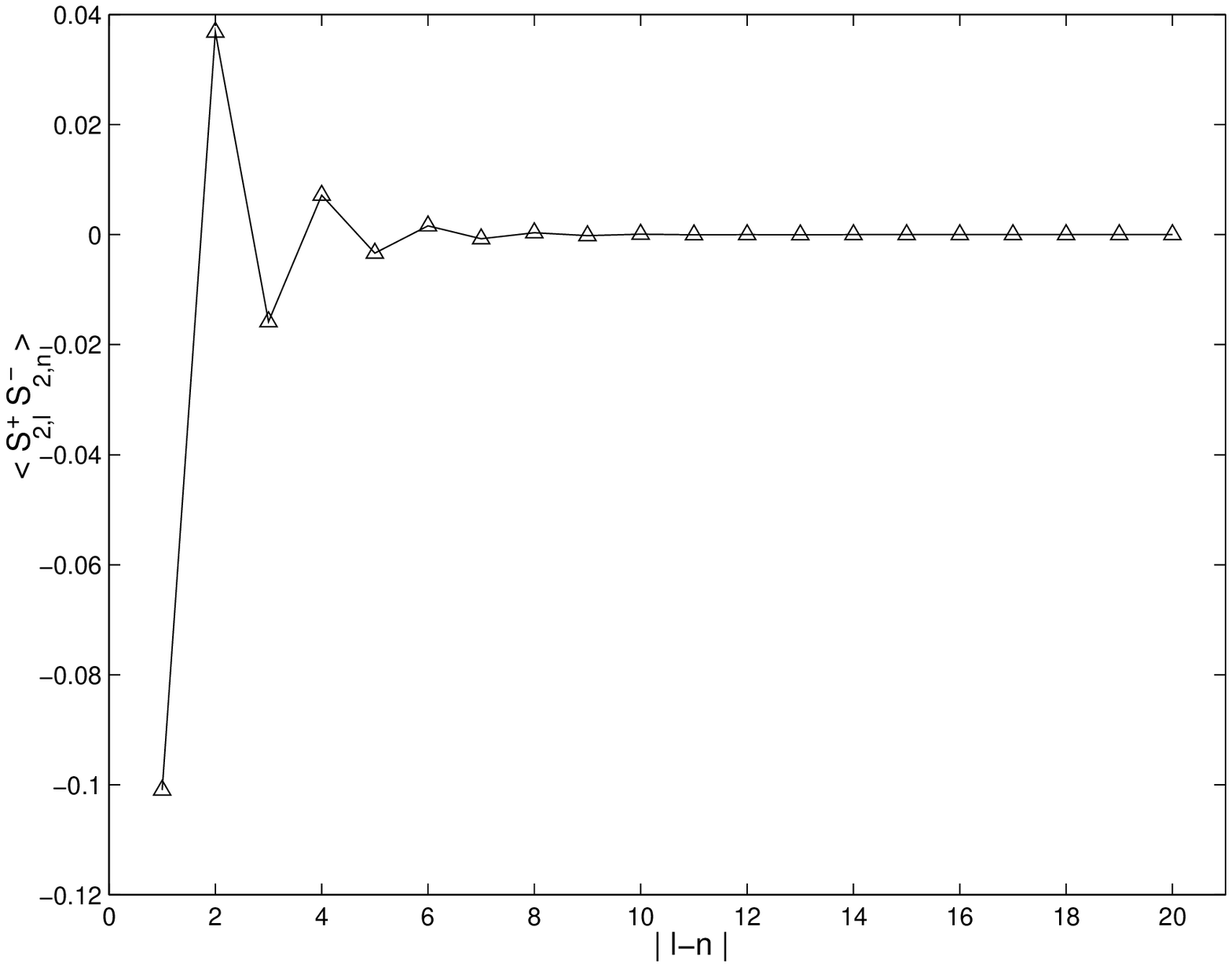,width=14cm}
\end{center}
\vspace*{1.0cm}
\centerline{Fig. 24}
\label{24corr22pm}
\end{figure}

\newpage
\clearpage

\begin{figure}
\begin{center}
\epsfig{figure=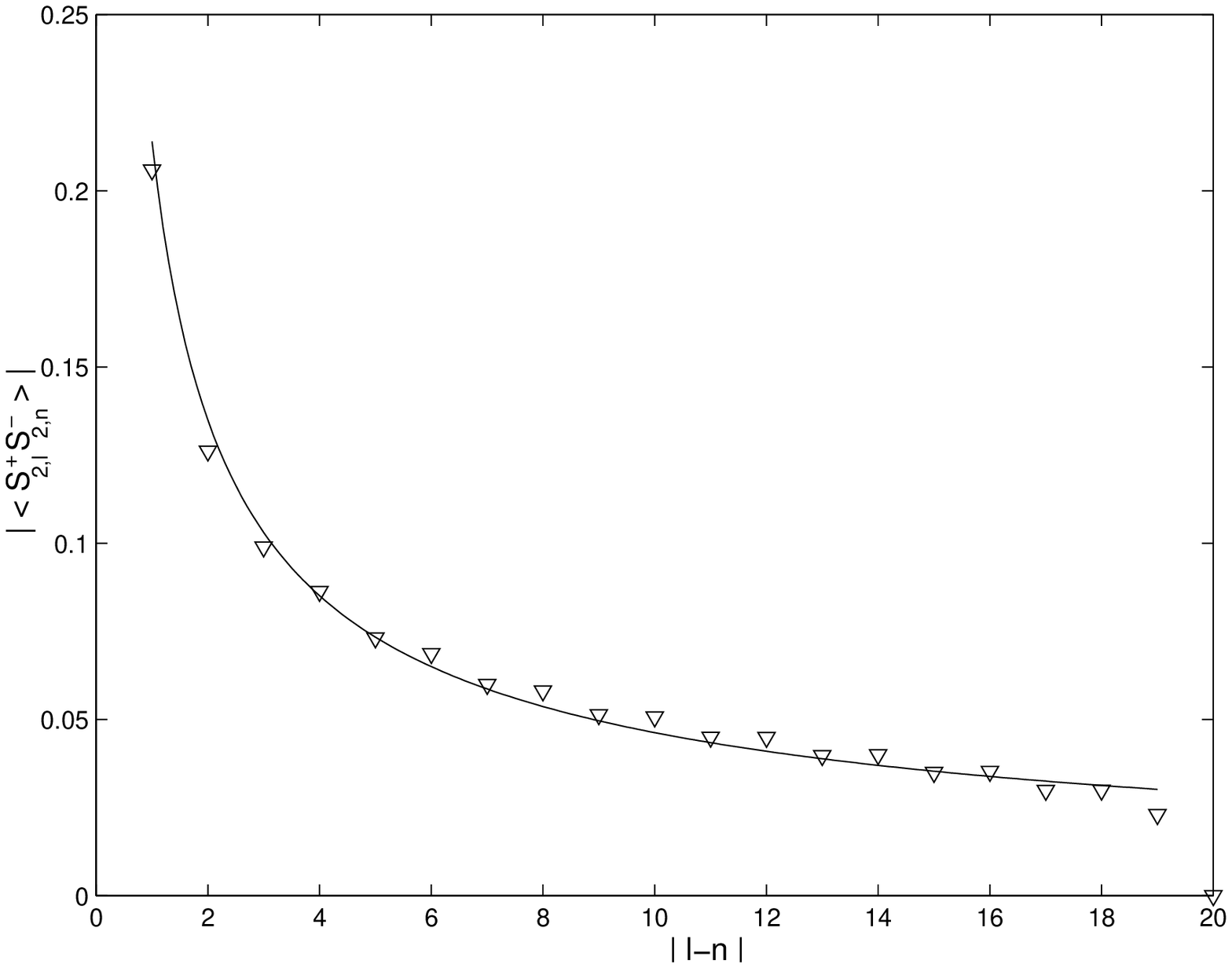,width=14cm}
\end{center}
\vspace*{1.0cm}
\centerline{Fig. 25}
\label{25corr12pm}
\end{figure}

\newpage
\clearpage

\begin{figure}
\begin{center}
\epsfig{figure=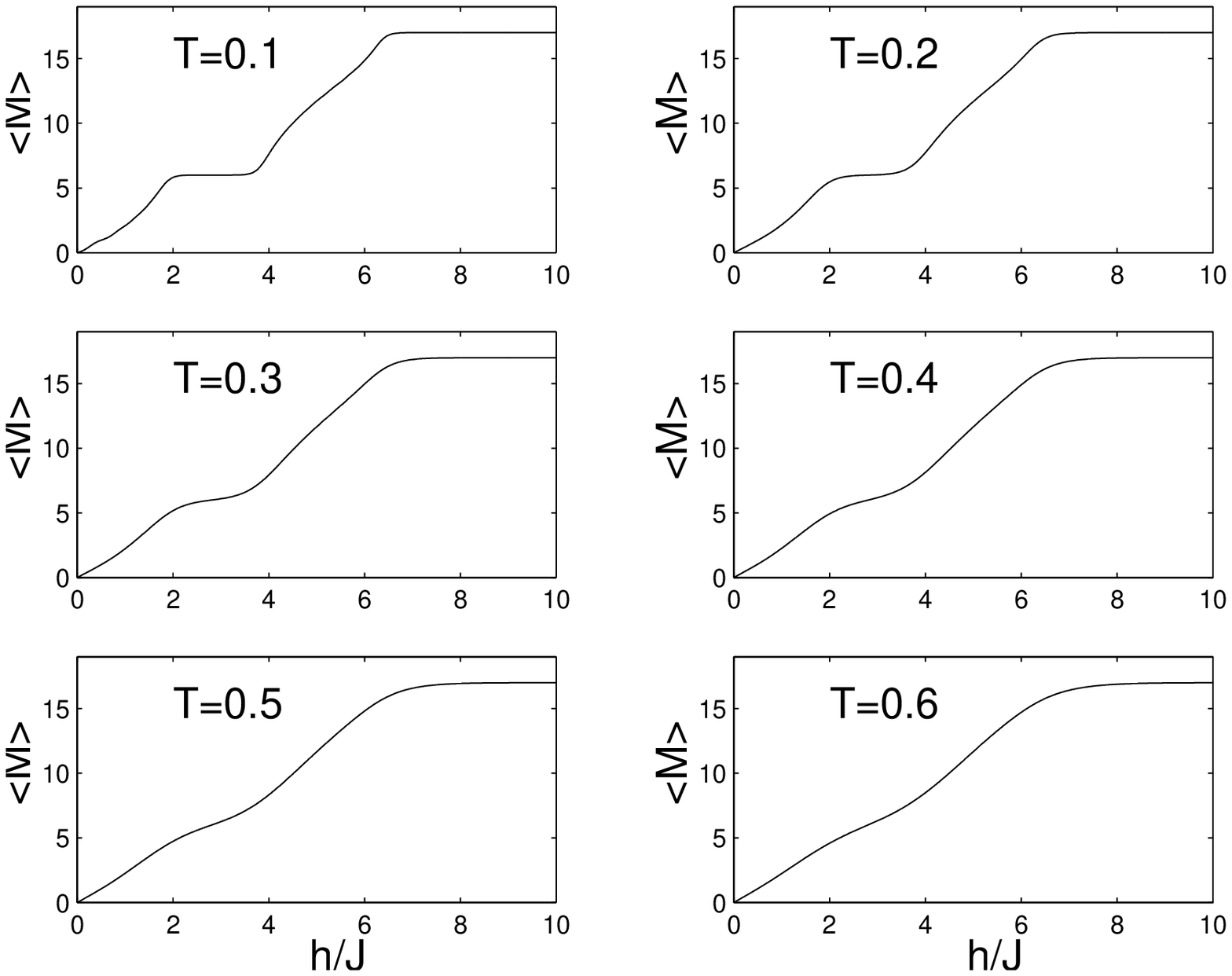,width=14cm}
\end{center}
\vspace*{1.0cm}
\centerline{Fig. 26}
\label{26mo36}
\end{figure}

\newpage
\clearpage

\begin{figure}
\begin{center}
\epsfig{figure=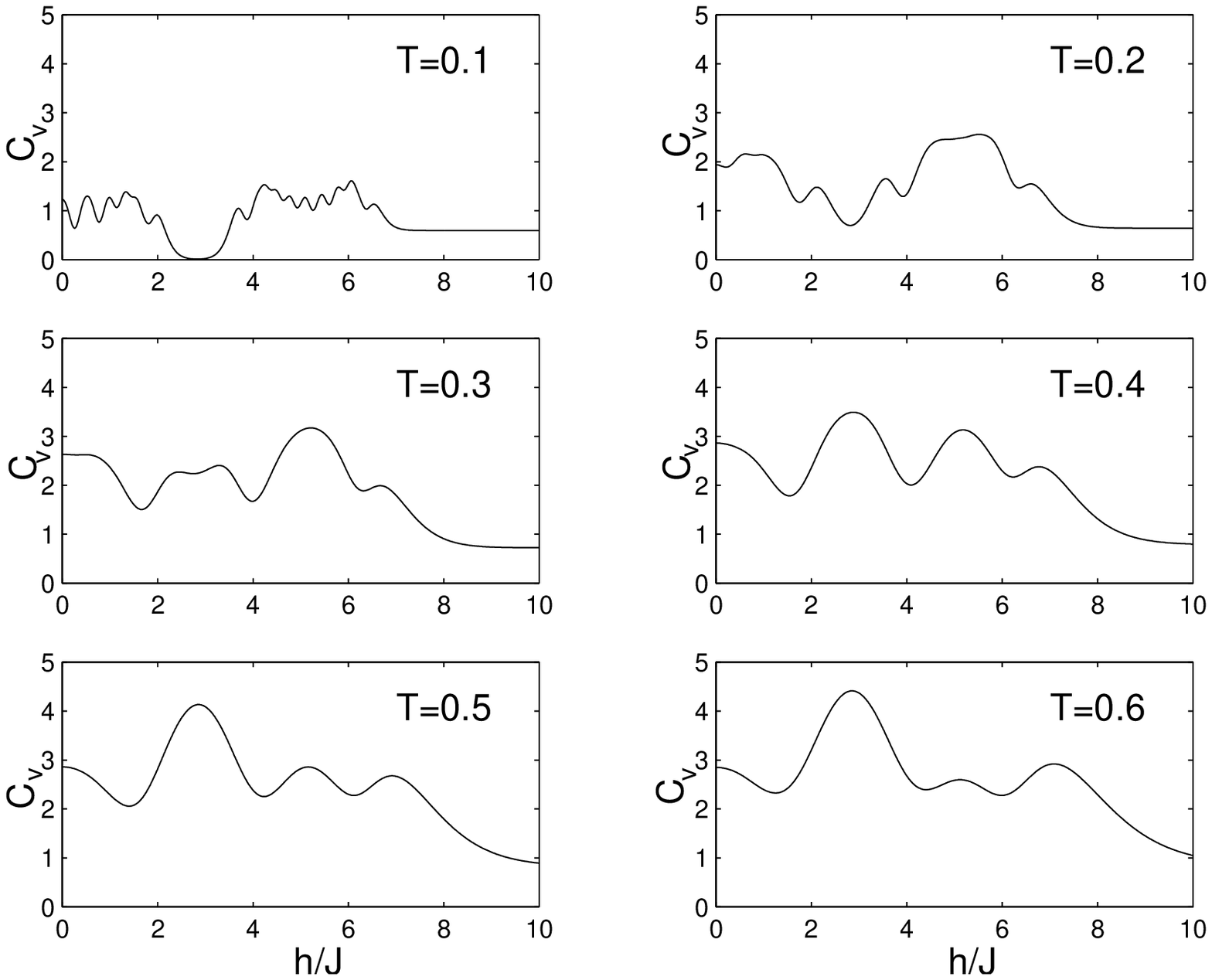,width=14cm}
\end{center}
\vspace*{1.0cm}
\centerline{Fig. 27}
\label{27cvo36}
\end{figure}

\newpage
\clearpage

\begin{figure}
\begin{center}
\epsfig{figure=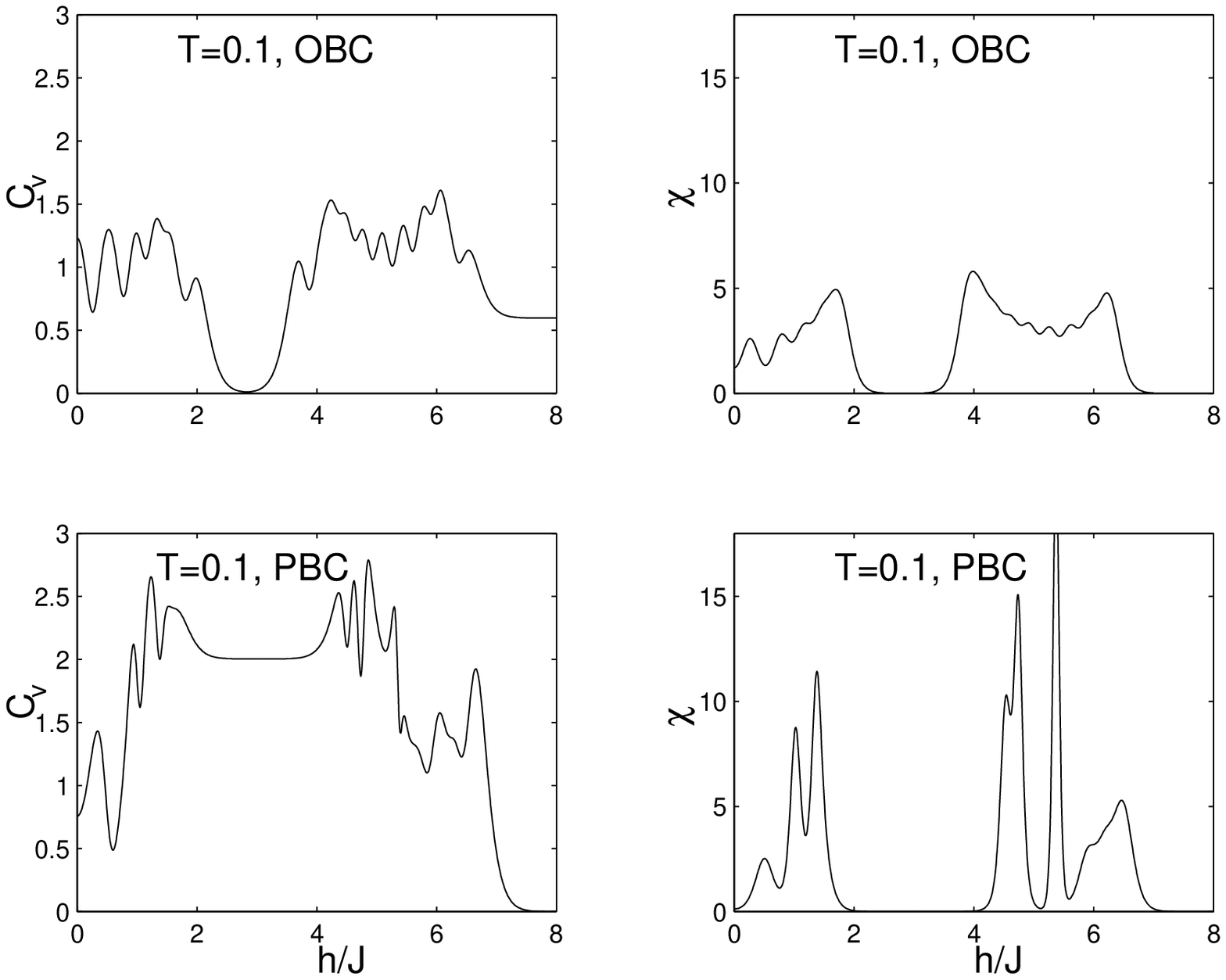,width=14cm}
\end{center}
\vspace*{1.0cm}
\centerline{Fig. 28}
\label{28cvchi36}
\end{figure}

\newpage
\clearpage

\begin{figure}
\begin{center}
\epsfig{figure=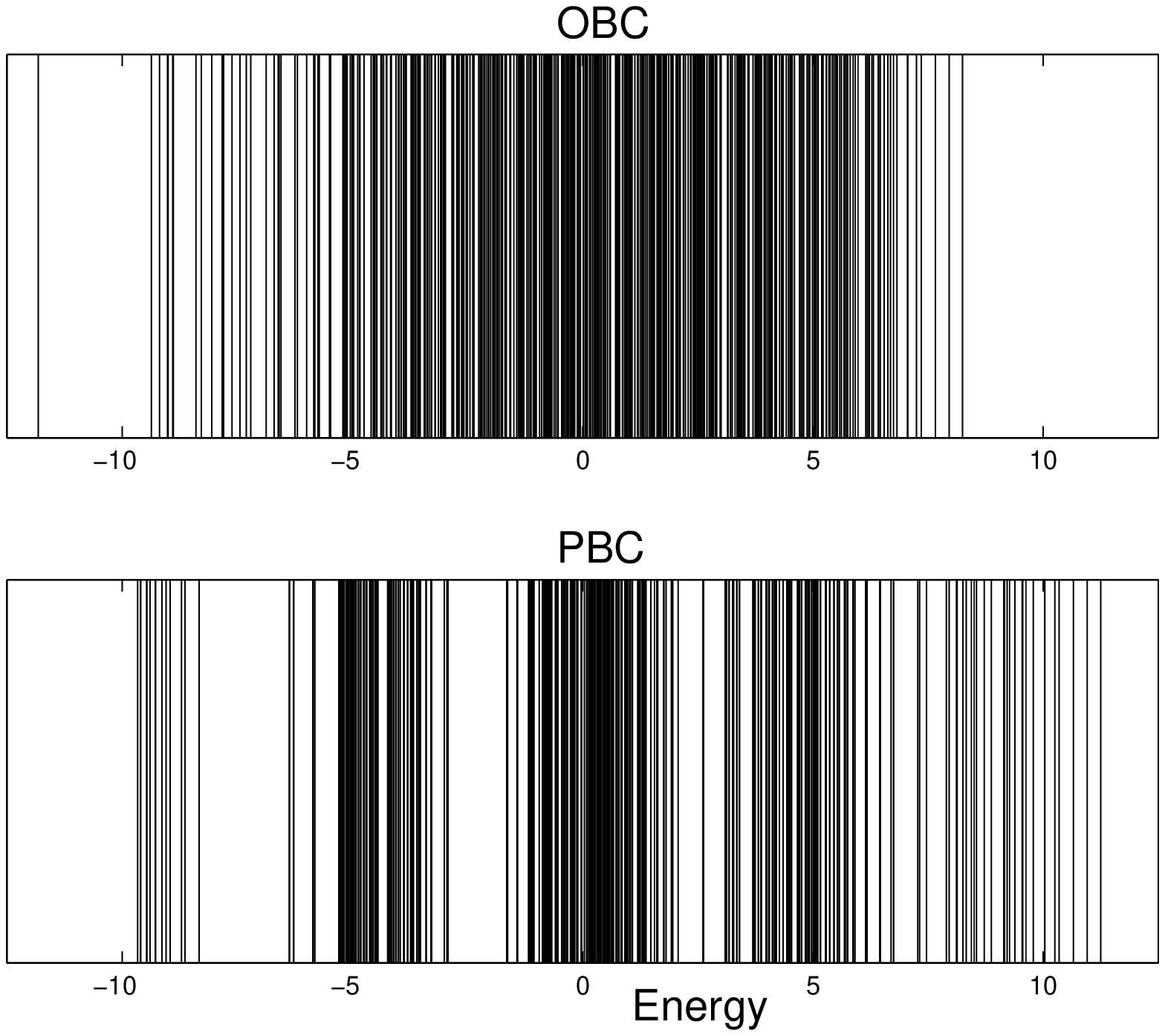,width=14cm}
\end{center}
\vspace*{1.0cm}
\centerline{Fig. 29}
\label{29nrg12}
\end{figure}


\begin{thebibliography}{99}

\bibitem{landau} L. D. Landau and E. M. Lifshitz, {\it Qauntum Mechanics: 
Non-Relativistic Theory}, 3rd Ed. (Pergamon, 1977).

\bibitem{mattis} D. C. Mattis, {\it The Theory of Magnetism}, vol. 1 
(Springer-Verlag, Berlin, 1987); R. M. White, {\it Quantum Theory of 
Magnetism}, 2nd Ed. (Springer-Verlag, Berlin, 1983).

\bibitem{sinha} B. Sinha and S. Ramasesha, Phys. Rev. B {\bf 48}, 16410 (1993).

\bibitem{bray} J. N. Bray, L. V. Interrante, J. S. Jacobs and J. C. Bonner, in
{\it Extended Linear Chain Compounds}, Vol. III, edited by J. S. Miller 
(Plenum, New York, 1982).

\bibitem{rado} G. T. Rado and H. Suhl eds., {\it Magnetism}, (Academic Press,
New York, 1963).

\bibitem{klein} Z. G. Soos and S. Ramasesha, in {\it Valence Bond Theory and 
Chemical Structure}, edited by D. J. Klein and N. Trinajstic (Elsevier, 
Amsterdam, 1990) p. 81.

\bibitem{raghu} C. Raghu, I. Rudra, D. Sen and S. Ramasesha, cond-mat/0101274,
to appear in Phys. Rev. B (2001).

\bibitem{rettrup} S. Rettrup, J. Comput. Phys. {\bf 45}, 100 (1982).

\bibitem{srzgs} S. Ramasesha and Z. G. Soos, J. Chem. Phys. {\bf 98}, 4015 
(1993).

\bibitem{deneef} T. De Neef, Phys. Lett. A {\bf 47}, 51 (1974).

\bibitem{blote} H. J. Blote, Physica B, {\bf 93}, 93 (1978).

\bibitem{botet} R. Botet and R. Jullien, Phys. Rev. B {\bf 27}, 613 (1983); R.
Botet, R. Jullien and M. Kolb, {\it ibid.}, {\bf 28}, 3914 (1983).

\bibitem{bonner} J. B. Parkinson and J. C. Bonner, Phys. Rev. B {\bf 32},
4703 (1985).

\bibitem{moreo} A. Moreo, Phys. Rev. B {\bf 35}, 8562 (1987).

\bibitem{taka1} M. Takahashi, Phys. Rev. Lett. {\bf 62}, 2313 (1989); Phys. 
Rev. B {\bf 48}, 311 (1993).

\bibitem{goli} O. Golinelli, Th. Jolicoeur and R. Racaze, Phys. Rev. B {\bf 
50}, 3037 (1994).

\bibitem{chud} E. M. Chudnovsky, Science {\bf 274}, 938 (1996).

\bibitem{wern} W. Wernsdorfer and R. Sessoli, Science {\bf 284}, 133 (1999).

\bibitem{chio} I. Chiorescu, W. Wernsdorfer, A. M\"uller, H. B\"ogge and B. 
Barbara, Phys. Rev. Lett. {\bf 84}, 15 (2000).

\bibitem{qtm} {\it Quantum Tunneling of Magnetization - QTM'94}, edited by L.
Gunther and B. Barbara, NATO ASI Ser. E, Vol. 301 (Kluwer, Dordrecht, 1995);
J. R. Friedman, M. P. Sarachik, J. Tejada and R. Ziolo, Phys. Rev. Lett. {\bf
76}, 3830 (1996); L. Thomas, F. Lionti, R. Ballou, D. Gatteschi, R. Sessoli
and B. Barbara, Nature {\bf 383}, 145 (1996) and references therein;
W. Wernsdorfer and R. Sessoli, Science {\bf 284}, 133 (1999).

\bibitem{mn12ac} D. N. Hendrickson, G. Christou, E. A. Schmitt, B. Libby, J. 
S. Baskin, S. Wang, H. L. Tsai, J. B. Vincent, W. P. D. Boyd, J. C. Haffman,
K. Folting, Q. Li, W. E. Streib, J. Am. Chem. Soc. {\bf 114}, 2455 (1992) and
references therein.

\bibitem{gat1} R. Sessoli {\it et al.}, J. Am. Chem. Soc. {\bf 115}, 1804
(1993).

\bibitem{egap} M. Hennion, L. Pardi, I. Mirebeau, E. Surad, R. Sessoli and
A. Caneschi, Phys. Rev. B {\bf 56}, 8819 (1997); A. A. Mukhin, V. D. Travkin,
A. K. Zvezdin, S. P. Lebedev, A. Caneschi and D. Gatteschi, Europhys. Lett.
{\bf 44}, 778 (1998).

\bibitem{mn12spnden} P. A. Reynolds, E. P. Gilbert, B. N. Figgis, Inorg. Chem.
{\bf 35}, 545 (1996).

\bibitem{fe8butterfly} S. M. Gorun, S. J. Lippard, Inorg. Chem. {\bf 27}, 149
(1988); W. H. Armstrong, M. E. Roth, S. J. Lippard, J. Am. Chem. Soc. {\bf 
109}, 6318 (1987).

\bibitem{delfs} C. D. Delfs, D. Gatteschi, L. Pardi, R. Sessoli, K. Weighardt
and D. Hanke, Inorg. Chem. {\bf 32}, 3099 (1993).

\bibitem{fe8jval} I. Tupitsyn and B. Barbara, cond-mat/0002180, to appear in 
{\it Magnetoscience - From Molecules to Materials}, edited by J. Miller and 
M. Drillon (Wiley VCH Verlag GmbH, 2000).

\bibitem{fe8spnden} Y. Pontillon {\it et al.}, J. Am. Chem. Soc. {\bf 121}, 
5342 (1999).

\bibitem{hald1} F. D. M. Haldane, Phys. Lett. {\bf 93A}, 464 (1983); Phys.
Rev. Lett. {\bf 50}, 1153 (1983).

\bibitem{buye} W. J. L. Buyers, R. M. Morra, R. L. Armstrong, M. J. Hogan,
P. Gerlach and K. Hirakawa, Phys. Rev. Lett. {\bf 56}, 371 (1986); J. P. 
Renard, M. Verdaguer, L. P. Regnault, W. A. C. Erkelens, J. Rossat-Mignod
and W. G. Stirling, Europhys. Lett. {\bf 3}, 945 (1987); S. Ma, C. Broholm,
D. H. Reich, B. J. Sternlieb and R. W. Erwin, Phys. Rev. Lett. {\bf 69}, 
3571 (1992).

\bibitem{dago} E. Dagotto and T. M. Rice, Science {\bf 271}, 618 (1996).

\bibitem{eccl} R. S. Eccleston, T. Barnes, J. Brody and J. W. Johnson, Phys. 
Rev. Lett. {\bf 73}, 2626 (1994).

\bibitem{azum} M. Azuma, Z. Hiroi, M. Takano, K. Ishida and Y. Kitaoka, Phys. 
Rev. Lett. {\bf 73}, 3463 (1994).

\bibitem{chab1} G. Chaboussant, P. A. Crowell, L. P. Levy, O. Piovesana, A.
Madouri and D. Mailly, Phys. Rev. B {\bf 55}, 3046 (1997).

\bibitem{hase} M. Hase, I. Terasaki and K. Uchinokura, Phys. Rev. Lett. {\bf 
70}, 3651 (1993); M. Nishi, O. Fujita and J. Akimitsu, Phys. Rev. B {\bf 50}, 
6508 (1994); G. Castilla, S. Chakravarty and V. J. Emery, Phys. Rev. Lett. 
{\bf 75}, 1823 (1995).

\bibitem{chab2} G. Chaboussant, Y. Fagot-Revurat, M.-H. Julien, M. E. Hanson,
C. Berthier, M. Horvatic, L. P. Levy and O. Piovesana, Phys. Rev. Lett. {\bf
80}, 2713 (1998).

\bibitem{oshi} M. Oshikawa, M. Yamanaka and I. Affleck, Phys. Rev. Lett. {\bf
78}, 1984 (1997).

\bibitem{cabr1} D. C. Cabra, A. Honecker and P. Pujol, Phys. Rev. Lett. {\bf 
79}, 5126 (1997).

\bibitem{cabr2} D. C. Cabra, A. Honecker and P. Pujol, Phys. Rev. B {\bf 58},
6241 (1998). See also the web site
http://thew02.physik.uni-bonn.de/$\sim$honecker/roc.html

\bibitem{hida} K. Hida, J. Phys. Soc. Jpn. {\bf 63}, 2359 (1994).

\bibitem{kura} T. Kuramoto, J. Phys. Soc. Jpn. {\bf 67}, 1762 (1998).

\bibitem{tots} K. Totsuka, Phys. Lett. A {\bf 228}, 103 (1997); Phys. Rev. B 
{\bf 57}, 3454 (1998).

\bibitem{tone1} T. Tonegawa, T. Nishida and M. Kaburagi, cond-mat/9712297.

\bibitem{saka} K. Sakai and M. Takahashi, Phys. Rev. B {\bf 57}, R3201 (1998); 
Phys. Rev. B {\bf 57}, R8091 (1998).

\bibitem{chit} R. Chitra and T. Giamarchi, Phys. Rev. B {\bf 55}, 5816 (1997).

\bibitem{schu} H. J. Schulz, G. Cuniberti and P. Pieri, in {\it Field 
Theories for Low Dimensional Condensed Matter Systems}, edited by G. Morandi,
A. Tagliacozzo and P. Sodano (Springer, Berlin, 2000); H. J.
Schulz, in {\it Proceedings of Les Houches Summer School LXI}, edited by E. 
Akkermans, G. Montambaux, J. Pichard and J. Zinn-Justin (Elsevier, Amsterdam,
1995).

\bibitem{affl1} I. Affleck, in {\it Fields, Strings and Critical Phenomena}, 
edited by E. Brezin and J. Zinn-Justin (North-Holland, Amsterdam, 1989); I. 
Affleck and F. D. M. Haldane, Phys. Rev. B {\bf 36}, 5291 (1987).

\bibitem{rao2} S. Rao and D. Sen, J. Phys. Condens. Matter {\bf 9}, 1831 
(1997).

\bibitem{fn} We use the word `phase' only for convenience to distinguish 
between regions with different modulations of the two-spin correlation 
function. Our model actually has no phase transition from Neel to spiral even 
at zero temperature.

\bibitem{affl2} I. Affleck, Nucl. Phys. B {\bf 265}, 409 (1986).

\bibitem{kato} Y. Kato and A. Tanaka, J. Phys. Soc. Jpn. {\bf 63}, 1277 
(1994); K. Totsuka, Y. Nishiyama, N. Hatano, and M. Suzuki, J. Phys. Condens.
Matter {\bf 7}, 4895 (1995).

\bibitem{rao1} S. Rao and D. Sen, Nucl. Phys. B {\bf 424}, 547 (1994).

\bibitem{alle} D. Allen and D. Senechal, Phys. Rev. B {\bf 51}, 6394 (1995). 

\bibitem{gogo} A. O. Gogolin, A. A. Nersesyan and A. M. Tsvelik, {\it
Bosonization and Strongly Correlated Systems} (Cambridge University Press,
Cambridge, 1998).

\bibitem{shan1} R. Shankar, Lectures given at the BCSPIN School, Kathmandu,
1991, in {\it Condensed Matter and Particle Physics}, edited by Y. Lu, J. 
Pati and Q. Shafi (World Scientific, Singapore, 1993).

\bibitem{vond} J. von Delft and H. Schoeller, Ann. der Physik {\bf 4}, 225
(1998).

\bibitem{bouc} J. P. Boucher and L. P. Regnault, J. de Phys. I {\bf 6}, 1939
(1996).

\bibitem{wilson} K. G. Wilson, Rev. Mod. Phys. {\bf 47}, 773 (1975).

\bibitem{oldrg} J. W. Bray and S. T. Chui, Phys. Rev. B {\bf 19}, 4876 (1979);
J. E. Hirsh, Phys. Rev. B {\bf 22}, 5259 (1980).

\bibitem{white1} S. R. White and R. M. Noack, Phys. Rev. Lett. {\bf 68}, 3487
(1992).

\bibitem{white2} S. R. White, Phys. Rev. Lett. {\bf 69}, 2863 (1992); Phys. 
Rev. B {\bf 48}, 10345 (1993).

\bibitem{mcweeny} R. McWeeny and B. T. Sutcliffe, {\it Methods of Molecular
Quantum Mechanics} (Academic Press, London, 1969).

\bibitem{mx} Y. Anusooya, S. K. Pati and S. Ramasesha, J. Chem. Phys. {\bf 
106}, 1 (1997).

\bibitem{white3} S. R. White, Phys. Rev. Lett. {\bf 45}, 5752 (1992).

\bibitem{skp1} R. Chitra, S. K. Pati, H. R. Krishnamurthy, D. Sen and S.
Ramasesha, Phys. Rev. B {\bf 52}, 6581 (1995); S. K. Pati, R. Chitra, D. Sen, 
H. R. Krishnamurthy and S. Ramasesha, Europhys. Lett. {\bf 33}, 707 (1996).

\bibitem{white4} S. R. White and D. A. Huse, Phys. Rev. B {\bf 48}, 3844
(1993); E. S. Sorensen and I. Affleck, Phys. Rev. Lett. {\bf 71}, 1633 (1993).

\bibitem{white5} E. S. Sorensen and I. Affleck, Phys. Rev. B {\bf 49}, 13235
(1994); Phys. Rev. B {\bf 49}, 15771 (1994).

\bibitem{suzu} U. Schollw\"ock and Th. Jolicoeur, Europhys. Lett. {\bf 30},
493 (1995); Y. Nishiyama, K. Totsuka, N. Hatano and M. Suzuki, J. Phys. Soc.
Jpn. {\bf 64}, 414 (1995).

\bibitem{scho} U. Schollw\"ock, Th. Jolicoeur, and T. Garel, Phys. Rev. B {\bf 
53}, 3304 (1996).

\bibitem{burs} R. J. Bursill, T. Xiang and G. A. Gehring, J. Phys. A {\bf 
28}, 2109 (1994); R. J. Bursill, G. A. Gehring, D. J. Farnell, J. B. Parkinson,
T. Xiang and C. Zeng, J. Phys. C {\bf 7}, 8605 (1995).

\bibitem{chen} S. R. White, R. M. Noack and D. J. Scalapino, Phys. Rev. Lett.
{\bf 73}, 886 (1994); M. Azzouz, L. Chen and S. Moukouri, Phys. Rev. B {\bf 
50}, 6223 (1994).

\bibitem{naru} T. Narushima, T. Nakamura and S. Takada, J. Phys. Soc. Jpn.
{\bf 64}, 4322 (1995); U. Schollw\"ock and D. Ko, Phys. Rev. B {\bf 53}, 240
(1996).

\bibitem{qin} S. J. Qin, T. K. Ng and Z. B. Su, Phys. Rev. B {\bf 52}, 12844 
(1995).

\bibitem{pang} H. B. Pang, H. Akhlaghpour and M. Jarrel, Phys. Rev. B {\bf 
53}, 5086 (1996).

\bibitem{hallberg} K. A. Hallberg, Phys. Rev. B {\bf 52}, R9827 (1995).

\bibitem{skp2} S. K. Pati, S. Ramasesha, Z. Shuai and J. L. Br\'edas, Phys. 
Rev. B {\bf 59}, 14827 (1999).

\bibitem{caron1} L. J. Caron and S. Moukouri, Phys. Rev. Lett. {\bf 77}, 4640 
(1996).

\bibitem{skp3} S. K. Pati, S. Ramasesha and D. Sen, Phys. Rev. B {\bf 55}, 
8894 (1997); J. Phys. Condens. Matt. {\bf 9}, 8707 (1997). 

\bibitem{caron2} L. G. Caron and S. Moukouri, Phys. Rev. Lett. {\bf 76}, 4050 
(1996).

\bibitem{nish1} T. Nishino and K. Okunishi, J. Phys. Soc. Jpn. {\bf 64}, 4084
(1995).

\bibitem{nish2} T. Nishino and K. Okunishi, J. Phys. Soc. Jpn. {\bf 65}, 891 
(1996).

\bibitem{tone2} T. Tonegawa, M. Kaburagi, N. Ichikawa, and I. Harada, J. Phys. 
Soc. Jpn. {\bf 61}, 2890 (1992).

\bibitem{rama} S. Ramasesha and Z. G. Soos, Solid State Commun. {\bf 46}, 509 
(1983).

\bibitem{rao3} S. Rao and D. Sen, Phys. Rev. B {\bf 48}, 12763 (1993).

\bibitem{suth} B. Sutherland, Phys. Rev. B {\bf 12}, 3795 (1975).

\bibitem{stein76} For a review, see M. Steiner, J. Villain and C. G. Windsor,
{\it Adv. Phys.} {\bf 25}, 88 (1976).

\bibitem{kahn1} O. Kahn, {\it Structure and Bonding} {\bf 68}, 89 (Berlin).

\bibitem{kahn2} O. Kahn, Y. Pei, M. Verdaguer, J. P. Renard and J. Sletten, J.
Am. Chem. Soc. {\bf 110}, 782 (1988); P. Van Koningsbruggen, O. Kahn, K. 
Nakatani, Y. Pei, J. P. Renard, M. Drillon and P. Leggol, Inorg. Chem. {\bf 
29}, 3325 (1990).

\bibitem{kahn3} O. Kahn, Adv. Inorg. Chem. {\bf 43}, 179 (1995).

\bibitem{hall} K. A. Hallberg, P. Horsch and G. Martinez, Phys. Rev. B {\bf 
52}, R719 (1995); R. J. Bursill, T. Xiang and G. A. Gehring, J. Phys. A, {\bf
28}, 2109 (1994); Y. Kato and A. Tanaka, J. Phys. Soc. Jpn. {\bf 63}, 1277 
(1994).

\bibitem{kahn4} O. Kahn, {\it Molecular Magnetism} (VCH, New York, 1993).

\bibitem{taka2} M. Takahashi, Phys. Rev. B {\bf 36}, 3791 (1987).

\bibitem{tand} K. Tandon, S. Lal, S. K. Pati, S. Ramasesha and D. Sen, Phys. 
Rev. B {\bf 59}, 396 (1999).

\bibitem{kawa} K. Kawano and M. Takahashi, J. Phys. Soc. Jpn. {\bf 66}, 4001
(1997).

\bibitem{rudra} I. Rudra, S. Ramasesha and D. Sen, Phys. Rev. B {\bf 64},
014408 (2001).

\end{thebibliography}
\end{document}